\newcommand{\hmpc}{\mbox{ } {h}^{-1}~\rm{ Mpc}}
\newcommand{\hgpc}{\mbox{ } {h}^{-1}~\rm{ Gpc}}
\newcommand{\hkpc}{\mbox{ } {h}^{-1}~\rm{ kpc}}
\newcommand{\hmsol}{\mbox{ } {h}^{-1}~{M}_{\odot}}
\newcommand{\hsmsol}{\mbox{ } {h}_{70}^{-1}~{M}_{\odot}}
\newcommand{\msol}{\mbox{ } {M}_{\odot}}

\newcommand{\mpc}{{\rm Mpc}}

\newcommand{\mg}{\mbox{ }{\rm \mu G}}
\newcommand{\ghz}{\mbox{ }{\rm GHz}}
\newcommand{\mhz}{\mbox{ }{\rm MHz}}
\newcommand{\aveb}{\langle B \rangle}
\newcommand{\pmvir}{P_{\rm 1.4 GHz}-M_v}
\newcommand{\whz}{~{\rm W}~{\rm Hz}^{-1}}

\newcommand{\eqref}[1]{Equation (#1)}
\newcommand{\figref}[1]{Figure #1}
\newcommand{\tabref}[1]{Table #1}

\documentclass[floatfix]{emulateapj}
\usepackage{epsfig,subfigure}

\bibpunct[; ]{(}{)}{;}{a}{}{,}

\begin{document}
\title{A first estimate of radio halo statistics from large-scale cosmological simulation}

\author{P.~M.~Sutter} \email{psutter2@illinois.edu}
\affil{Department of Physics, University of Illinois at Urbana-Champaign, Urbana, IL 61801\\
UPMC Univ Paris 06, UMR7095, Institut d'Astrophysique de Paris, F-75014, Paris, France \\
CNRS, UMR7095, Institut d'Astrophysique de Paris, F-75014, Paris, France
\\
Center for Cosmology and Astro-Particle Physics, Ohio State University, Columbus, OH 43210\\
}

\author{P.~M.~Ricker} \email{pmricker@illinois.edu}
\affil{Department of Astronomy,
       University of Illinois at Urbana-Champaign,
             Urbana, IL 61801\\
		National Center for Supercomputing Applications,
      University of Illinois at Urbana-Champaign,
            Urbana, IL 61801}
            
\begin{abstract} 
We present a first estimate based on a cosmological gasdynamics simulation of
galaxy cluster radio halo counts to be expected in forthcoming low-frequency
radio surveys. Our estimate is based on a FLASH simulation of the $\Lambda$CDM
model for which we have assigned radio power to clusters via a model that
relates radio emissivity to cluster magnetic field strength, intracluster
turbulence, and density. We vary several free parameters of this model and find
that radio halo number counts 
vary by up to a factor of two for average magnetic fields ranging from 
0.2 to 3.1~$\mu$G.
However, we
predict significantly fewer low-frequency radio halos than expected from
previous semi-analytic estimates, although this discrepancy could be
explained by frequency-dependent radio halo probabilities as predicted in 
reacceleration models. 
We find that upcoming surveys will have
difficulty in 
distinguishing models because of large uncertainties and low number
counts. Additionally, according to our modeling 
we find that expected number counts can be degenerate
with both reacceleration and hadronic secondary models of cosmic ray
generation. We find that relations between radio power and mass and X-ray
luminosity may be used to distinguish models, and by building mock radio sky
maps we demonstrate that surveys such as LOFAR may have sufficient resolution
and sensitivity to break this model degeneracy by imaging many individual
clusters.  

\end{abstract} 


\maketitle

\section{Introduction}
\label{sec:rh_introduction}

Although diffuse radio halos were discovered in clusters of galaxies more than
50 years ago~\citep{LARGE1959}, complete statistical information about them has
   only been forthcoming within the past decade, owing to their rarity, steep
spectra, and low surface brightnesses. Radio surveys using the Very Large Array
~\citep[VLA;][]{GIOVANNINI1999,Cohen2007,Giovannini2009a}, the Westerbork Synthesis
Radio Telescope~\citep[WSRT;][]{Kempner2001}, and the Giant Metrewave Radio
Telescope~\citep[GMRT;][]{Venturi2009} have detected $\sim 30$ radio halos at
redshifts up to $z \sim 0.5$, along with a variety of smaller-scale radio
features in clusters~\citep{Kempner2004}.  Only about $1/3$ of massive
($>10^{15} \msol$) clusters are known to host radio halos, and the halos
themselves are not associated with any particular member galaxy, but rather
dispersed throughout the intracluster medium~\citep[ICM;][]{Feretti2004}. For
clusters that do host halos, strong correlations are seen between radio power
and X-ray luminosity~\citep{Liang2000,Bacchi2003,Cassano2006,Brunetti2007},
halo mass~\citep{Cassano2006}, and gas velocity dispersion~\citep{Cassano2008}.
Also, observations indicate a strong connection between the presence of a halo
and morphological evidence for recent 
mergers~\citep{Buote2001, Brunetti2009, Cassano2010c},
although some exceptions do exist~\citep{Russell2011}. Indeed, recent
simulations of merging clusters suggest that the fraction of turbulent clusters
is roughly equal to the fraction of clusters hosting radio
halos~\citep{Vazza2010a}.

The proximate cause of diffuse radio halos is synchrotron emission
by high-energy electrons in galaxy cluster magnetic fields, but the means of
generating and accelerating these electrons remains an open question, since
these electrons have relatively short ($\sim 0.1$~Gyr) lifetimes.
\citet{Dennison1980} proposed that cosmic-ray (CR) electrons are produced as
secondary particles by collisions of $>1$~GeV CR protons with
ambient thermal ICM protons. The CR protons can be accelerated by shocks and
 diffuse throughout the cluster; because of their larger 
mass, they have much longer synchrotron lifetimes than the electrons. 
 This naturally explains the
diffuse, cluster-wide properties of radio
halos~\citep{Pfrommer2008,Blasi1999a}.~\citet{Kushnir2009} 
and~\citet{Keshet2010b} discuss a way in
which the correlation between radio and X-ray surface brightness can be
explained by hadronic secondary-type models, although this approach requires an
extremely strong magnetic field ($\sim 5~\mg$ at radii $\sim 1$~Mpc for
$z=0.2-0.4$), which conflicts with some estimates of cluster 
magnetic fields~\citep{Bonafede2011}.  
Gamma-ray observations place limits on the abundance of
high-energy protons, since 
in addition to producing charged pions that decay into the secondary 
electrons, the proton-proton collisions produce neutral pions, which 
decay into gamma-ray photons~\citep{Blasi1999,Wolfe2008}.
These observations indicate that hadronic CRs may
contribute at most $5$-$10$\% of the total pressure support in
clusters~\citep{Ackermann2010}.  
While
gamma-ray observations have not strictly ruled out this model, recent estimates
indicate a tension between the magnetic field strengths required for this
scenario and those observed in clusters~\citep{Jeltema2011}.
In addition, the hadronic secondary
 model has difficulty
explaining the 
shape of the radio spectrum within the Coma 
cluster~\citep{Donnert2010b,Donnert2010c}.  Also, the ability of the
high-energy protons to stream away from their sources can have significant
implications for the resulting radio emission~\citep{Enßlin2011}.  

A promising explanation for the acceleration of the CR electrons
is second-order Fermi acceleration by intracluster
turbulence~\citep{Schlickeiser1987,Petrosian2001,Brunetti2011}. The CRs
themselves must be injected into the ICM by radio galaxies~\citep{Jaffe1977} or
accelerated by merger shocks.  Since the lifetime of synchrotron-emitting GeV
electrons in the intracluster magnetic field ($\langle B\rangle \sim 1 - 10\
\mu$G) is at most $\sim 0.1$~Gyr~\citep{Kuo2003,Brunetti2009}, these electrons
must be reaccelerated by some process that operates in a more diffuse fashion.
Hence, a local acceleration mechanism is favored.  The existence of some
steep-spectrum low-frequency radio halos may support the reacceleration
model and disfavor hadronic models, 
since the spectral index of the reaccelerated electrons depends 
on the turbulent Mach number~\citep{Brunetti2008}.
Future low-frequency surveys are necessary
to fully test this picture.  Assuming that ICM turbulence locally accelerates
CRs to produce clusterwide radio halos, we expect that the radio
emission should correlate spatially with the turbulent pressure $\langle\rho
v^2\rangle$.  Indeed, this expectation is consistent with
observations using \emph{Chandra} temperature maps~\citep{Govoni2004}.  
However, the 
fraction of CR electrons that are reaccelerated primaries rather than
hadronic secondaries
remains to be determined~\citep{Brunetti2011, Brown2011}.

Within the next decade and a half the number of known radio halos should
increase dramatically owing to the development of sensitive low-frequency radio
surveys.
Examples of the operating and planned instruments include the LOw Frequency
ARray (LOFAR), GMRT, the Karoo Array Telescope (KAT), and the Square Kilometer
Array (SKA). LOFAR, for example, will be sensitive to radio frequencies between
20 and 240 MHz and will be able to detect sources as faint as $0.4-110$ mJy at
   $15-240$ MHz~\citep{ROTTGERING2003,Rottgering2006,Jarvis2007}.  These
characteristics are ideal for detecting and counting radio halos such as the
Coma radio halo ($\sim$640~mJy at 1.4~GHz with a spectral 
index $\sim 1.3$;~\citealt{Deiss1997}) as far away as a redshift of 0.75.  
Moreover, if the CR electron population is dominated by reaccelerated 
primaries, then many more radio halos may be detected at low frequencies owing 
to their steep spectra~\citep{Cassano2008}.

Counts of cluster radio halos, in addition to probing the evolution of cluster
merger activity, also potentially provide an additional means for using
clusters to constrain cosmological parameters. Unlike other methods for using
clusters as cosmological probes that are based on their mass function or gas
fraction, this measure is linked to their ``instantaneous'' formation activity
rather than their time-integrated numbers. In principle its dependence on the
cosmological volume element $dV/dz$ and the growth factor of linear density
fluctuations $D_{+}(z)$ should also be different from and thus complementary to
the more traditional measures.  Additionally, if the CR electrons
responsible for the halos are accelerated by shocks and/or turbulence generated
by mergers, it is reasonable to expect that recently merged clusters would
display the most radio activity.  Thus determining the abundances, spectral
distributions, and other characteristics of radio halos as functions of
redshift could provide information about the evolution of clusters and their
merging activity over time.

In this paper we present results from a numerical simulation of cluster
formation intended to study the form and evolution of the radio halo population
as might be observed in a typical LOFAR survey.  We apply a model of radio
power that is generalized to include both hadronic and reacceleration 
CR-generation mechanisms. We specifically choose this model to be as broad
possible to allow immediate comparisons based on the same simulation. The means
employed in the conversion of cluster density and velocity information into a
simulated LOFAR radio sky are somewhat rudimentary given the uncertainties in
the physics responsible for radio halos and the small scales on which it likely
operates.  However, our results are the first based on combining a large-scale
cosmological gasdynamics simulation with observed features of radio halos, and
they show that future simulations with higher resolution and more realistic
physics should enable straightforward comparisons with results from
low-frequency radio observatories. While earlier analytical studies have
involved more sophisticated models of CR generation, such as including spectral
steepening effects, these have relied on the X-ray luminosity function combined
with the known correlation between radio power and X-ray
luminosity~\citep{Ensslin2002}, the Press-Schechter mass
function~\citep{Cassano2006}, or Monte Carlo realizations~\citep{Cassano2010},
whereas our work is based on direct access to the internal state of the
simulated clusters.~\citet{Donnert2010b} introduced a simulation of the Local
Group including magnetic field injection and CR generation, but these results
may be sensitive to the assumptions made about the magnetic field injection and
are limited in volume. Our simulation, while involving simpler physics, covers
a large ($1 \hgpc$) volume, which will allow us to gather reliable statistics
and produce mock whole-sky radio maps, which are difficult to produce
accurately with methods based on analytical mass functions. 

In Section~\ref{sec:rh_simulation} we provide details of the cosmological
simulation, while in Section~\ref{sec:rh_radio} we explain the procedure used
to associate a radio power with each cluster.  We explore the range of valid
models in Section~\ref{sec:rh_exploration} and use these results to produce
radio power relations in Section~\ref{sec:rh_relations}.
Section~\ref{sec:rh_counts} discusses the results in terms of radio halo counts
as functions of flux and redshift, and we present various example radio sky
maps in Section~\ref{sec:rh_skymap}.  Finally, we conclude in
Section~\ref{sec:rh_conclusions} with a discussion of future directions.

\section{The simulation}
\label{sec:rh_simulation}

We simulated structure formation using the $\Lambda$CDM cosmological model
within a periodic box spanning 1024$~h^{-1}$ comoving Mpc. We assumed a Hubble
constant $H_0 = 100h$~km~s$^{-1}$~Mpc$^{-1}$ with $h = 0.719$, a present-day
matter density parameter $\Omega_{m,0} = 0.262$, baryonic density parameter
$\Omega_{b,0} = 0.0437$, vacuum density parameter $\Omega_{\Lambda,0} = 0.738$,
and spatially flat geometry, as suggested by results from Wilkinson Microwave
Anisotropy Probe
data~\citep{Komatsu2011}.  Initial conditions for $1024^3$ dark matter
particles at a starting redshift $z_i = 66$ were generated using a version of
GRAFIC~\citep{Bertschinger2001} modified to accept power spectra generated by
CMBFAST~\citep{Seljak1996}.  We normalized the power spectrum using $\sigma_8 =
0.74$.  We included adiabatic gas dynamics for the baryons using a perfect-gas
equation of state with adiabatic index $\gamma = 5/3$ and mean particle mass
determined using interpolation from collisional ionization equilibrium tables
for primordial gas from~\citet{Sutherland1993}.  Although no additional physics
is included in this first calculation, we initialized the gas temperature at
$z_i$ to a constant value of 9100~K, corresponding to a preheating entropy of
250~keV~cm$^2$ at a redshift of 3.  This level of preheating is adequate to
reproduce the observed X-ray luminosity-temperature relation~\citep{Bialek2001}
for clusters of galaxies, although details of the scatter in this relation and
its correlation with other cluster properties such as the presence of cold
cores are not constrained to match observations.

We ran our simulation using the FLASH code version
3.3~\citep{Fryxell2000b,Dubey2008} using a new direct multigrid Poisson
  solver~\citep{Ricker2008a} with 1024$^3$ dark matter particles and a uniform
1024$^3$ base mesh.  The piecewise-parabolic method~\citep{COLELLA1984} was
used to solve the Euler equations of gas dynamics.  To achieve the resolution
necessary to estimate the level of turbulence within clusters, we used adaptive
mesh refinement within $100$ preselected regions. Each region was $50 \hmpc$ on
a side centered on a halo identified using a lower-resolution precursor run.
The halos were selected to uniformly sample the range of resolvable halos using
mass function weighting to ensure a representative sample in the full
simulation.  Within the preselected regions we used a dark matter particle
refinement criterion, allowing no more than $100$ dark matter particles within
a zone.  We refined to a maximum resolution of $32 \hkpc$.  Estimates of the
integral scale of turbulence in clusters suggests a power law spectrum from
spatial scales of $0.8$ to $8$~kpc, with no visible
turnover~\citep{Kuchar2011}, suggesting that the integral scale is 
larger than 8 kpc and
that our resolution should allow us to determine the level of 
turbulent dissipation.  Since the refined regions
were larger than the halos on which they were centered, we captured a total of
$131$ high-resolution clusters.  We ran the code on the Cray~XT5 machine at Oak
Ridge National Laboratory, where the simulation required approximately
450,000~CPU-hours on 16,000 processors. Output files containing both particle
and gas information were written beginning at $z=2.0$ at every $\Delta z =
0.25$ for the purposes of mock sky generation.

\subsection{Halo Finding and Virial Mass}
\label{sec:rh_hf}

\begin{figure}
  \centering
  \includegraphics[width=\columnwidth]{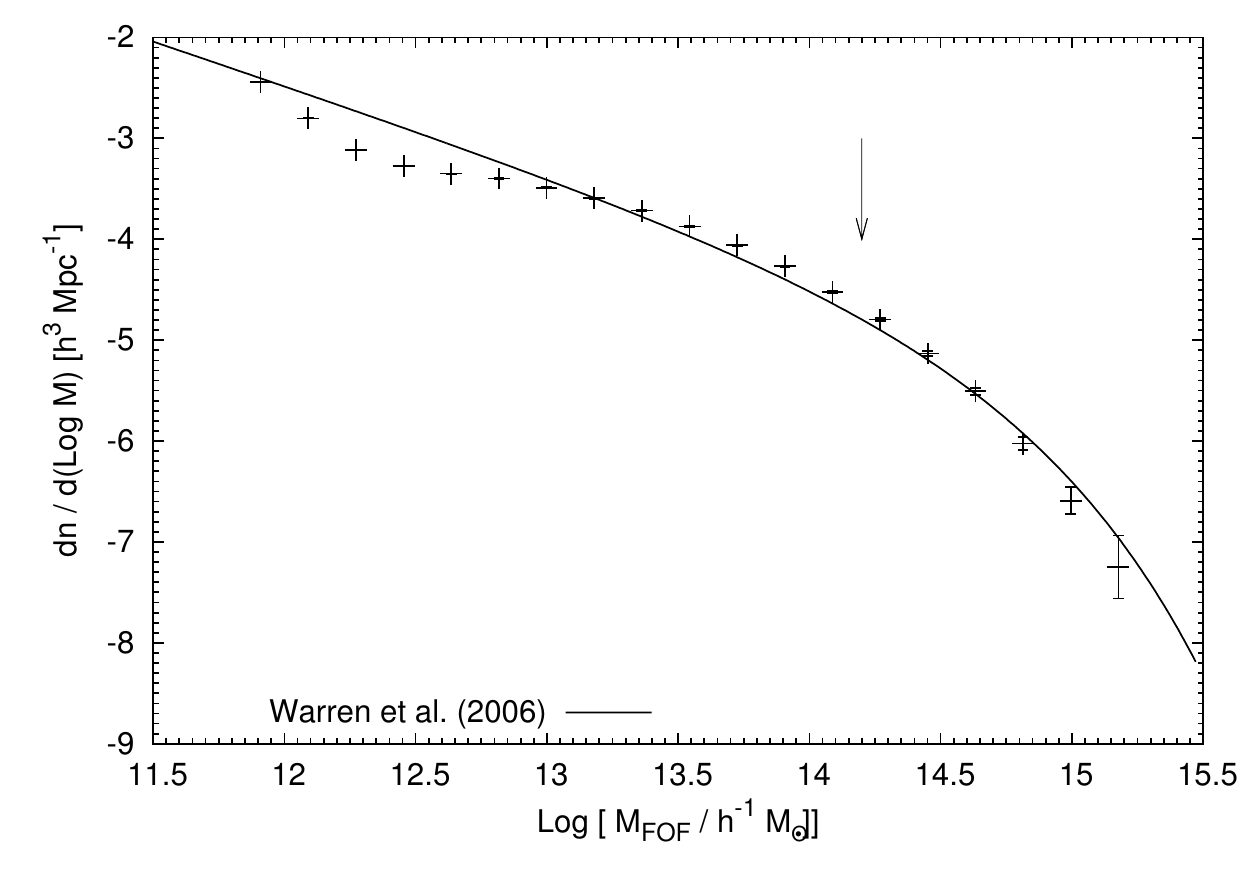}
  \caption[Mass function used in the radio halo simulation]
           {Mass function of all halos in the simulation volume 
           compared against the best fit of~\citet{Warren2006}. 
            Errors bars are given at $2\sigma$ and the vertical arrow denotes
            our FOF halo completeness limit.} 
  \label{fig:rh_massfunc} 
\end{figure}

We created halo catalogs from the simulation outputs using the
friends-of-friends (FOF) technique with a linking length parameter $b = 0.2$
and considered only halos with at least $3000$ particles (i.e. an FOF
dark matter mass of $2 \times 10^{14} \hmsol$).  Our base-grid spatial
resolution is sufficient to ensure accurate counts of halos with this many
particles throughout the range of redshifts we consider
here~\citep{Heitmann2005,Lukic2007}.  \figref{\ref{fig:rh_massfunc}} shows our
mass function for all halos in the simulation volume compared against the best
fit of~\citet{Warren2006}. We find fewer high-mass objects relative to the
Warren fit, but this is not unexpected~\citep[see, for example,][]{Knebe2011}.
Also, we tend to over-produce low mass objects, even below our resolvability
limit.  However, note that even though we produce too many low-mass objects
relative to the Warren fit below our resolution limit, our mass function still
turns away from the expected slope, and thus we cannot fully trust the number
counts below this threshold. There were $\sim 4000$ resolvable objects at
$z=0.0$.

To make comparisons with the observational analysis 
of~\citet[][hereafter CBS06]{Cassano2006}, 
we compute a spherical overdensity radius $R_v$ for each of our halos. For the 
high-resolution sample (i.e., the adaptively-refined halos within the 
$100$ predefined regions), we compute overdensities including both 
just dark matter and with dark matter plus gas. 
For the remaining fixed-resolution halos outside the predefined 
regions, we only include dark matter in the overdensity calculation 
since the gas data in these halos 
were poorly resolved. In Section~\ref{sec:rh_counts} we discuss our 
procedure for assigning radio power to these lower-resolution halos.
We use the same definition of overdensity 
as in CBS06, namely~\citet{Kitayama1996}:
\begin{equation}
  \Delta_c = 18 \pi^2 \left( 1 + 
             0.4093 \omega(z)^{0.9052}\right),
\end{equation}
where $\omega(z) \equiv \Omega_f(z)^{-1} -1$. Here,
\begin{equation}
  \Omega_f = \frac{\Omega_{m,0}(1+z)^3}{\Omega_{m,0}(1+z)^3 + \Omega_\Lambda}.
\end{equation} The virial mass, $M_v$, follows as $M_v = (4/3) \pi \Delta_c
\rho_m(z) R_v^3$, where $\rho_m(z)$ is the mean mass density:
\begin{equation}
\rho_m(z) = 2.87 \times 10^{11} \Omega_{m,0} (1+z)^3 \; h^2 \, \msol~\mpc^{-3}.
\end{equation} 
The most massive cluster in our simulation has a mass $M_v = 1.2 \times 10^{15}
\hmsol$.

\section{Simulating radio emission}
\label{sec:rh_radio}

We identify gas zones within $R_v$ for each halo and associate them with the
halo in which they are found. We create two-dimensional maps of projected
density and projected turbulent pressure, $m_i v_i^2$, where $m_i$ is the mass
in the cell $i$ and the average velocity is defined as the difference between
the measured velocity in the cell and the bulk velocity averaged over 300~kpc
regions, ${\bf v}_i \equiv {\bf v}_i - {\bf \bar{v}}_{\rm 300~kpc}$. This is a
strategy for removing bulk motions 
similar to~\citet{Vazza2010a} (an alternative approach to remove bulk
velocities from this calculation is discussed in~\citealt{Paul2011}).  We use
these projections in two ways: to create simulated surface brightness maps and
to construct total radio luminosities by integrating these quantities across
the entire projected cluster surface out to the virial radius.  The integrated
projected density is of course $M_v$ and we will designate the integrated
turbulent pressure as $\Gamma_v = \sum_{i} m_i v_i^2$.

Therefore we can create simulated radio surface brightness maps for our
clusters by normalizing maps of projected turbulent pressure and projected mass
using assumed radio luminosities and rest-frame spectra.  Because we may not
fully resolve intracluster turbulence, the total amount of turbulent pressure
in our clusters may be lower than the $\sim 10\%$ of hydrostatic pressure seen
in high-resolution simulations~\citep{Ricker2001,Ritchie2002,Mitchell2009}.
However, because the normalization of the radio power is supplied independently
(see below), all we require of the turbulent pressure maps is that they be
sufficiently diffuse and representative in spatial extent of clusters
containing radio halos. The detailed structure of the maps should not be
regarded as realistic. Because of the beam smearing described below, this fact
does not significantly affect our analysis. Also, while we cannot depend on
these simulated clusters to provide correct high-resolution X-ray and radio
surface brightness maps, we can still use them to identify broad features, such
large-scale shocks and the relative radial dependence of turbulence.

Using these integrated quantities, we construct a rest frame $1.4$~GHz radio 
power via
\begin{equation}
  P_{1.4 \ghz} = C_s B_s M_v^a \Gamma_v^c,
\label{eq:rh_model}
\end{equation}
where $C_s$ is a scaling constant, $M_v$ is the virial mass, 
$\Gamma_v$ is the virial turbulent pressure, and $B_s$ is the 
magnetic field parameter:
\begin{equation}
  B_s = \frac{B(M_v)^2}{(B(M_v)^2 + B_{\rm CMB}^2)^2},
\label{eq:rh_bm}
\end{equation}
where $B(M_v) \equiv \aveb (M_v/\langle M \rangle)^b$ and 
$B_{\rm CMB} \equiv 3.2(1+z)^2 \mg$ is the equivalent magnetic 
field strength of the cosmic microwave background. 
This formulation separates physical processes that generate 
CRs ($M_v$ and $\Gamma_v$) from those that contribute 
to radio emission ($B$ in the numerator) and CR losses 
due to emission ($B^2$ in the denominator) and inverse 
Compton scattering ($B_{\rm CMB}$).
The losses enter into this equation because they limit the 
maximum CR energy.
In this formalism, $M_v$ measures the total cluster mass and thus 
should scale with the dependence of CR generation on 
hadronic secondary processes, whereas $\Gamma_v$ measures the total 
cluster turbulence and thus should provide a measure of the 
reacceleration of CR electrons by that turbulence.
We will set $\langle M \rangle = 1.5 \times 10^{15} \msol$. There are thus 
five independent parameters: the average magnetic field $\aveb$, 
the scaling of magnetic field with cluster mass, $b$, the scaling 
of radio power with virial mass, $a$, the scaling of radio 
power with turbulent pressure, $c$, and an overall scaling 
parameter $C_s$. 
A summary of our model parameters is given 
in \tabref{\ref{tab:rh_modelParms}}.

This model is a generalization of the one derived
in~\citet{Cassano2005a}. Note that the analysis of hadronic secondary models
of~\citet{Dolag2000} identified the functional form of the magnetic field as $B
= B(M_v)^2 / (B(M_v)^2 + B_{\rm CMB}^2)$: i.e., the denominator is not squared.
Our model easily accommodates this scenario: when $B_{\rm CMB}$ dominates, this
will appear as a constant factor folded into $C_s$, and when $B(M_v)$ dominates
this will simply adjust the mass scaling factor $a$.  
Provided that the necessary adjustments to $C_s$ and $a$ are made, 
our model holds 
even for intermediate cases where $B \sim B_{\rm CMB}$, 
since we are largely in the regime where $M_v < \langle M \rangle$.
We are fixing the form of
the magnetic field dependence since the radio synchrotron power will always
depend on magnetic field pressure ($B^2$) independently of the CR
generation and acceleration mechanisms (see~\citet{Cassano2005a} for a
discussion).  We stress that although this model is relatively simple, it
allows us to explore a range of plausible acceleration mechanisms and examine
relative changes to luminosity functions, scaling relations, and other radio
properties.

\begin{table}
  \centering
  \caption{Parameters of the Radio Luminosity Model.}
\begin{tabular}{cc}
Parameter & Description \\
\hline
\hline
$C_s$ & Overall scaling \\
$\aveb$ & Average magnetic field \\
$b$ & Scaling of magnetic field with cluster mass \\
$a$ & Scaling of radio power with cluster mass \\
$c$ & Scaling of radio power with turbulent pressure \\
\hline
\end{tabular}
\label{tab:rh_modelParms}
\end{table}

\begin{figure}
  \centering
  \includegraphics[width=\columnwidth]{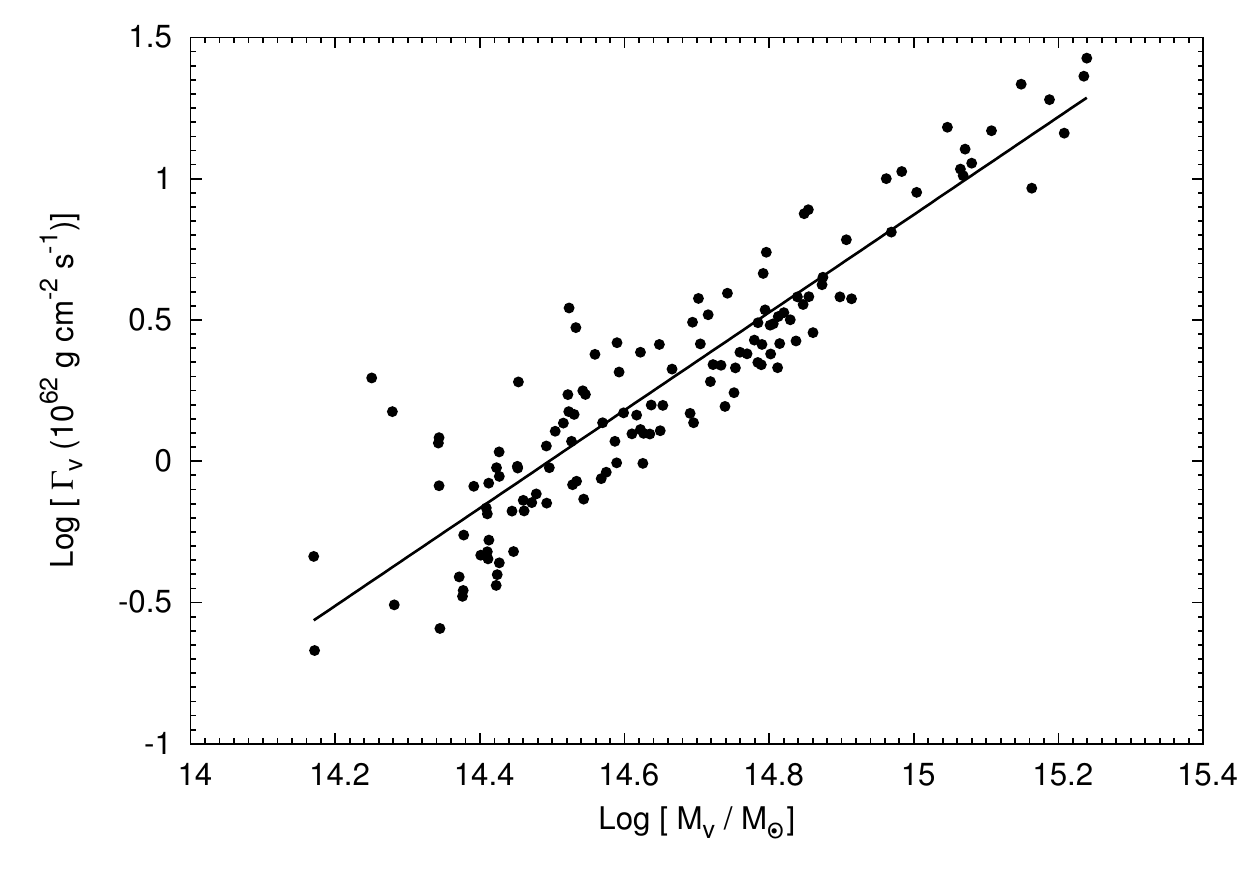}
  \caption[$\Gamma_v$ versus $M_v$.]
           {Total turbulent pressure, $\Gamma_v$, versus virial mass, $M_v$ 
           at $z=0$ for the high-resolution cluster sample. 
            Shown is a best-fit line in log space. The slope of the line is
            $\sim 1.7$.}
\label{fig:rh_gammamvir}
\end{figure}

This model allows us to explore both CR generation mechanisms or a
mixture of both. For example, the hadronic secondaries model should predict
radio power which scales with cluster mass, so $c=0$ in this case. A
reacceleration model is proportional to turbulent pressure, so $a$ would be
$0$. Note that the model of CBS06 is based on reacceleration, but only scales
with cluster mass. This is because $\Gamma_v$ roughly scales with $M_v$ with a
logarithmic slope of $1.7$, as shown in~\figref{\ref{fig:rh_gammamvir}}. 
Note that our scaling relation 
is roughly consistent with those derived from other 
simulations~\citep[e.g.,][]{Vazza2006}.
There is some scatter in this relation due to the merger history of 
a particular cluster: recent mergers produce stronger levels of turbulence, 
which tend to scatter the cluster higher in this relation.
Note
that the model of CBS06 corresponds here to $a=4/3$ and $c=0$.

\begin{figure*}
  \centering  
    {\includegraphics[scale=0.33]{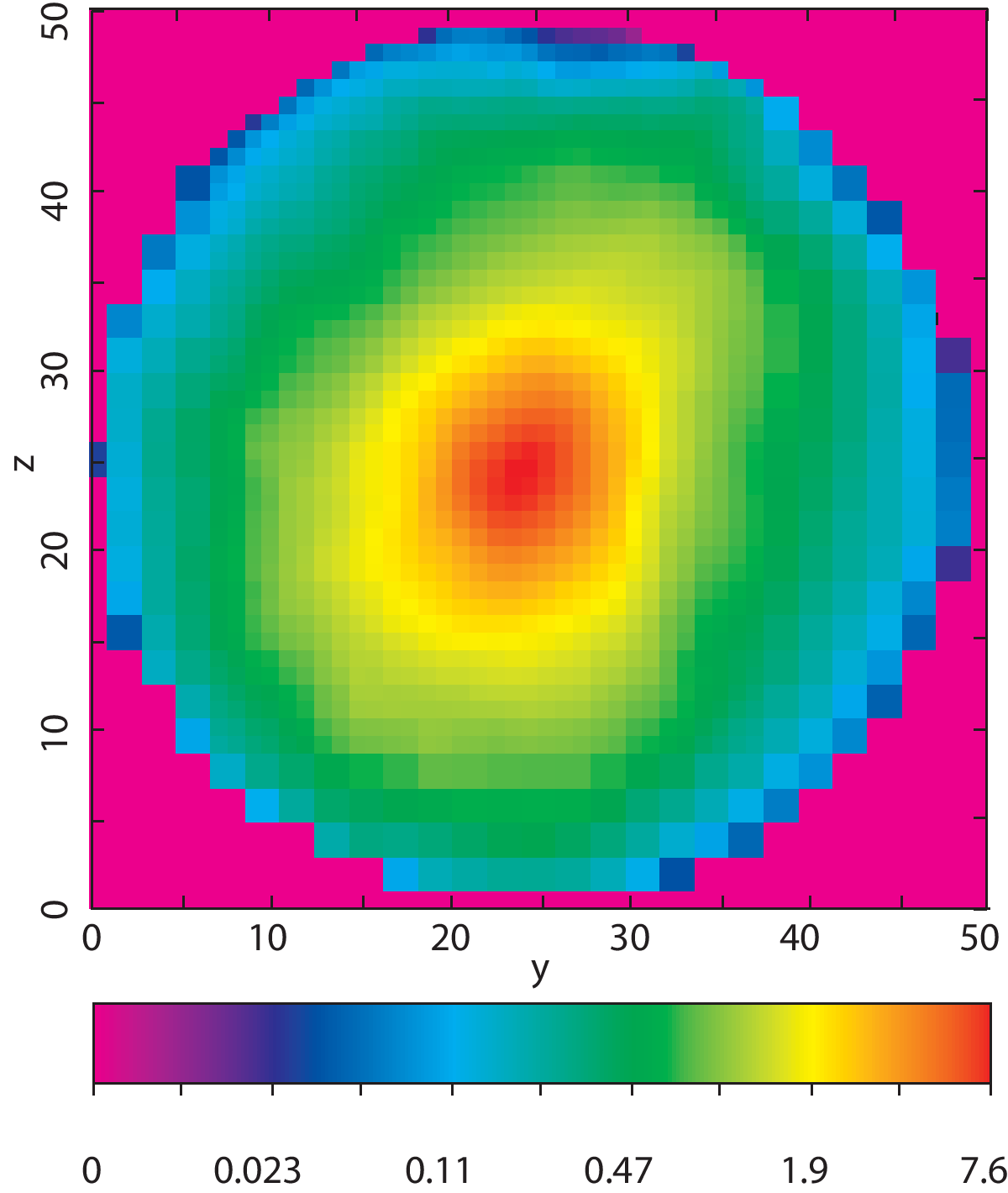}}
    {\includegraphics[scale=0.33]{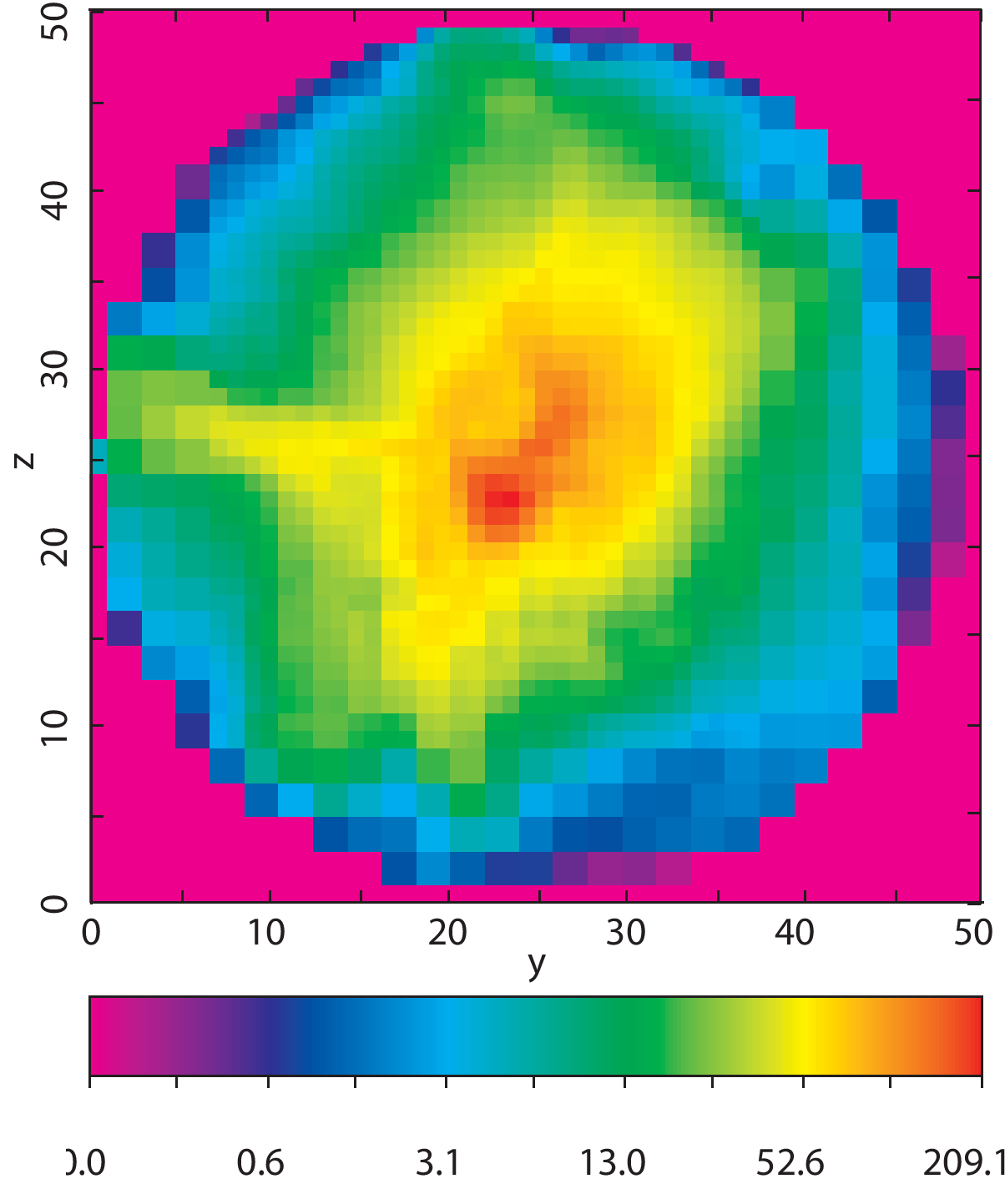}}
    {\includegraphics[scale=0.33]{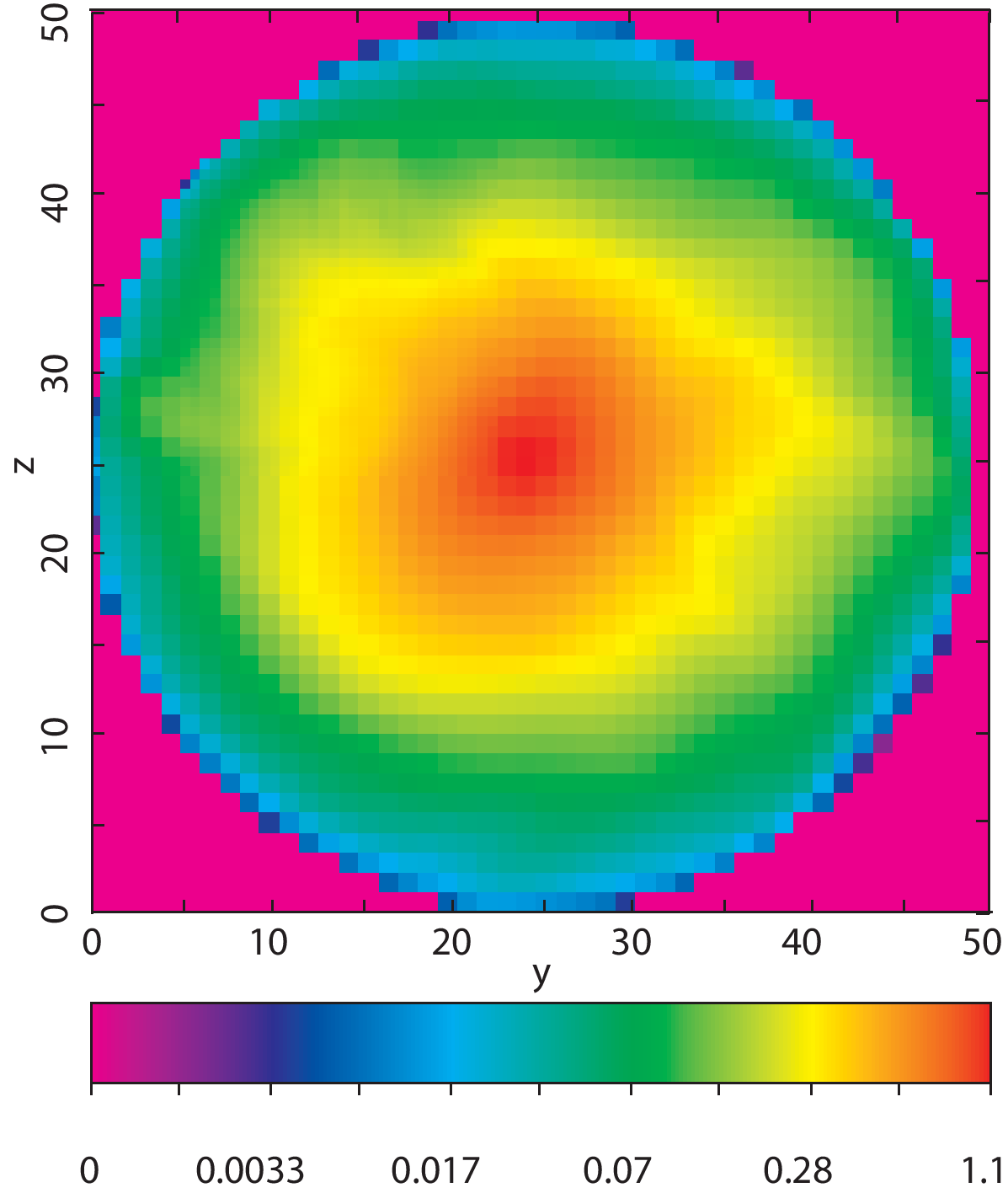}}
    {\includegraphics[scale=0.33]{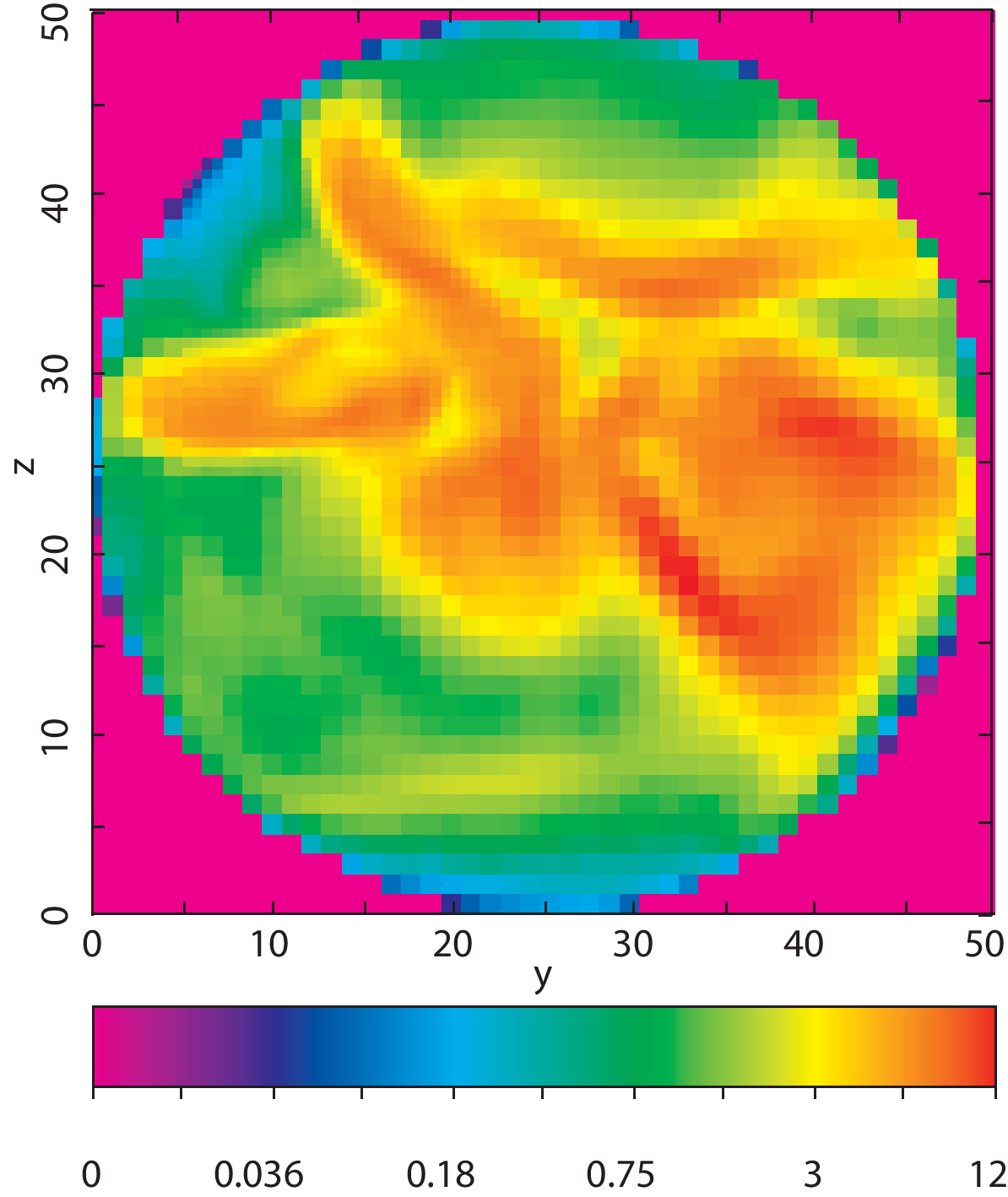}}
  \caption[Projected density and projected turbulent pressure]
           {Projected density (left-hand plots) and projected 
            turbulent pressure (right-hand plots) for a 
            two clusters (top and bottom rows). Projections are taken along 
            the $x$-direction within $R_v$ for each cluster. The units for 
            projected density are $10^{44}~{\rm g}~{\rm cm}^{-2}$ and 
            for projected turbulent pressure 
            are $10^{58}~{\rm g}~{\rm s}^{-2}$. The images are 
            normalized to a uniform grid $50$ cells on a side. The
            corresponding comoving length scale is indicated in the figure.}
\label{fig:rh_projgamma}
\end{figure*}

A degeneracy exists for calculations of total radio luminosity between models
that scale with turbulent pressure and those that scale with mass, since we may
freely exchange $c$ for $1.7a$ and vice-versa. However, a more detailed
examination of cluster atmospheres reveals striking differences.  Even with the
relatively low resolution of our simulation, and the resulting inability to
fully reproduce correct structures in the cluster atmospheres, we can identify
gross differences in the projected maps.  ~\figref{\ref{fig:rh_projgamma}}
shows projections of mass and turbulent pressure for two clusters. The mass
projections of both clusters are roughly spherical, as expected. However, the
turbulent pressure maps show more varying morphology. While the halos are
roughly equal in mass ($\sim 8 \times 10^{14} \hmsol$), one shows much greater
turbulent structure, indicating recent merger activity, which may explain the
scatter in~\figref{\ref{fig:rh_gammamvir}}.  Thus, even though different
mechanisms of CR generation may produce similar cluster counts (as we will see
below), high-resolution radio and X-ray imaging of clusters may help to
determine which mechanism dominates.

\begin{figure}
  \centering
  \includegraphics[width=\columnwidth]{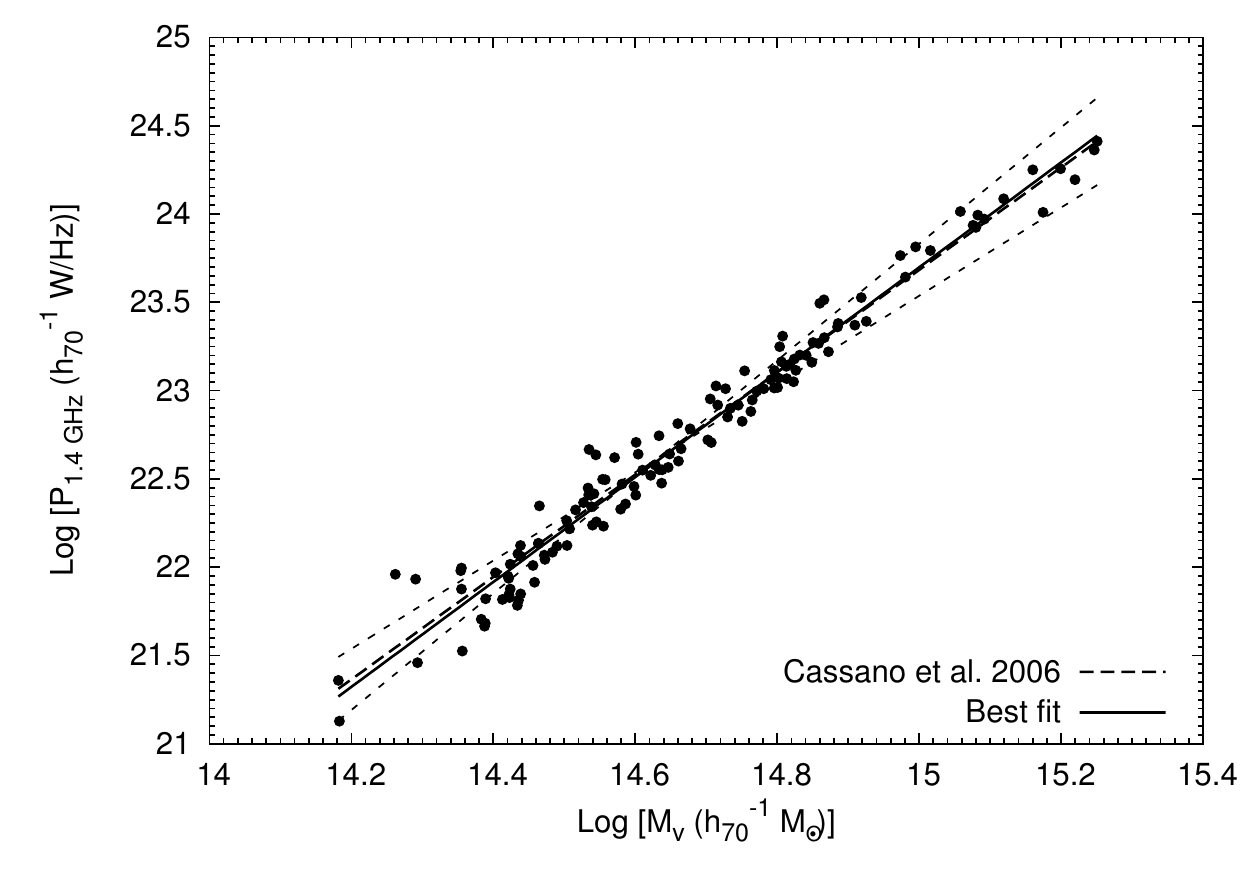}
  \caption[Radio halo luminosity versus virial mass.]
           {Radio halo luminosity versus virial mass for one example model
(points) with best fit above $10^{15} \hmsol$ (solid line) compared against the
observed best fit found in CBS06 (thick dashed line) and $1\sigma$
uncertainties (thin dashed lines). Points are the high-resolution cluster
sample.  The example uses parameters $\langle B \rangle = 2.0~\mu$G, $b=1.0$,
$a=0.0$, and $c=0.7$.} \label{fig:rh_radmass} \end{figure}

Since we do not include in our simulations any detailed CR 
generation mechanisms, and because we want to keep our model 
as general as possible, we must fix the scaling 
parameter $C_s$ by using observations. This scaling will then combine any extra 
constants and parameters not included in our analysis. For a given 
set of model parameters, we set $C_s$ by assigning a radio 
luminosity to the most massive
cluster in our simulation. This cluster has a mass 
$\sim 2\times 10^{15} \hmsol$, which fits within the observed 
radio halo mass range of $2 \times 10^{15} - 6 \times 10^{15} \hmsol$.
We do this 
with the $P_{1.4} - M_v$ relation found in CBS06, which is based 
on combining the observed correlation of radio halo power and X-ray luminosity 
with the correlation between X-ray luminosity and mass:
\begin{equation}
 \begin{array}{l}
  \log \left[ \frac{P_{1.4}}{3.16 \times 10^{24} h_{70}^{-1} 
     {\rm W}~{\rm Hz}^{-1} } \right] = \\
     (2.9 \pm 0.4) \log \left[ \frac{M_v}{10^{15} \hsmsol} \right]
     - (0.814 \pm 0.147)
\end{array}
\label{eq:rh_radmass}
\end{equation} 
We then apply this same constant scaling to all remaining high-resolution 
halos in the sample. While the scaling may contain some additional 
dependence on mass or turbulent pressure not accounted for in our 
parameterization, this can easily be accommodated in our study by adding to 
(or subtracting from) the parameters $a$ and $c$. An example of a 
particular model compared against the observed relation 
is shown in~\figref{\ref{fig:rh_radmass}}. Our best-fit relation in this 
plot and throughout this paper 
uses only the clusters within the observed mass range.

Since very few radio halos 
have been observed beyond a redshift of $\sim 0.4$, and
available statistics do not strongly constrain evolution in this relation, we
will fix the scaling at $z=0$ and apply the same scaling to higher-redshift
clusters.  We will also assume power-law energy spectra with a spectral index
of 1.2, consistent with low-redshift observations~\citep{Feretti2004}.
Finally, we do not include in our model the relationship between synchrotron
break frequency and the presence of a radio halo, which can be used to
calibrate models to the observed fraction of clusters hosting radio halos
(CBS06). We will discuss the potential impacts of this assumption 
in the conclusion.

\section{Exploration of valid models}
\label{sec:rh_exploration}

To constrain our model choices we make selections for the model parameters,
assign radio powers to the clusters using the procedure described above, find
the best-fit line to our derived $P_{1.4}-M_v$ data above a mass threshold of
$10^{15}~\hmsol$, and compare the best-fit slope and normalization to the
observed values. We only accept model choices that produce fits that lie
within $1\sigma$ of the observed relation.  This is a strategy similar to the
one employed by CBS06: except that we are applying a test
by enforcing the known relation to somewhat lower radio powers than they
consider. We do this so that we can capture enough halos ($\sim 10$) to generate
sufficient statistics for our best-fit lines. Obviously, we could just select
two models that span the valid range and analyze their difference, but we wish
to explore the relationships among the various model parameters and the
separate consequences of varying each one.

\begin{figure} \centering
  \subfigure[$a+c$ as a function of $\aveb$ and $b$.] {\includegraphics[width=\columnwidth]{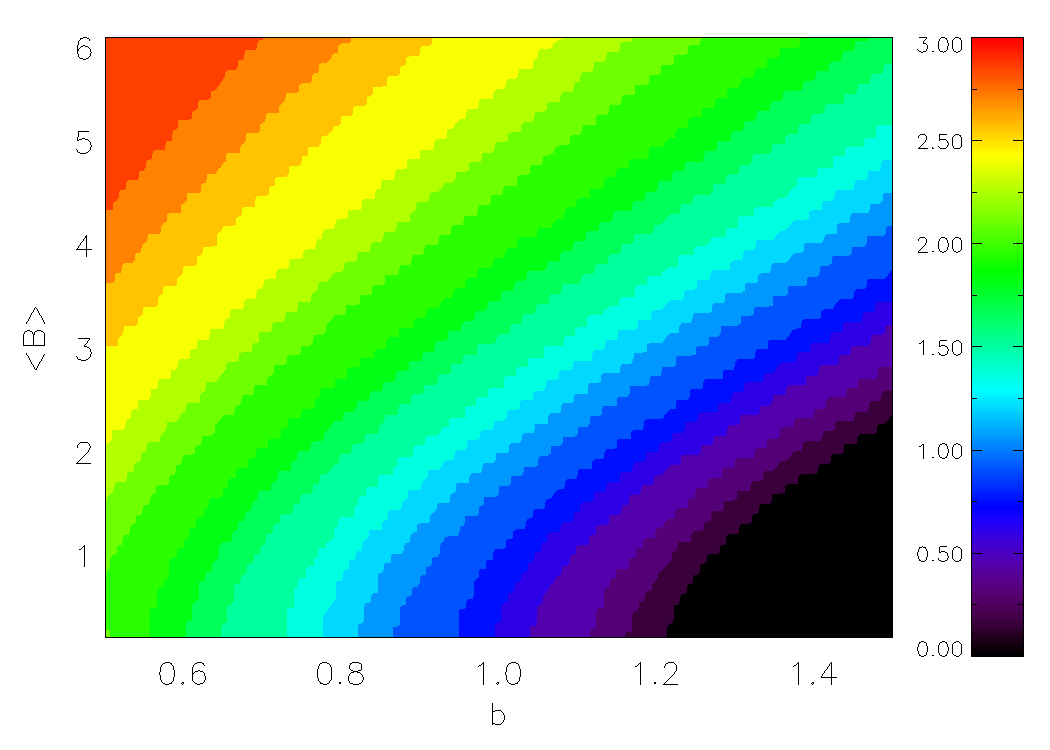}}
  \subfigure[$\aveb$ as a function of $b$ and $a+c$.] {\includegraphics[width=\columnwidth]{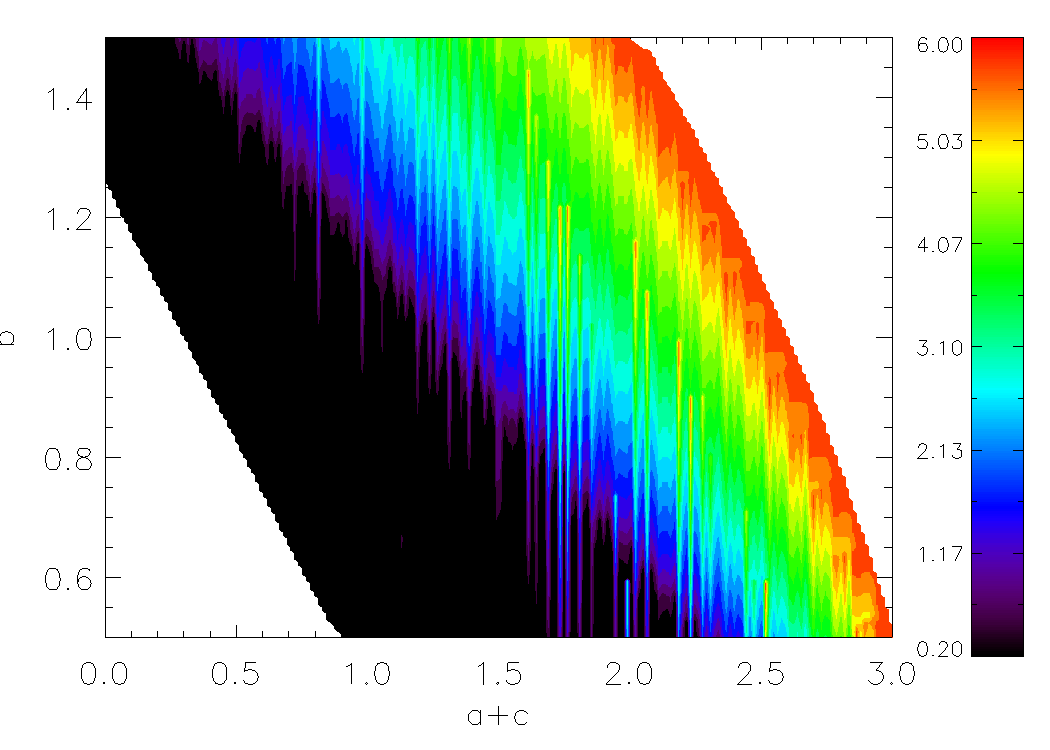}}
  \subfigure[$b$ as a function of $a+c$ and $\aveb$.] {\includegraphics[width=\columnwidth]{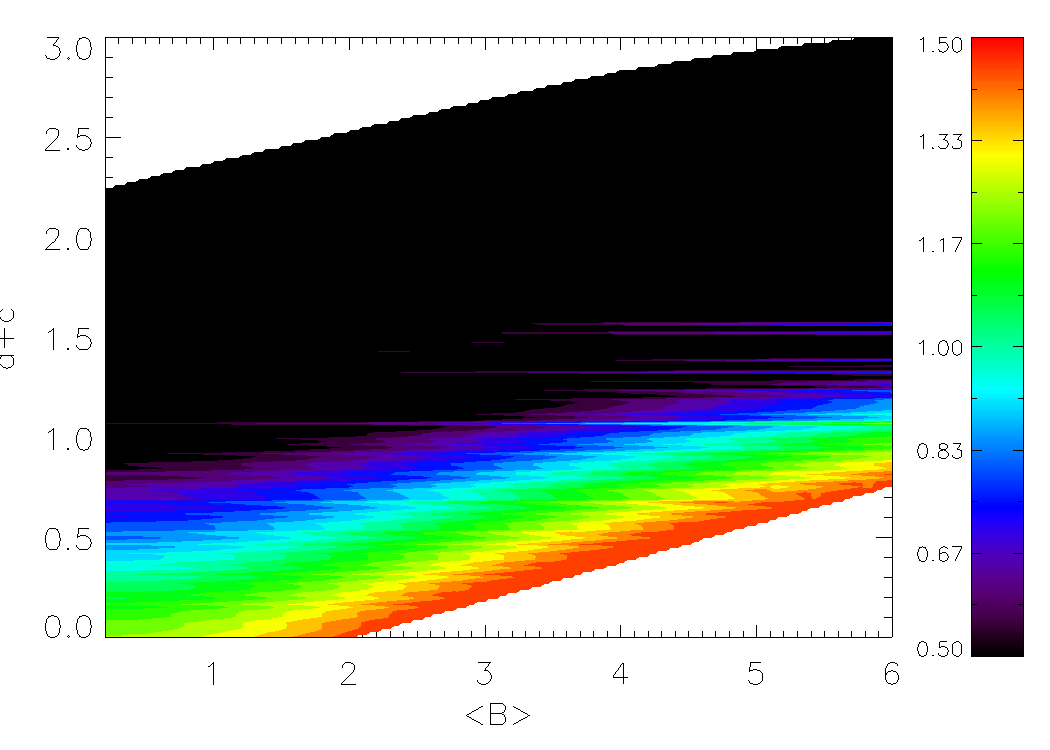}}
  \caption[Allowable radio halo model contours]{Contours of allowed radio halo 
           model parameters. Allowed models are determined by fitting a line 
           to our $P_{1.4}-M_v$ data above $M^{15}\hmsol$ and ensuring 
           that the slope and normalization are within $1\sigma$ of the known
relation. $\aveb$ is given in units of ${\rm \mu G}$.}
\label{fig:rh_modelContours} 
\end{figure}

\figref{\ref{fig:rh_modelContours}} shows colormap plots of allowable models.
We vary $\aveb$ from $0.2$ to $6.0 \mg$, $b$ from $0.5$ to $1.5$, $a$ from
$0.0$ to $5.0$, and finally $c$ from $0.0$ to $3.0$.  We could explore even
larger values of $a$ and $c$, but as we will discuss below $1\sigma$
uncertainties in the measured $\pmvir$ relation place upper limits on the
scaling of $a$ and $c$ at these chosen maximum values.  
We also assume a positive
correlation between radio power and $M_v$ and $\Gamma_v$.  While we allow the
mass and turbulent pressure scaling parameters to vary all the way to $0$, we
constrain the scalings associated with magnetic fields.  We constrain the
average cluster magnetic field strength from $0.2 \mg$, which is set by
observed upper limits on hard X-ray emission (CBS06), to $6.0 \mg$, which is a
reasonable upper limit from rotation measure
observations~\citep[e.g.][]{Bonafede2011}.  The restrictions on $b$
come from the simulations of~\citet{Dolag2002}, which followed the adiabatic
compression of seed magnetic fields as clusters formed.  They found a scaling
$B \propto M^{1.33}$. We allow some uncertainty in this value, but do not allow
a compete lack of scaling of magnetic field with cluster mass.  For simplicity,
we have combined the mass and turbulent pressure values as $a+c$, so $a+c$ is
varied from $0.0$ to $9.0$.  The contours for each individual parameter show
structures similar to those for this combined parameter.  In these plots we are
showing the \emph{maximum} allowed value for a given point on each contour
plot. All values less than the plotted value are also allowed. For example, for
$b=1.0$ and $\aveb=3.0 \mg$ the allowable values for $a+c$ are from $0.0$ to
$\sim 1.5$.  

We find that very strong magnetic fields are only allowed if the scalings with
virial mass and turbulent pressure are very steep. In these cases strong radio
power in low mass objects due to high $\aveb$ is offset by significantly lower
radio power associated with $M_v$ or $\Gamma_v$. If the scaling of magnetic
field strength with cluster mass is above unity, then it is difficult to fit
strong magnetic fields at high mass within the observed relations.  We find
several regions forbidden in our models: strong magnetic fields coupled with
low $a+c$, and very low or very high $a+c$ and $b$ values.  

We see interesting structures in the contours: steps and wiggles in the $a+c$
plots, and striations in the others. These are due to the scatter that develops
in the $P_{1.4}-M_v$ relations and the resulting variations of the best-fit
lines. Because of this variation, we do not see monotonically increasing (or
decreasing) behavior in the contour plots, especially at extreme values.
Surprisingly, we find that $a=c=0.0$ is allowed, but only at low $\aveb$ and
high $b$. This is because of the implicit mass dependence in the calculation of
the cluster magnetic field strength, $B(M_v)$.  Also, the model used in CBS06
is \emph{forbidden} in our analysis, since we are enforcing the known relation
to slightly lower cluster masses.  However, this relation is not 
observationally verified at lower masses, and we can allow their model choice
when restricting ourselves to the mass ranges they consider.

We use these contours to guide our selection of models for further study. We
wish to adequately sample the space of allowable models and explore the limits
allowed by observational constraints. We also wish to explore the effects of
holding one parameter constant and varying the others to their extreme allowed
values. To aid analysis, we collect our choices into six model groups,
enumerated in~\tabref{\ref{tab:rh_rhModels}}. In this table we list the values
chosen for a particular parameter set and a unique designation for that set
used in further plots.

\begin{table}
  \centering
  \caption{Model Groups and Parameter Sets.}
\begin{tabular}{ccccc}
Designation & $ \langle B \rangle ({\rm \mu G})$ & $ b $ & $ a $ & $ c $ \\
\hline
\hline
\\
\multicolumn{5}{c}{Model Group 1: Fixed $c$, Varying Magnetic Field} \\
\hline
1A & 0.2 & 0.650 & 0.000 & 0.700  \\
1B & 0.5 & 1.000 & 0.000 & 0.700  \\
1C & 1.5 & 0.800 & 0.000 & 0.700  \\
1D & 1.5 & 1.100 & 0.000 & 0.700  \\
1E & 3.1 & 1.470 & 0.000 & 0.700  \\
\\
\multicolumn{5}{c}{Model Group 2: Fixed Magnetic Field} \\
\hline
2A & 2.0 & 1.000 & 0.000 & 0.510  \\
2B & 2.0 & 1.000 & 0.000 & 0.930  \\
\\
\multicolumn{5}{c}{Model Group 3: Exchanging $a$ and $c$} \\
\hline
3A & 2.0 & 1.000 & 1.300 & 0.000  \\
3B & 2.0 & 1.000 & 0.650 & 0.325  \\
3C & 2.0 & 1.000 & 0.000 & 0.650  \\
\\
\multicolumn{5}{c}{Model Group 4: Extreme Allowed Magnetic Fields, Fixed $c$} \\
\hline
4A & 0.2 & 1.000 & 0.000 & 0.700  \\
4B & 3.1 & 1.000 & 0.000 & 0.700  \\
\\
\multicolumn{5}{c}{Model Group 5: Extreme Allowed Magnetic Fields} \\
\hline
5A & 0.2 & 0.500 & 0.000 & 0.875  \\
5B & 6.0 & 1.500 & 2.000 & 0.000  \\
\\
\multicolumn{5}{c}{Model Group 6: Extreme Allowed Scalings} \\
\hline
6A & 0.2 & 1.260 & 0.000 & 0.010  \\
6B & 6.0 & 0.600 & 0.000 & 1.550  \\
\hline
\end{tabular}
\label{tab:rh_rhModels}
\end{table}

In Model Group 1 we set $a$ to $0.0$, fix $c=0.7$, and vary the magnetic field
parameters as widely as possible from a minimum of $\aveb=0.2$ to $3.1~\mg$. We
also vary the scaling parameter associated with the magnetic field, $b$. Each
$\aveb$ is coupled with a unique $b$, except for $\aveb=1.5 \mg$, where we
examine $b=0.8$ and $b=1.1$, which are the minimum and maximum allowed values
for this particular configuration. We explore the opposite behavior in Model
Set 2 by fixing the magnetic field parameters to $\aveb=2.0 \mg$ and simple
linear scaling $b=1.0$ while having no explicit $M_v$ dependence and studying
the maximum and minimum allowed values for turbulent pressure scaling, $c$. We
chose this value of the magnetic field so that we could get the maximum
difference in $c$. We keep the same magnetic field configuration for Model Set
3, but here we exchange $a$ and $c$ using the measured relation
(\figref{\ref{fig:rh_gammamvir}}).  In this Model Set we fix the quantity
$a+2c$.  This allows us to hold the magnetic field fixed while going from a
hadronic-like CR model ($c=0$) to a reacceleration model ($a=0$).  We designed
this Model Set to verify that our results are robust to even exchanges of $a$
and $c$ using the measured relation, which they should be.  In Model Set 4 we
fix $b=1.0$, $a=0.0$, and $c=0.7$ and examine the extreme allowed average
magnetic field. We chose these values of $b$, $a$, and $c$ such that we could
get the maximum change in $\aveb$.  We repeat this test in Model Set 5, but now
allow $b$, $a$, and $c$ to vary to accommodate the extreme values studied of
$\aveb$.  Finally in Model Set 6 we pick two model parameter sets that
represent extremes of all four parameters.

\section{Radio power relations}
\label{sec:rh_relations}

We begin our analysis by using our sample of $131$ high-resolution
clusters to examine the relationship between radio luminosity and virial mass
and X-ray luminosity. To construct X-ray luminosities, we use the {\tt mekal}
plasma emissivity model supplied with the XSPEC package (Arnaud 1996). We then
build a composite X-ray spectrum for each cluster and use that spectrum to
generate the $0.1-2.4$~keV rest-frame luminosity within the spherical radius 
$R_v$ for each cluster.
Since our simulation does not include cooling and central active galactic 
nuclei feedback, our
cluster temperatures and hence X-ray luminosities are uniformly higher than
observed (see~\citet{Stanek2010} for a discussion of such effects).  However,
the slopes of our relations are still within observed limits, and we can still
study the relative differences among models and their evolution with redshift.

\begin{figure*}
  \centering
  {\includegraphics[width=0.48\textwidth]{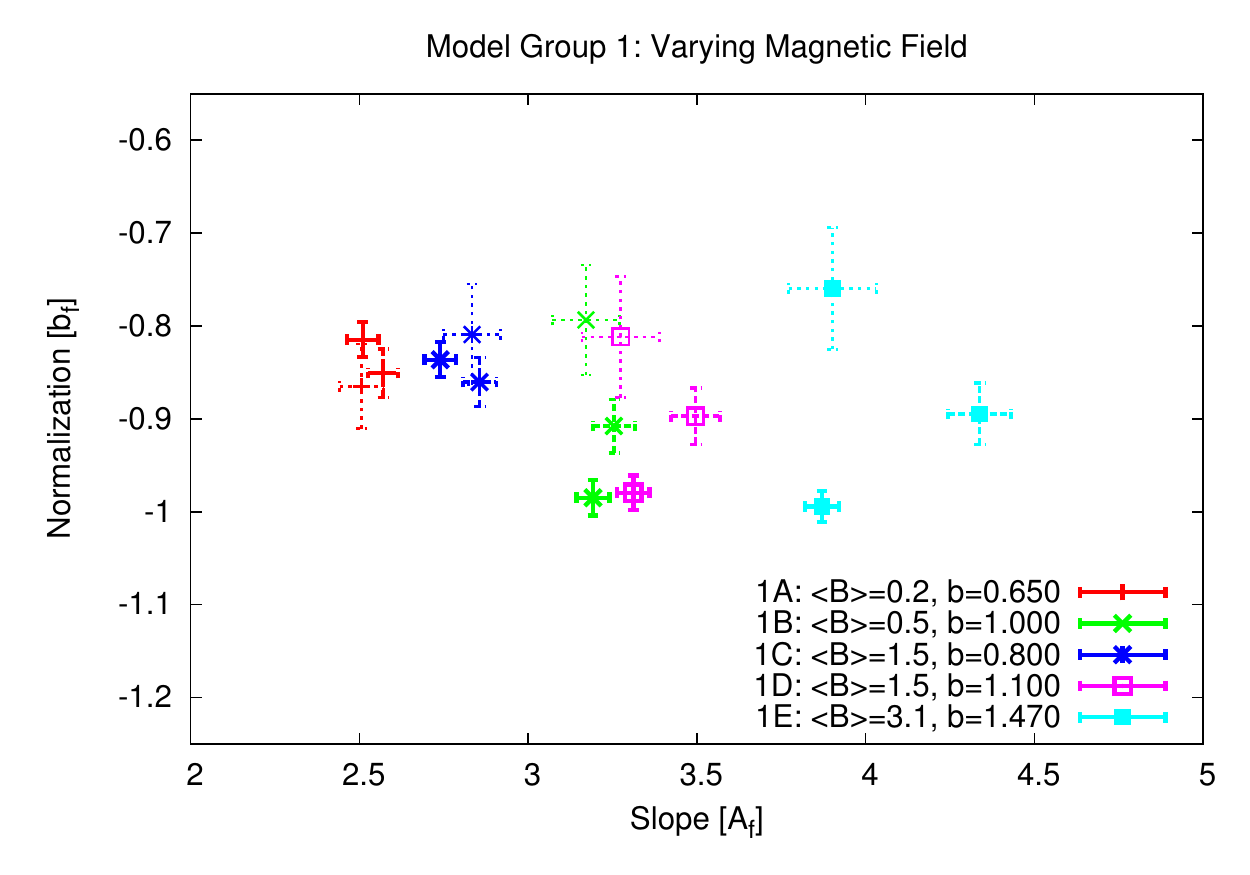}}
  {\includegraphics[width=0.48\textwidth]{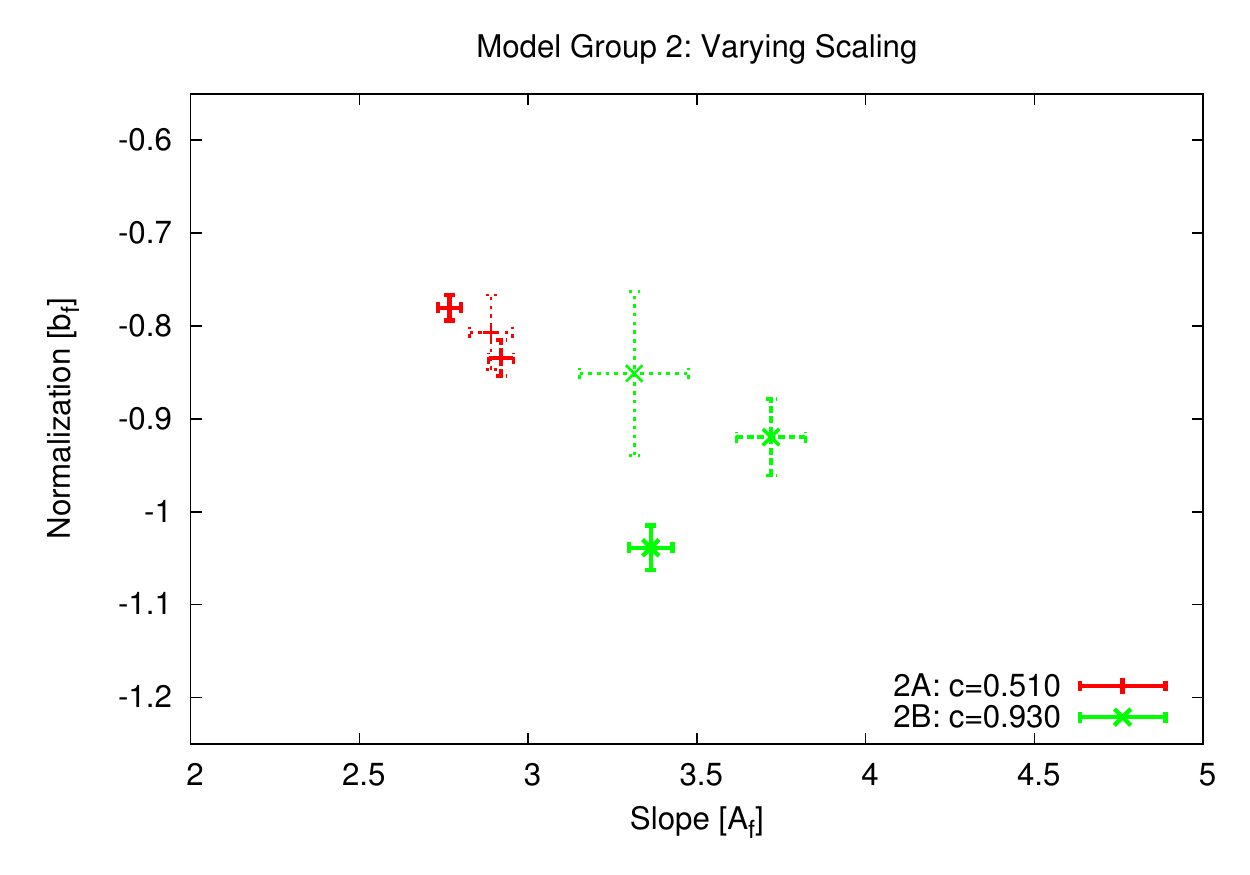}}
  {\includegraphics[width=0.48\textwidth]{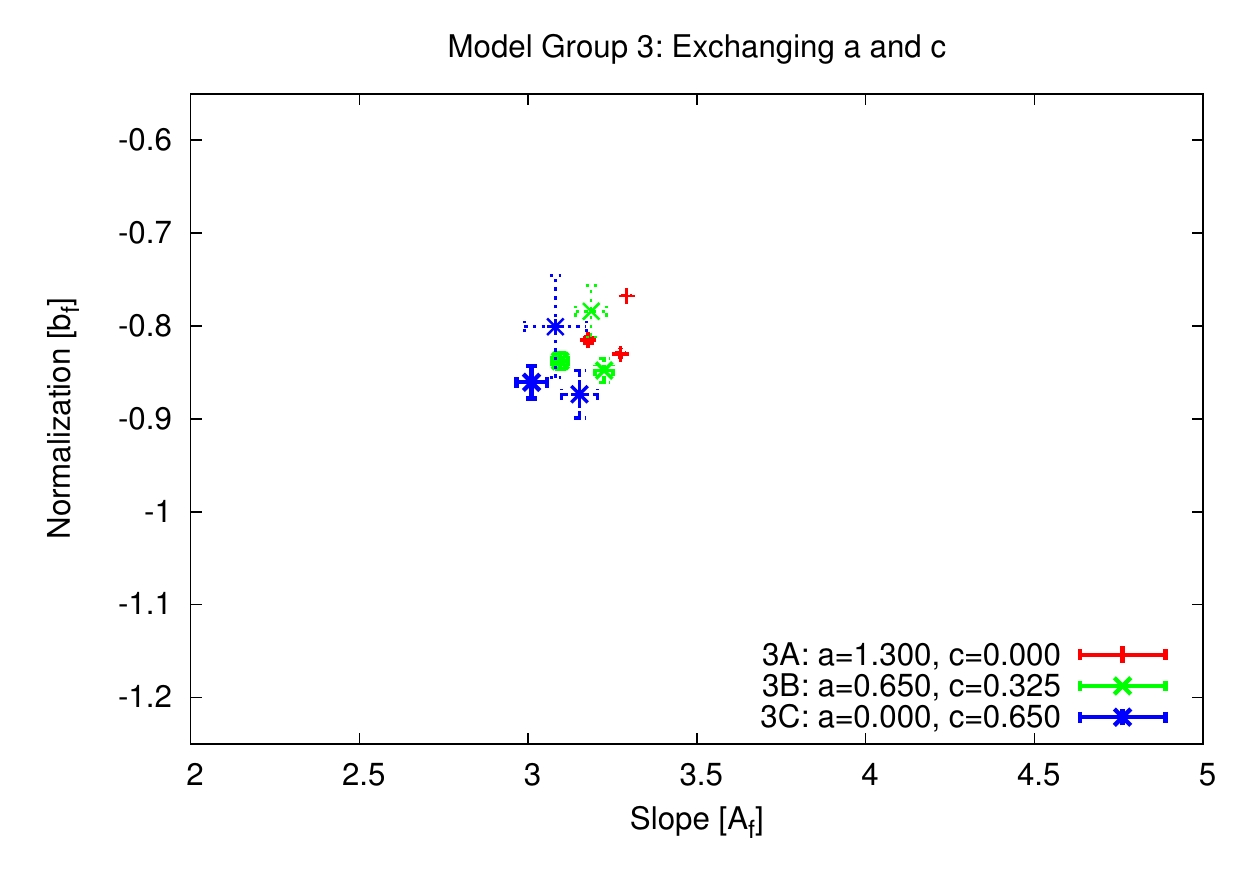}}
  {\includegraphics[width=0.48\textwidth]{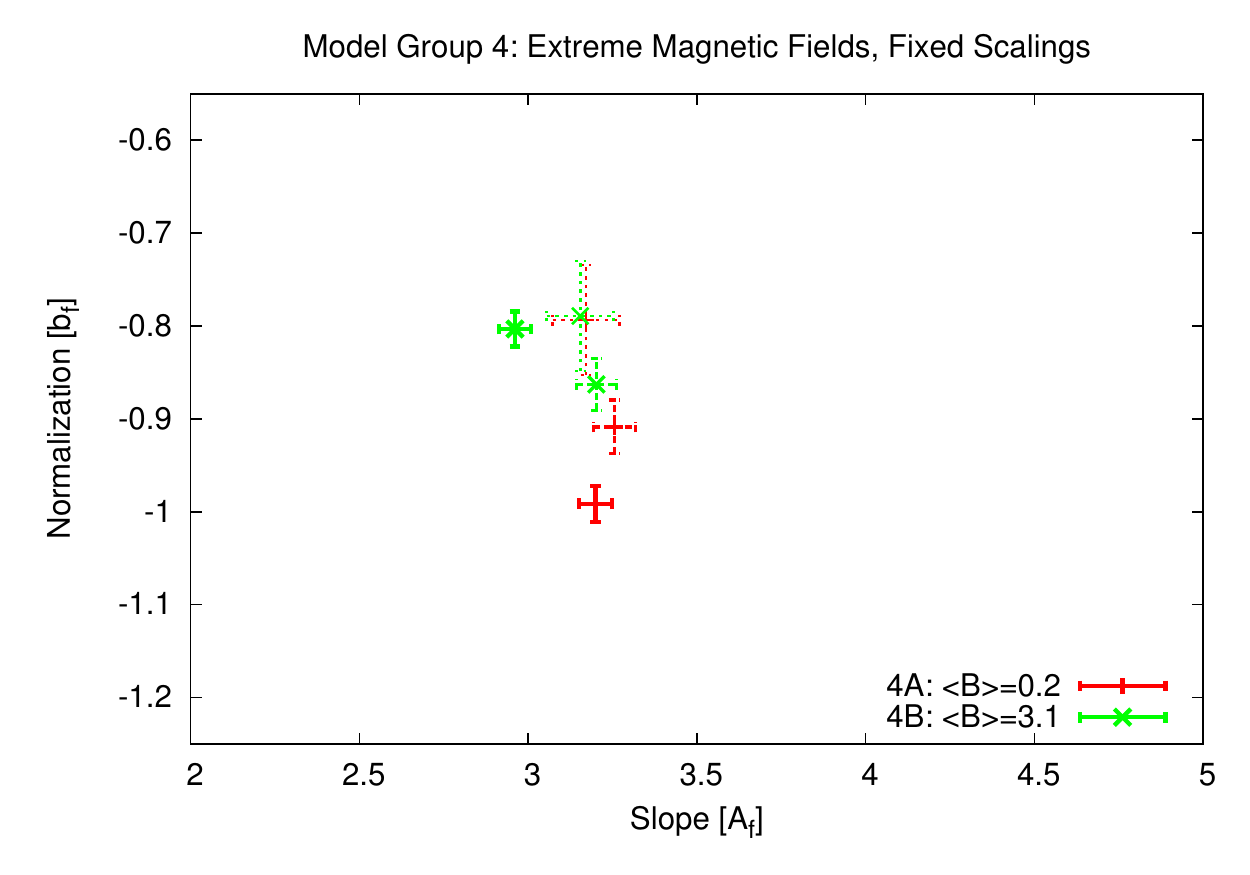}}
  {\includegraphics[width=0.48\textwidth]{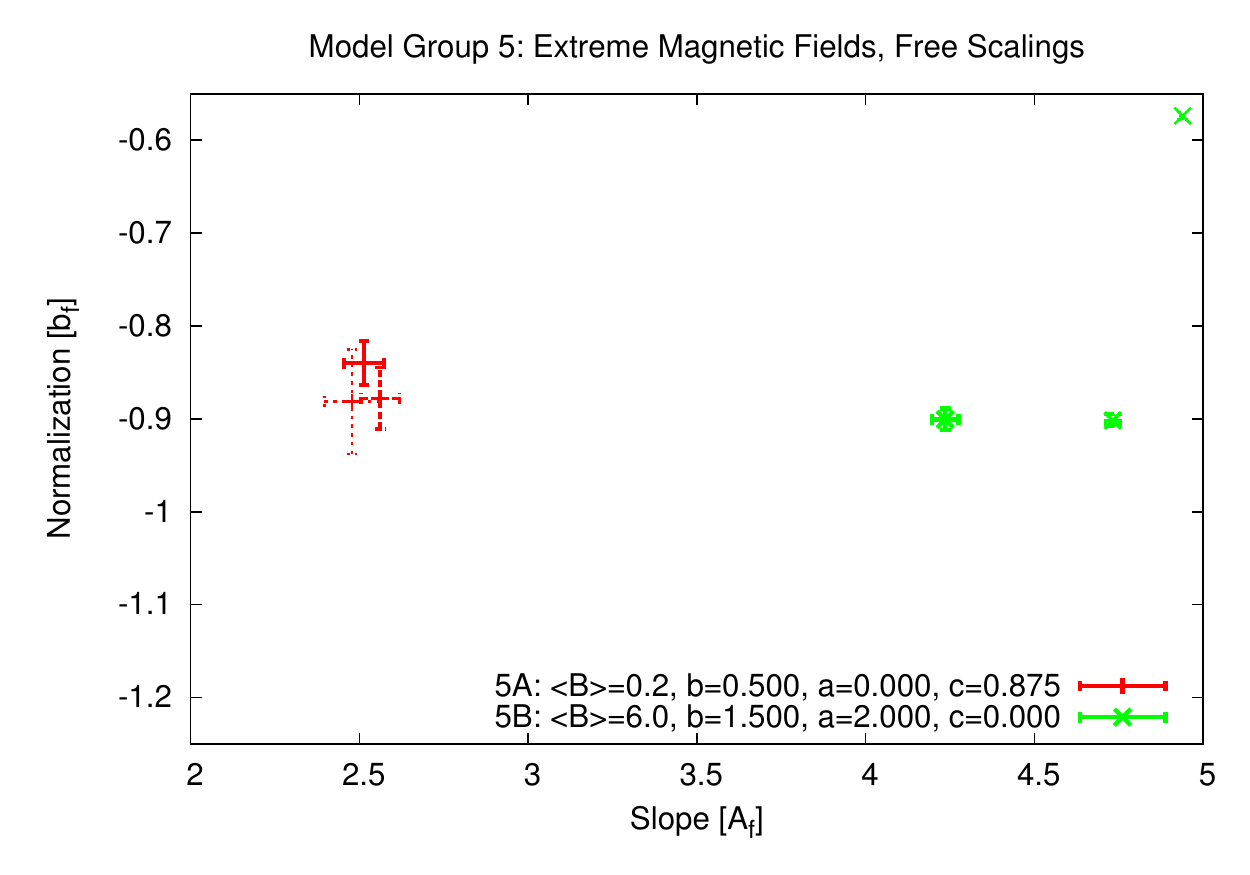}}
  {\includegraphics[width=0.48\textwidth]{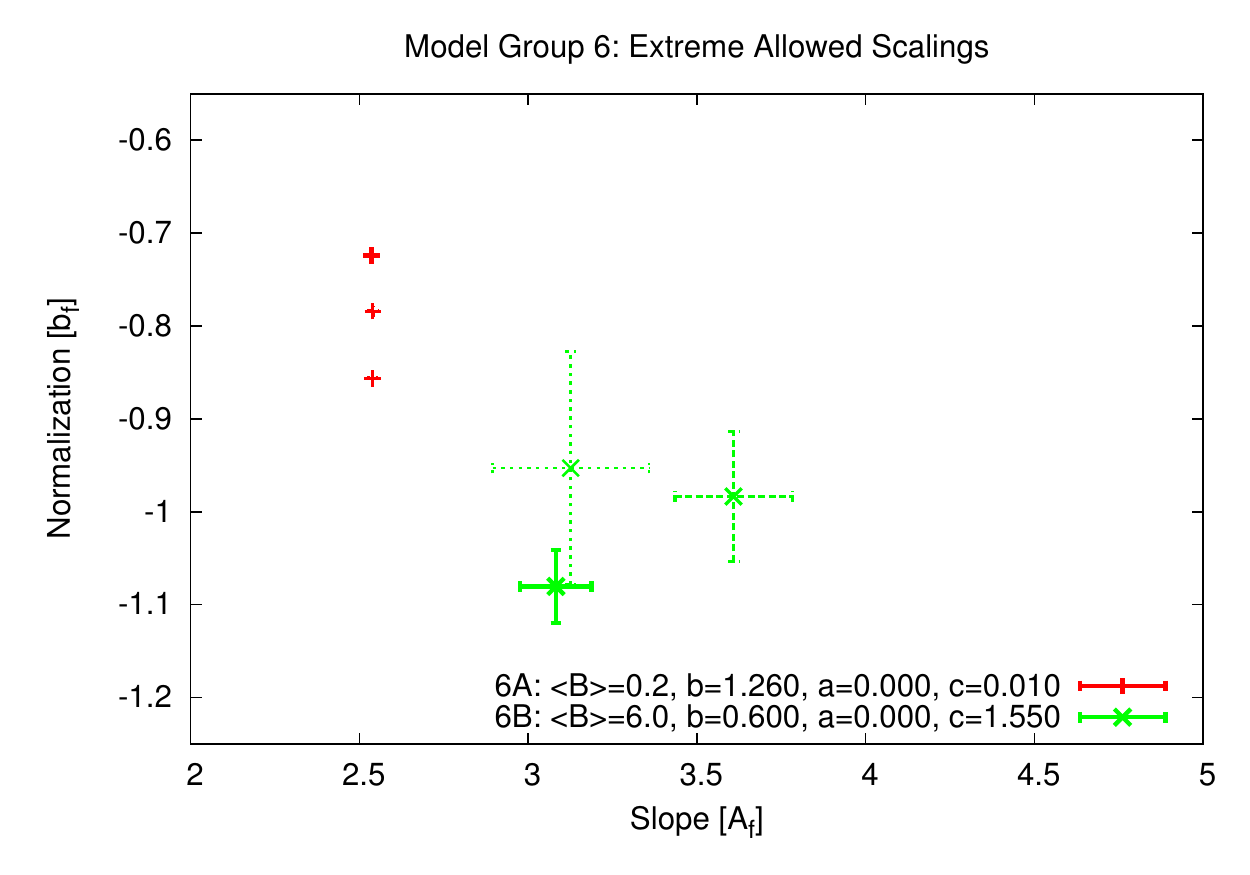}}
  \caption[Radio power - virial mass best fits for each radio halo model]
           {Best fits for the $P_{1.4}-M_v$ relation for each radio halo 
            parameter set.
            Each best fit to~\eqref{\ref{eq:rh_radmassFits}} generates a
            slope and normalization, which we represent as a point with 
            $1\sigma$ error bars. Solid lines are the best fit at $z=0.0$, 
            thick dashed lines are $z=0.25$, and thin dashed lines are $z=0.5$.
            Note that 
            the points for Model Set 5B are from 
            left ($z=0.0$) to right ($z=0.5$) and
            the points for Model Set 6A are from 
            top ($z=0.0$) to bottom ($z=0.5$).
            We have identified each model with its designation 
            from~\tabref{\ref{tab:rh_rhModels}} and the portions of the model
that change in the given model group. $\aveb$ is given in units of ${\rm \mu
G}$.  } \label{fig:rh_radMassFits} 
\end{figure*}

In \figref{\ref{fig:rh_radMassFits}} we show the best-fit slope and 
normalization for each model, grouped by model group, for the 
$P_{1.4}-M_v$ relation generalized from (\eqref{\ref{eq:rh_radmass}}):
\begin{equation}
  \log \left[ \frac{P_{1.4}}{3.16 \times 10^{24} h_{70}^{-1} 
     {\rm W}~{\rm Hz}^{-1} } \right] = 
     A_f \log \left[ \frac{M_v}{10^{15} \hsmsol} \right]
     + b_f
\label{eq:rh_radmassFits}
\end{equation} 
where $A_f$ and $b_f$ (note that we have added the subscript $f$ for ``fit'' 
to distinguish these from the parameters used in our radio power model) are the
slope and normalization, respectively. We fix the scaling parameter $C_s$ at
each redshift for each model. In essence this assumes that the $\pmvir$
relation holds even at high redshift. For each model we show three points: one
each for $z=0.0$, $0.25$, and $0.5$. Above redshift $0.5$ we do not have enough
halos above the minimum mass threshold to generate meaningful statistics. Note
that the observational uncertainties essentially fill the entire plotting
space, meaning that our derived fits are largely consistent with 
observations at $z=0.0$, even when considering the full mass range in our 
high-resolution sample.

For Model Group 1, where we vary only the magnetic field parameters, we see a
progressive steepening of the slope with higher $b$ values, as expected. Note
especially the differences between models 1C and 1D, which have identical
values of $\aveb$. The values of $b$ and $\aveb$ also jointly affect the
normalization of the $\pmvir$ relation, with smaller values $b$ generally
leading to lower normalizations. The redshift evolution of the models in this
Model Group shows diverse behavior. For models with $b<1.0$ the normalization
tends to decrease with increasing redshift, but not very significantly. This
makes sense as the clusters are in general uniformly smaller at higher
redshift. However, the slope increases at $z=0.25$, which perhaps suggests
greater variance in the turbulent properties of the clusters.  The
uncertainties in the values for redshifts $0.25$ and $0.5$ make them difficult
to compare against each other. With $b>1.0$, the trend reverses and the
normalization tends to increase with high redshift. 

In Model Group 2, where we keep the magnetic field fixed and vary the scaling
with turbulent pressure, we see that, as expected, larger values of $c$ lead to
steeper slopes and lower normalizations in the best-fit relation.  Since our
scatter is related to the turbulent pressure, models with higher values of $c$
will have correspondingly larger uncertainties.  We see similar redshift
dependence for Model Set 2A as in Model Group 1, but the Model Set 2B displays
reverse behavior (increasing normalization with redshift), although the
uncertainties are so large as to make firm statements difficult.  The steep
dependence on turbulent pressure overwhelms the general mass dependence, so
that even though the clusters are in general smaller at higher redshift
(leading to a lower normalization), the turbulence in the most massive cluster
(which is used to fix the normalization) increases, negating the mass effect.

For Model Group 3 we see that exchanging $a$ for $c$ in the radio power model
does lead to small differences in the slope and normalization. When we set
$c=0$ (so that there is no dependence on turbulent pressure), we essentially
eliminate the uncertainties. The small differences are due to the fact that our
scaling relation between $\Gamma_v$ and $M_v$ is only a best-fit approximation,
and that scatter in that relation can affect the resulting $\pmvir$ relation.
All the points, however, are within $2\sigma$ of each other. All of 
these models
show identical redshift dependence.

We see the drastic effects of changing the assumed average magnetic field in
Model Group 4. Higher magnetic fields lead to higher normalizations and flatter
slopes. However, the redshift evolutions exhibit opposite trends, such that at
$z=0.5$ the effects of the magnetic field are indistinguishable from each
other. For weak magnetic fields, the CRs are dominated by their
interactions with the CMB, and the equivalent pressure of the CMB increases
with redshift, lowering the synchrotron power at higher $z$. For strong fields,
the dependence on cluster mass is more explicit, and at higher redshifts the
clusters are, in general, smaller and the mass distribution has a steeper
slope.

In Model Group 5 we see that the dependence on the scalings overwhelms the
dependence on the average magnetic field. Even though Model Set 5A has the same
$\aveb$ as Set 4A, the dependence on $b$, $a$, and $c$ forces a much flatter
slope. Thus we may conclude that the effects of average magnetic field are
degenerate with the scaling parameters, although the parameters taken
individually can lead to significant differences. Note that Model Set 5A 
is the only set that is inconsistent with current observations of the 
$\pmvir$ relation, although this discrepancy only occurs when including 
all masses within our high-resolution sample. Similar behaviors are
displayed by Model Group 6, although the error bars are so small because of the
weak dependence on the turbulent pressure.

\begin{figure*}
  \centering
  {\includegraphics[width=0.49\textwidth]{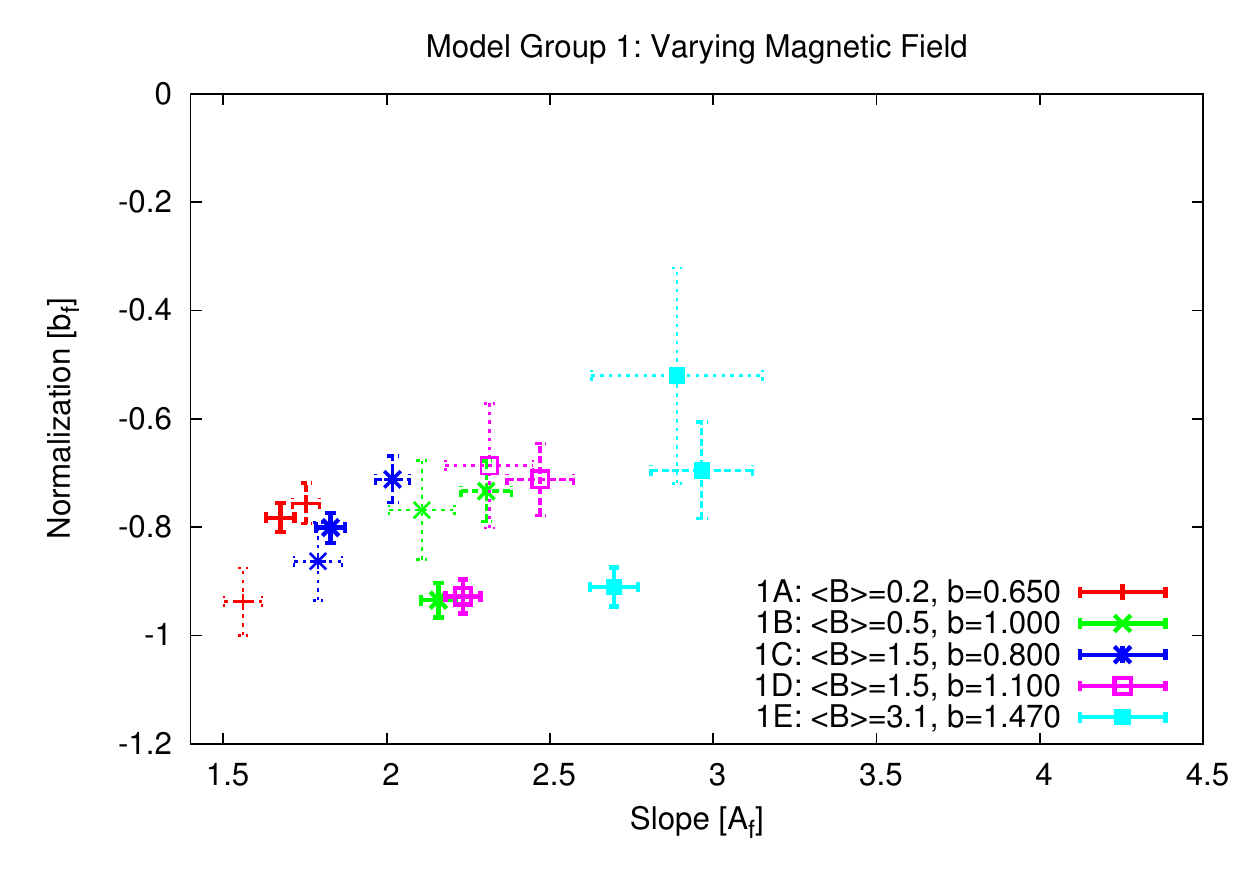}}
  {\includegraphics[width=0.49\textwidth]{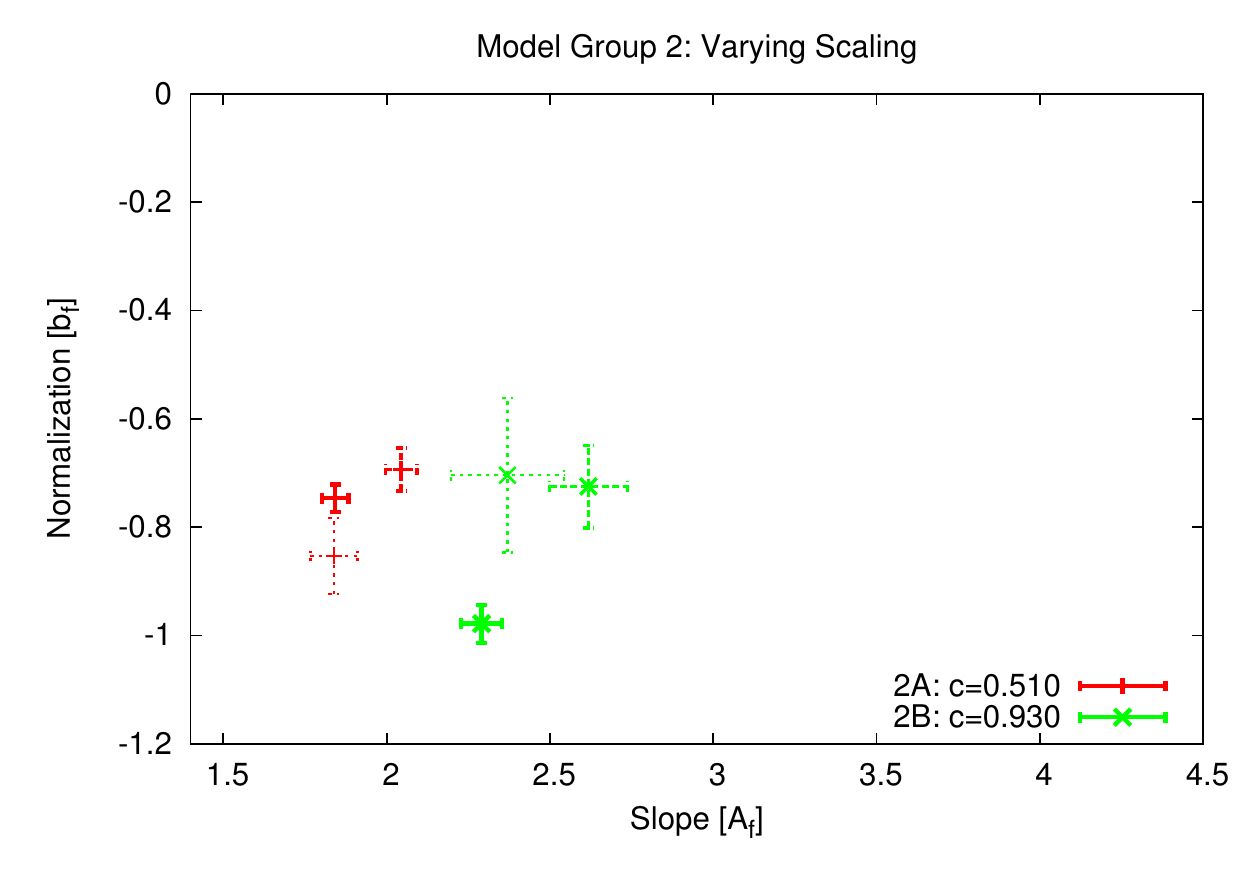}}
  {\includegraphics[width=0.49\textwidth]{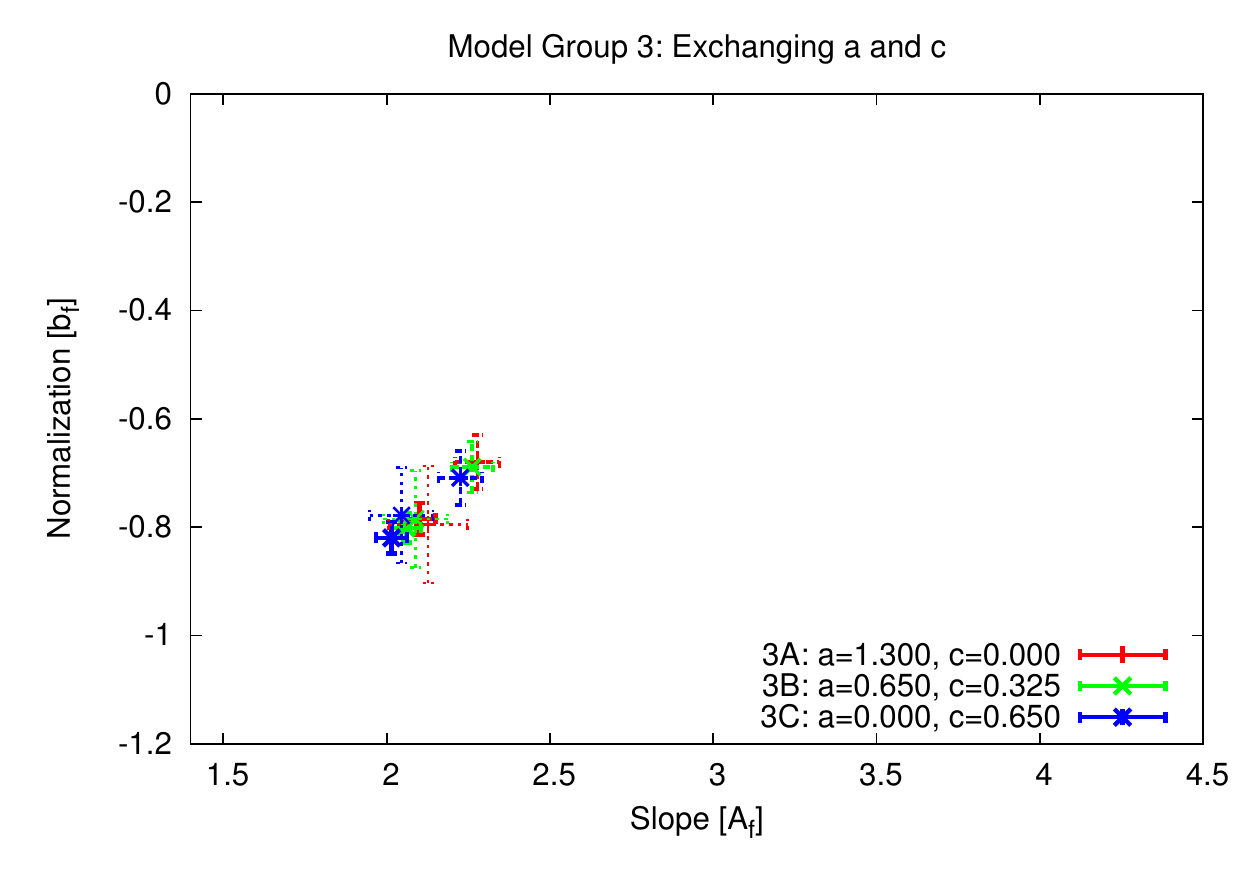}}
  {\includegraphics[width=0.49\textwidth]{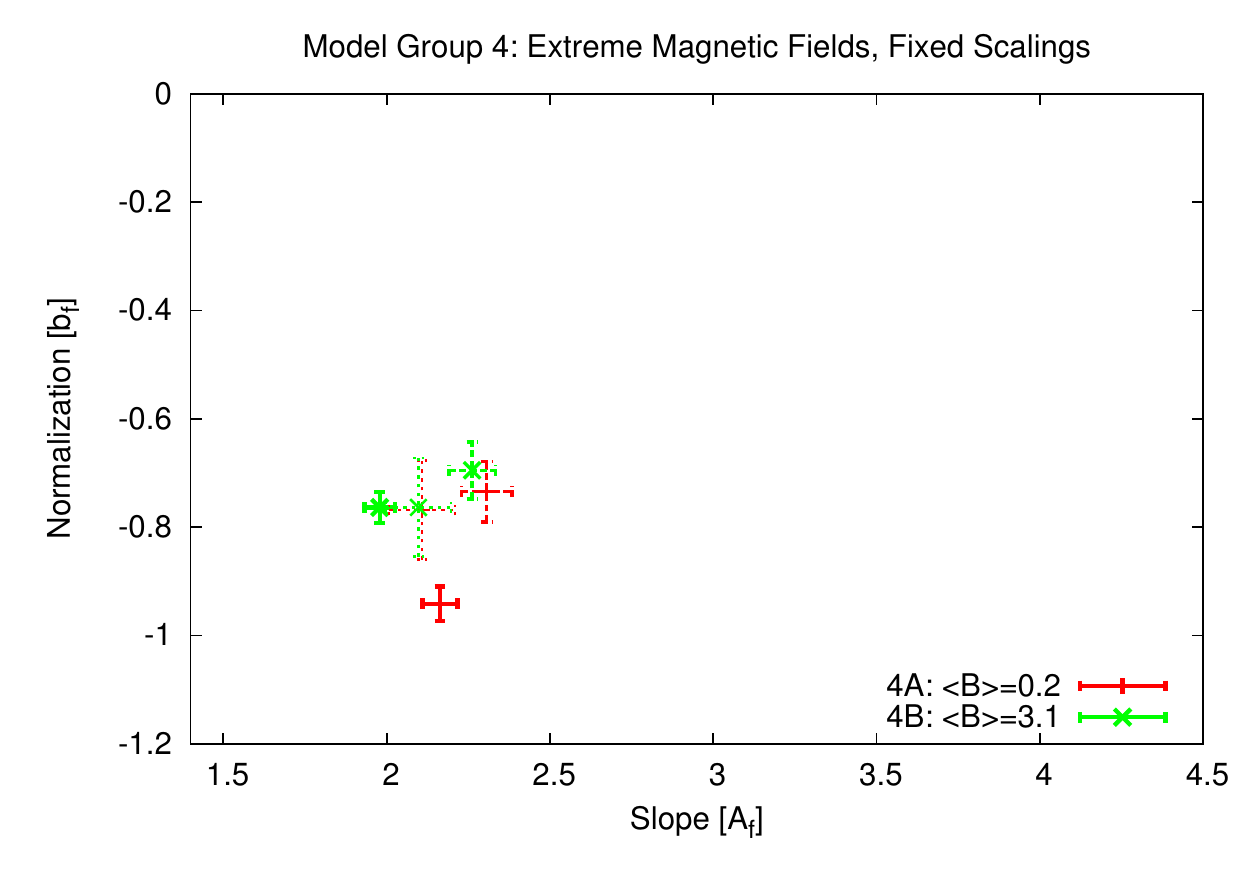}}
  {\includegraphics[width=0.49\textwidth]{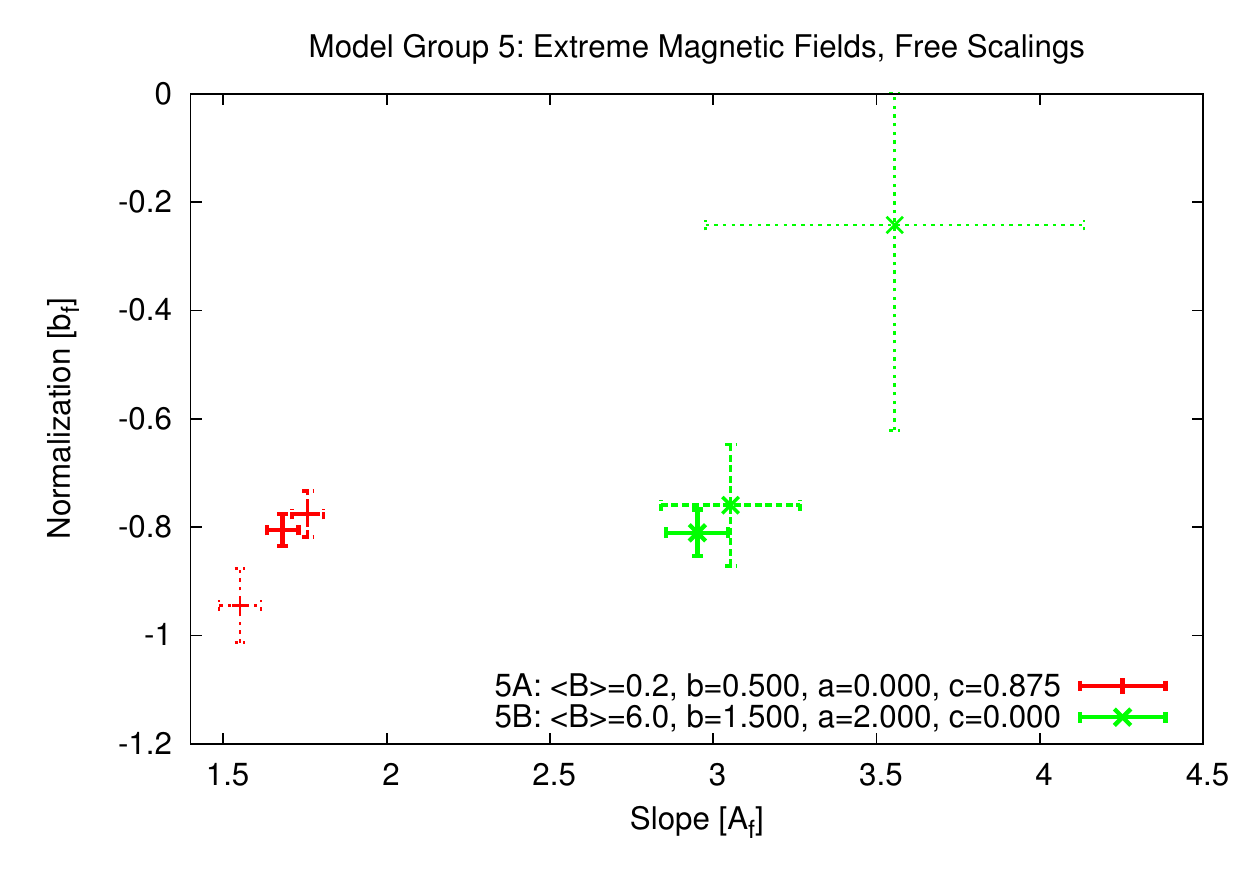}}
  {\includegraphics[width=0.49\textwidth]{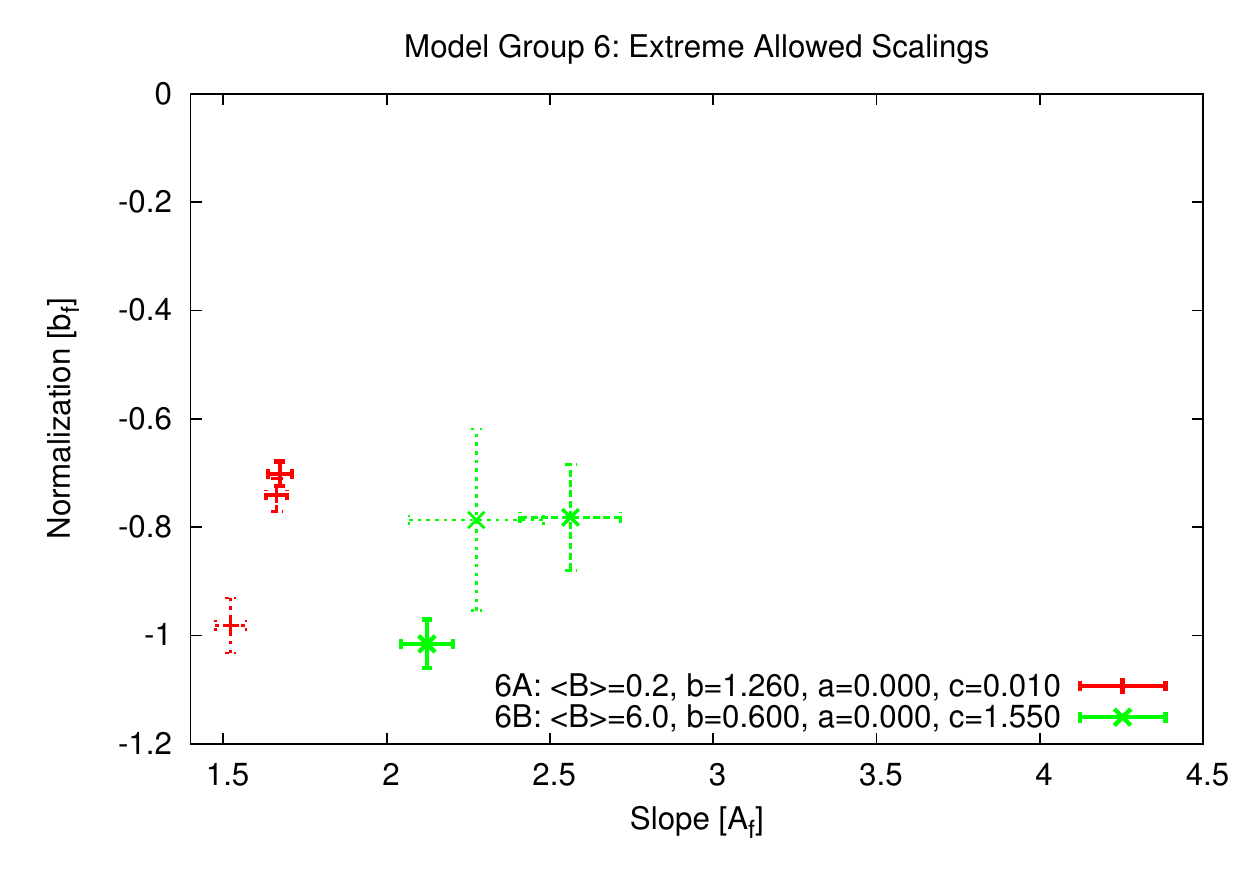}}
  \caption[Radio power - X-ray luminosity best fits for each radio halo model]
           {Best fits for the $P_{1.4}-L_x$ relation for each radio halo 
            parameter set. Methods and colors 
            are identical to~\figref{\ref{fig:rh_radMassFits}}.
            We have identified each model with its designation
            from~\tabref{\ref{tab:rh_rhModels}} and the portions of the model
that change in the given model group. $\aveb$ is given in units of ${\rm \mu
G}$.} \label{fig:rh_radXrayFits}  
\end{figure*}

In \figref{\ref{fig:rh_radXrayFits}} we repeat the above analysis for the 
$P_{1.4}-L_x$ relation:
\begin{eqnarray}
  \log \left[ \frac{P_{1.4}}{3.16 \times 10^{24} h_{70}^{-1} 
     {\rm W}~{\rm Hz}^{-1} } \right]  =  \\
    A_f \log \left[\frac{L_x}{10^{45} h_{70}^{-1} {\rm ergs}~{\rm s}^{-1}} 
     \right] + b_f. \nonumber
\label{eq:rh_radxrayFits}
\end{eqnarray}  
We find similar behaviors as in the $\pmvir$ relation plots above, except that 
our error bars are generally larger due to scatter from our estimates of 
$L_x$. Since X-ray luminosities are generally easier than virial masses 
to compute from observations, observations of many more radio halos 
may reduce the statistical uncertainties to such a level as to potentially 
distinguish the allowed scalings and dependencies. The redshift evolution 
of the $P_{1.4}-L_x$ relation in particular may provide a way of 
determining the dominant components of radio power and the average 
magnetic strength of clusters, as we have discussed above for 
the $P_{1.4}-M_v$ relation.. 

In general, if the observational uncertainties in $A_f$ and $b_f$ are reduced
by approximately a factor of two, many degeneracies in the model parameters
will be eliminated. We note that we are basing this analysis on our sample of
only $131$ clusters. While this is significantly more than the current known
number of radio halos, it is still far fewer than we expect to see with
instruments such as LOFAR, as we will see below. While more objects could
reduce the uncertainty, the precise amount of error also depends on the
intrinsic scatter in the observed relations, which can be affected by biases.
We also assume a power-law relationship holds between radio power and cluster
mass even to low-mass clusters, which may not be the case once significant
numbers of low-luminosity halos are detected. 

\section{Luminosity functions and radio halo counts}
\label{sec:rh_counts}

We now turn to a discussion of these models in terms of total counts of all
objects in the simulation. To do this, we must assign a rest-frame radio
luminosity to each cluster in the simulation box, even if it is not at high
resolution. We accomplish this by combining the derived $\Gamma_v-M_v$
relations described above with a relationship between $M_v$, which includes
gas, and $M_{v,{\rm DM}}$, which only includes dark matter.  Our smallest
clusters are only a few zones across and thus do not contain enough gas zones
to accurately capture the contribution of the gas to $M_v$.  However, we
defined our FOF halo completeness limit so that we can always get reliable
evaluations of $M_{v,{\rm DM}}$ (i.e., $R_v \geq \Delta x$, where $\Delta x$ is
the resolution of our pre-refinement uniform grid). This gives us an
interpolated value of $\Gamma_v$ for each fixed-resolution cluster, which we
then feed into~\eqref{\ref{eq:rh_model}} to generate a radio power for that
cluster. 

\begin{figure}
  \centering
  \includegraphics[width=\columnwidth]{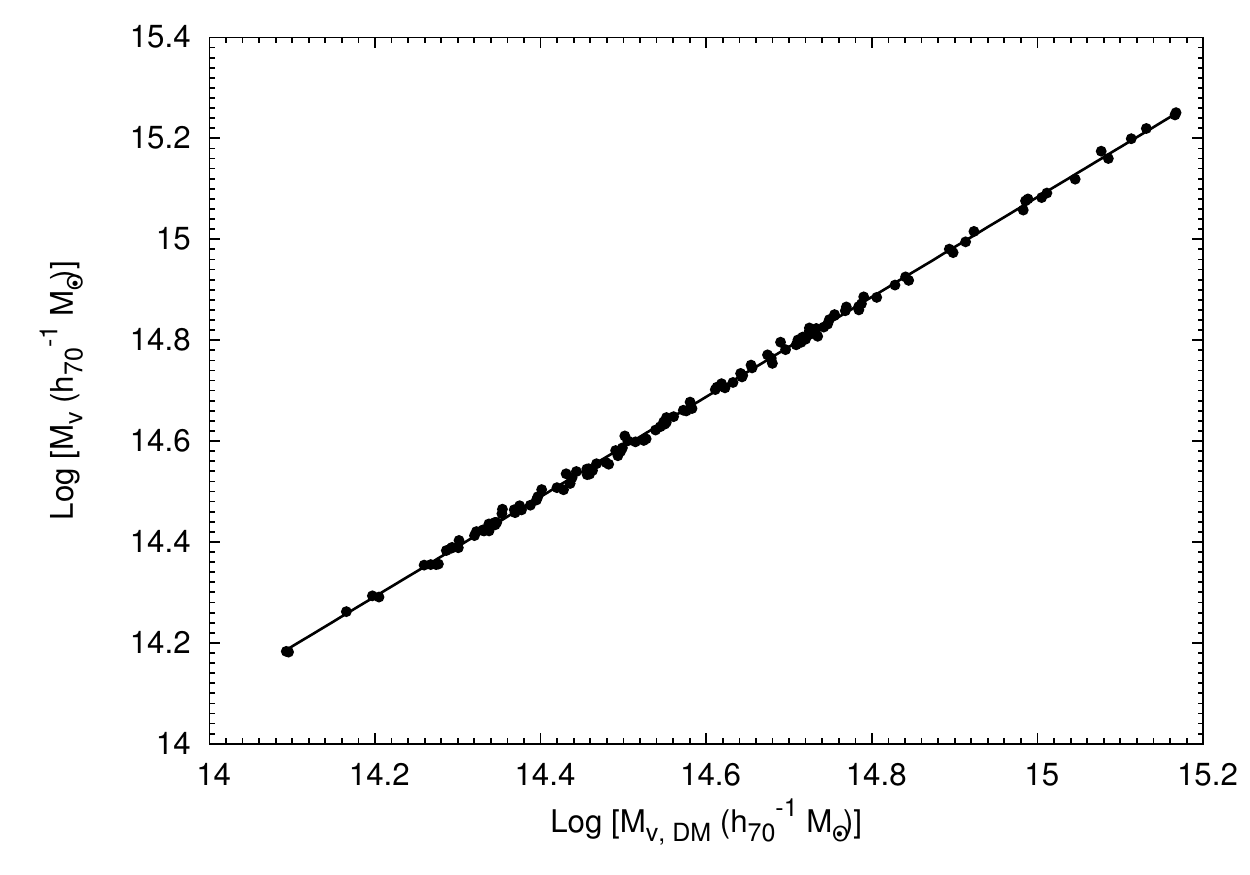}
  \caption[$M_v$ - $M_{v,{\rm DM}}$ relation.]
           {Data for (points) and best fit to (solid line) the 
           $M_v$ - $M_{v,{\rm DM}}$ relation. Points are 
           the high-resolution cluster sample.}
\label{fig:rh_mvirm200}
\end{figure}

We find a very tight correlation between $M_v$ and $M_{v,{\rm DM}}$ 
for our high-resolution sample, as shown 
in~\figref{\ref{fig:rh_mvirm200}}. We fit a line to these data and found the 
correlation to be
\begin{equation}
  \log \left[ \frac{M_v}{10^{15} \hsmsol }\right] = 
     0.99 \log \left[ \frac{M_{v,{\rm DM}}}{10^{15} \hsmsol } \right]
     + 0.08.
\label{eq:rh_mvirm200}
\end{equation} 
This relationship implies a uniform gas fraction consistent 
with other simulations~\citep[e.g.,][]{Stanek2010}.
We use this fit to extract an $M_v$ for each fixed-resolution cluster 
(i.e., those outside the refinement regions) which is then used to compute its
equivalent radio power using the $\Gamma_v-M_v$ relation. While equivalent to
directly interpolating from a $\Gamma_v-M_{v,{\rm DM}}$ relation, we found that
this procedure produces less scatter and hence more reliable interpolations.
We do \emph{not} add additional scatter to the interpolation procedure.  For
the analysis below we will include Poissonian uncertainty where appropriate,
and this dwarfs any uncertainty introduced by scatter. By binning our data for
luminosity functions, the main effect of scatter is to simply
move clusters around within a given luminosity bin, and any clusters that are
scattered into a luminosity bin are roughly offset by clusters scattered out of
the same bin, especially at low luminosities where the function is 
relatively flat.

Not every cluster hosts a radio halo, and for simplicity we will only assign
radio halos to a random sample of $5\%$ of our 
clusters. This is roughly in line with
observations, which indicate that $\sim 1/3$ of clusters above $2 \times
10^{15} \hmsol$ and only $2-5\%$ of smaller clusters host radio
halos~\citep{Cassano2008}.  We do this rather than employing a radio halo 
probability that is a step function in mass 
because we do not have enough high-mass
($> 10^{15} \hmsol$) halos to calibrate our number counts based on the
intrinsic scatter in our derived $\pmvir$ relation.  Note that in the context
of hadronic secondary models, this enforced fraction implies that not every
cluster is sufficiently magnetized to generate radio emission. More
sophisticated techniques to calibrate number counts exist, such as using the
synchrotron break frequency, $\nu_b$, which is used in the identification of
radio halos (see CBS06 for a discussion).  However, again we do not have enough
high-mass objects to use this approach.  With enough high-mass halos, the
precise calibration would depend on our choice of parameters $\aveb$, $b$, $a$,
and $c$. Thus we may be under-counting the number of radio halos at $1.4 \ghz$.
However, we can still gather useful results as to the relative effects of
varying radio halo models.  Also, our analysis will include results at $150
\mhz$, where $\nu_b$ is much lower and hence our results are more valid. To
calculate the radio luminosity at $150 \mhz$, we assume a simple power law with
spectral index $1.2$. Our smallest resolvable cluster has a $1.4 \ghz$ radio
luminosity of $\sim
2 \times 10^{21}~{\rm W}~{\rm Hz}^{-1}$.

\begin{figure*}
  \centering
  {\includegraphics[width=0.49\textwidth]{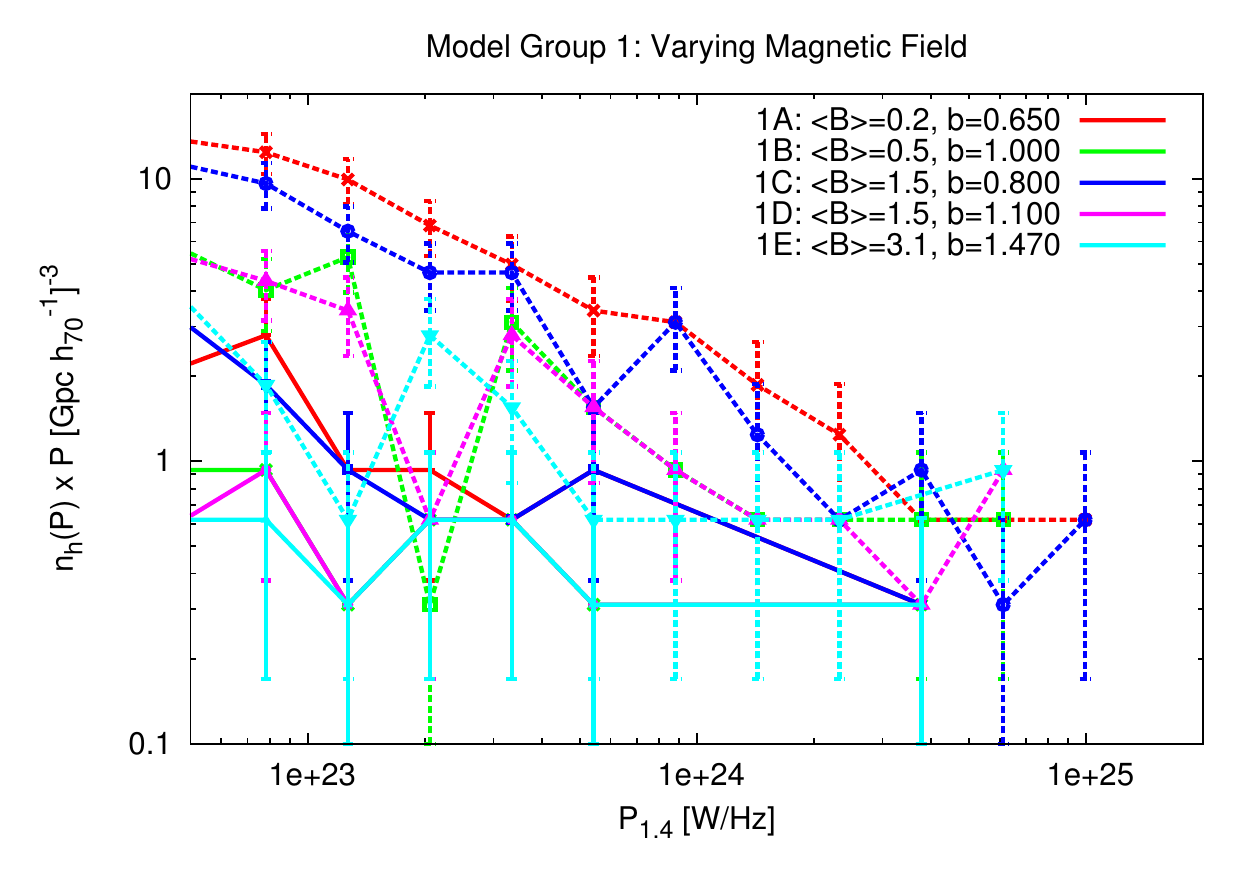}}
  {\includegraphics[width=0.49\textwidth]{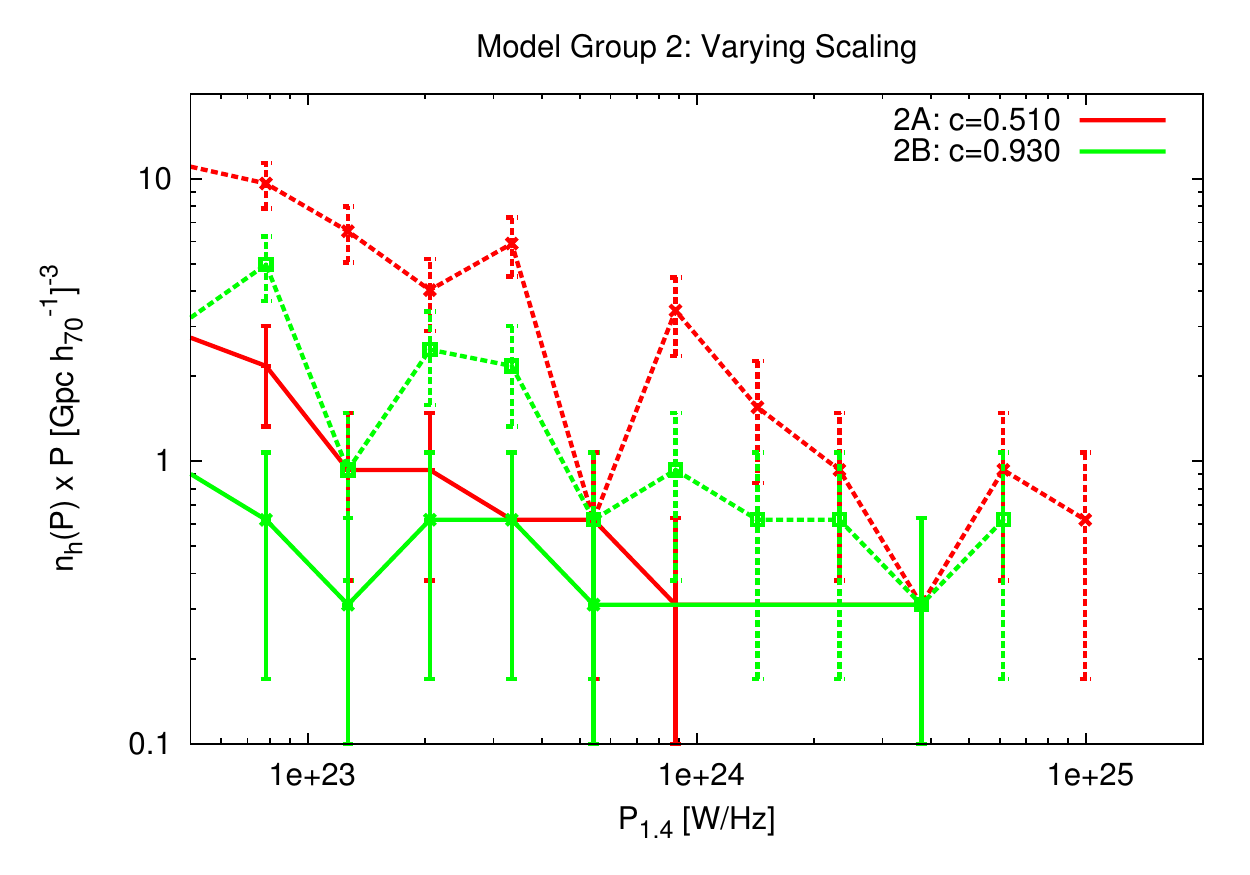}}
  {\includegraphics[width=0.49\textwidth]{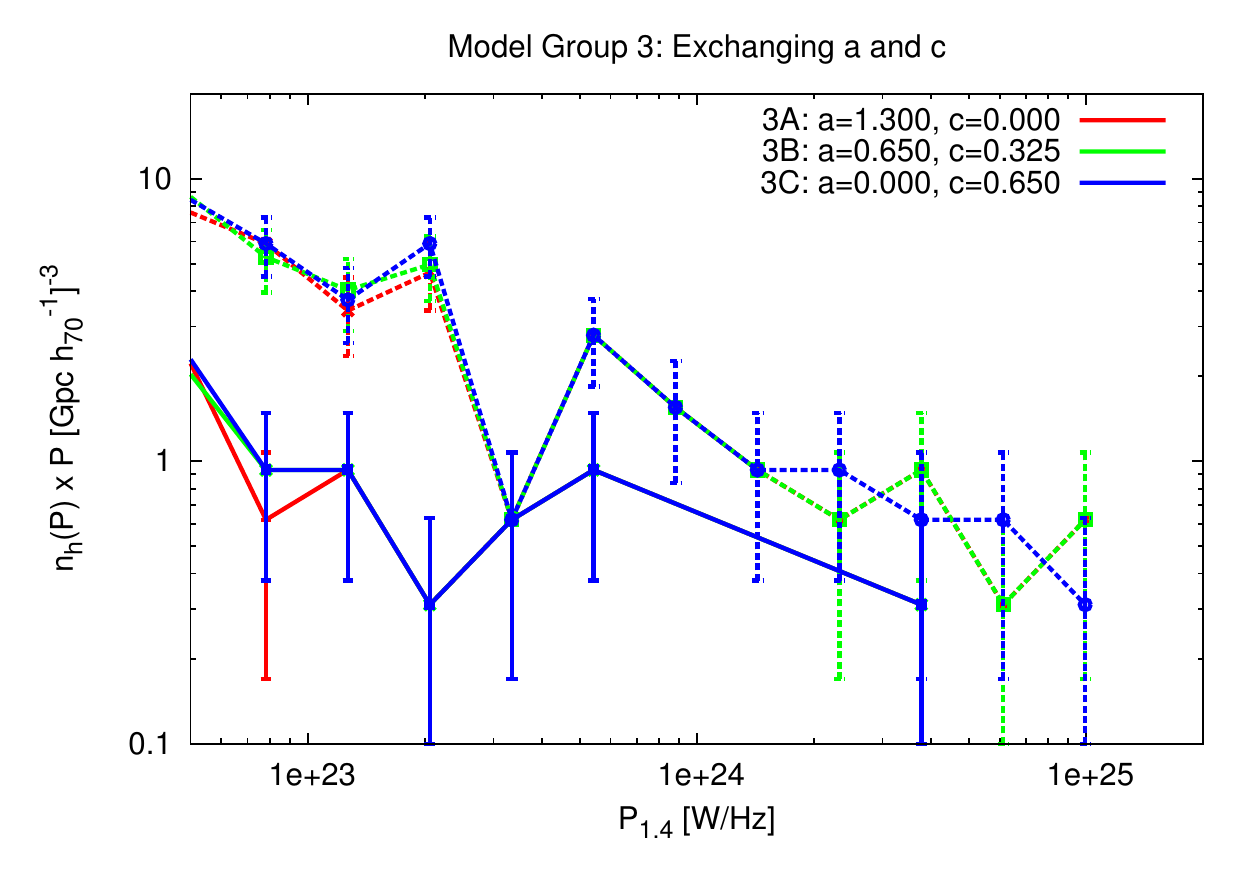}}
  {\includegraphics[width=0.49\textwidth]{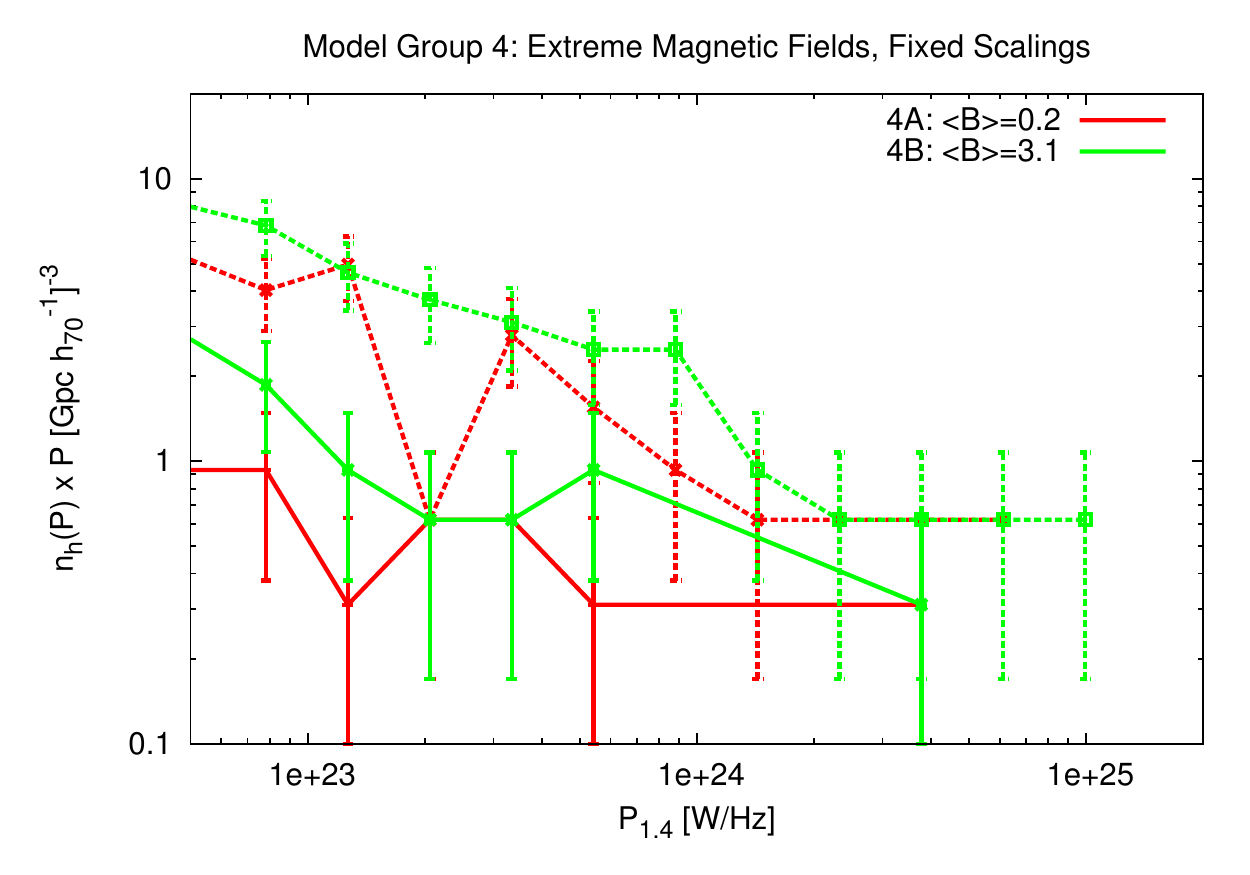}}
  {\includegraphics[width=0.49\textwidth]{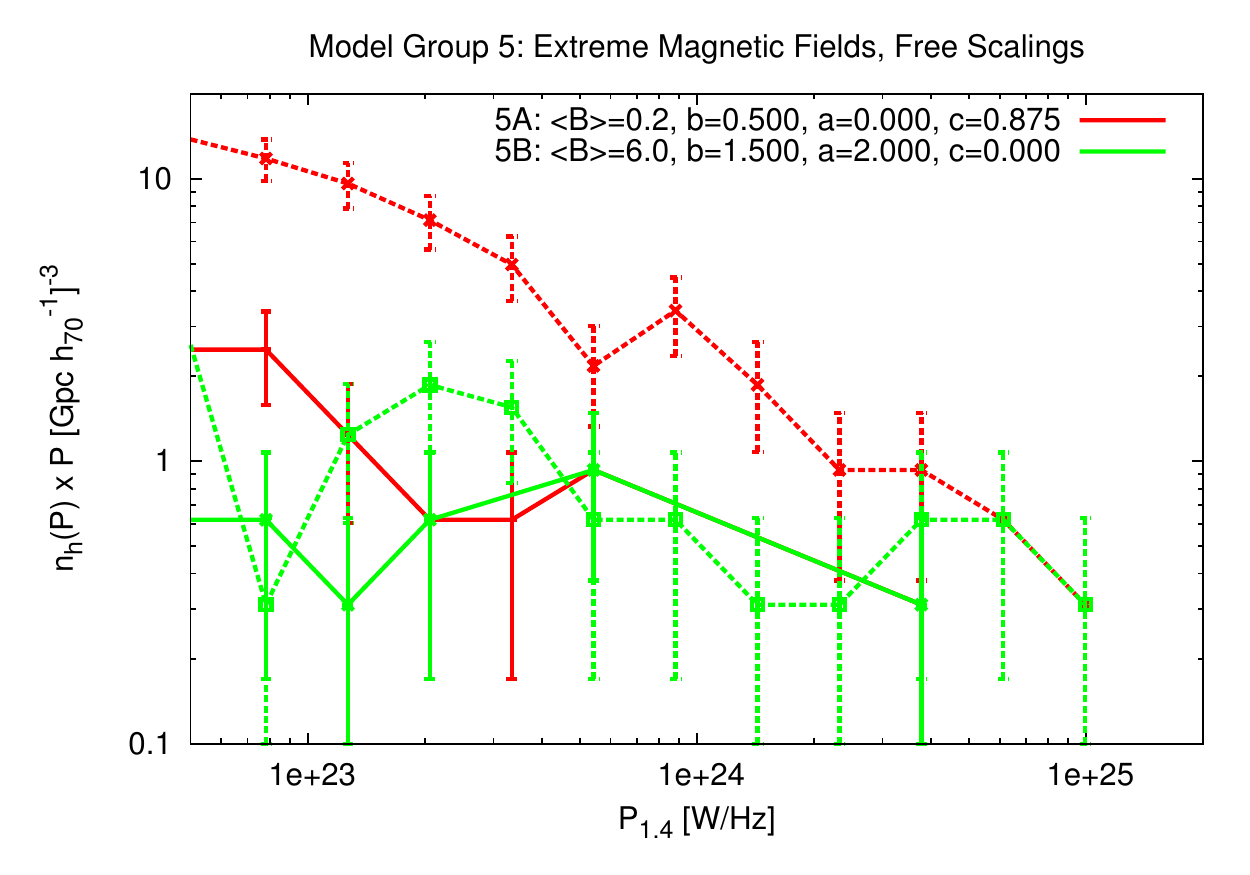}}
  {\includegraphics[width=0.49\textwidth]{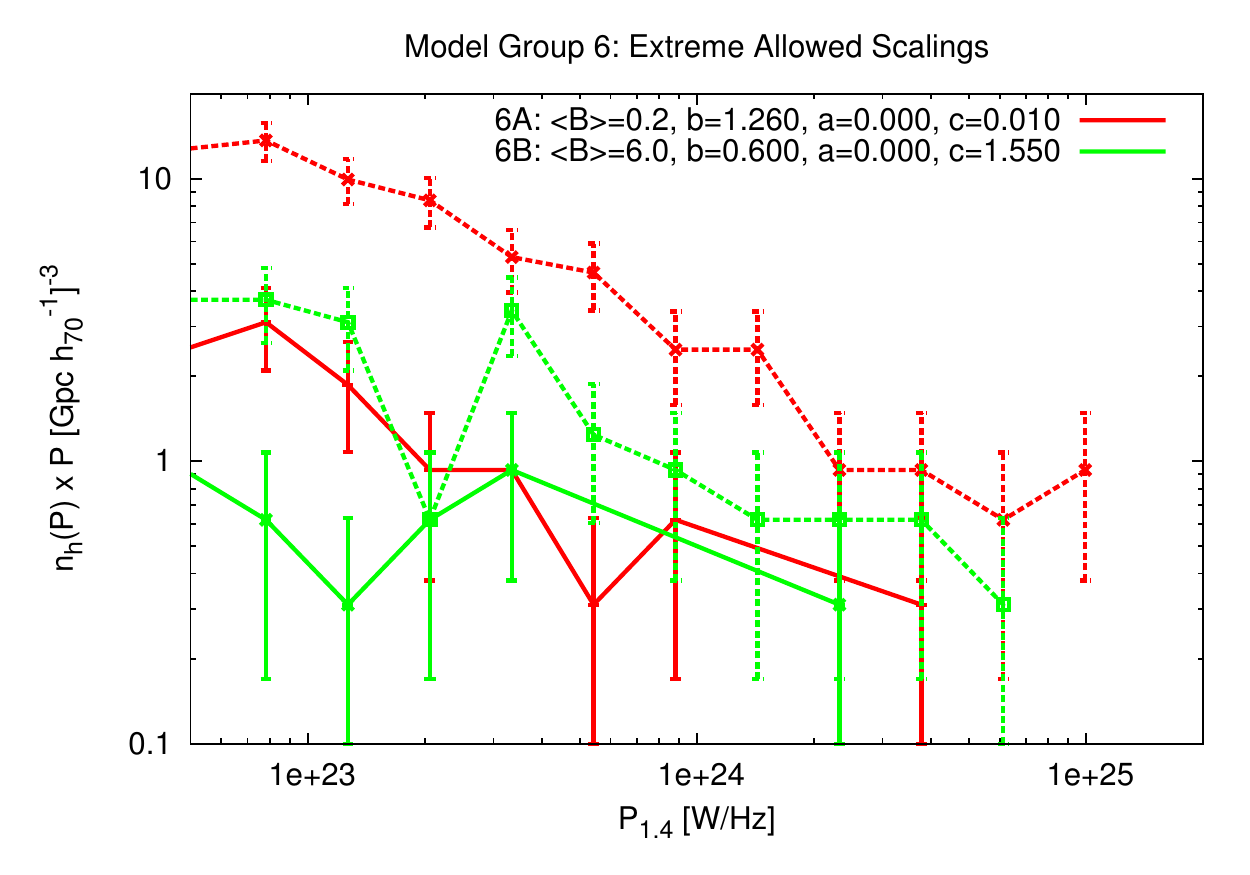}}
  \caption[Radio halo luminosity functions at $z=0.0$.]
           {Radio halo luminosity functions at $z=0.0$. Solid lines are 
            luminosity functions at $1.4 \ghz$ and dashed lines are at 
            $150 \mhz$. Error bars indicate $1\sigma$ Poisson uncertainties.
              We have identified each model with its designation 
                        from~\tabref{\ref{tab:rh_rhModels}} and the portions of
the model that change in the given model group.  $\aveb$ is given in units of
${\rm \mu G}$.} \label{fig:rh_radlumfuncz0.0} 
\end{figure*}

\figref{\ref{fig:rh_radlumfuncz0.0}} shows our calculated radio halo
luminosity functions at redshift $z=0.0$ at $1.4 \ghz$ and $150 \mhz$. At high
masses our luminosity function is well below the estimated values
of~\citet{Ensslin2002}.  However, this is not unexpected due to our limited
simulation volume and cosmic variance, and the fact that they assume a 
Press-Schechter mass function with a fixed $1/3$ fraction of clusters hosting 
radio halos. For Model Group 1, in which we fix the dependence on cluster mass
and turbulent pressure but vary the average magnetic field strength and scaling
of magnetic field with cluster mass, we see a bifurcation in the luminosity
function at low luminosities: models with $b<1.0$ produce up to a factor of two
more low-luminosity radio halos than those models with $b>1.0$.  The
distinction is much more significant at $150$~MHz, where more objects allow for
smaller uncertainties (assuming perfect detector sensitivity --- a point
that we will address later in the discussion of number counts). 
However, despite the difference in the $\pmvir$ relations,
models 1A and 1C are largely indistinguishable from each other, as are models
1B, 1D, and 1E. Number counts at low luminosity are significantly reduced when
the magnetic field scaling $b$ is greater than unity.

We see in Model Group 2, in which we keep the magnetic field values fixed, that
steeper scalings with turbulent pressure produce many fewer radio halos at both
$1.4$~GHz and $150$~MHz. The models at $1.4$~GHz are only statistically
distinguishable at the lowest luminosities, but the models at 
$150$~MHz are easily
separable throughout almost the entire range of radio luminosities. Similar
conclusions can be made regarding Model Group 4, which brackets the extreme
allowed magnetic fields with fixed scalings, where strong differences in the
assumed average magnetic field strength lead to somewhat distinguishable
differences. These behaviors persist for Model Groups 5 and 6, in which all
parameters are allowed to vary, with the general rule being that it is
difficult to separate these models when only relying on $1.4$~GHz halo counts.
The only models that remain inseparable are those in Model Group 3, in which
our exchanges of $a$ and $c$ lead to degenerate radio halo counts at both
frequencies, as expected.

To accumulate total and binned counts we set a vantage point in the center of
our computational domain and find halos whose locations lie on the light cone
emanating from this position. To do this we use saved checkpoint files
at $z=0.0$, $0.25$, $0.5$, $0.75$, and $1.0$.  
Moving outwards in small ($\Delta z = 0.05$) redshift
slices covering the same range, we locate the nearest output (in redshift) to
each slice and use the center-of-mass peculiar velocities of the clusters found
in that output to estimate new positions in the slice under consideration.  We
then compute the flux as $P/(4 \pi d_L^2)$, where $d_L$ is the luminosity
distance of the cluster. We do this at both $1.4 \ghz$ and $150 \mhz$ assuming
a spectral index of $1.2$.

\begin{figure*}
  \centering
  {\includegraphics[width=0.49\textwidth]{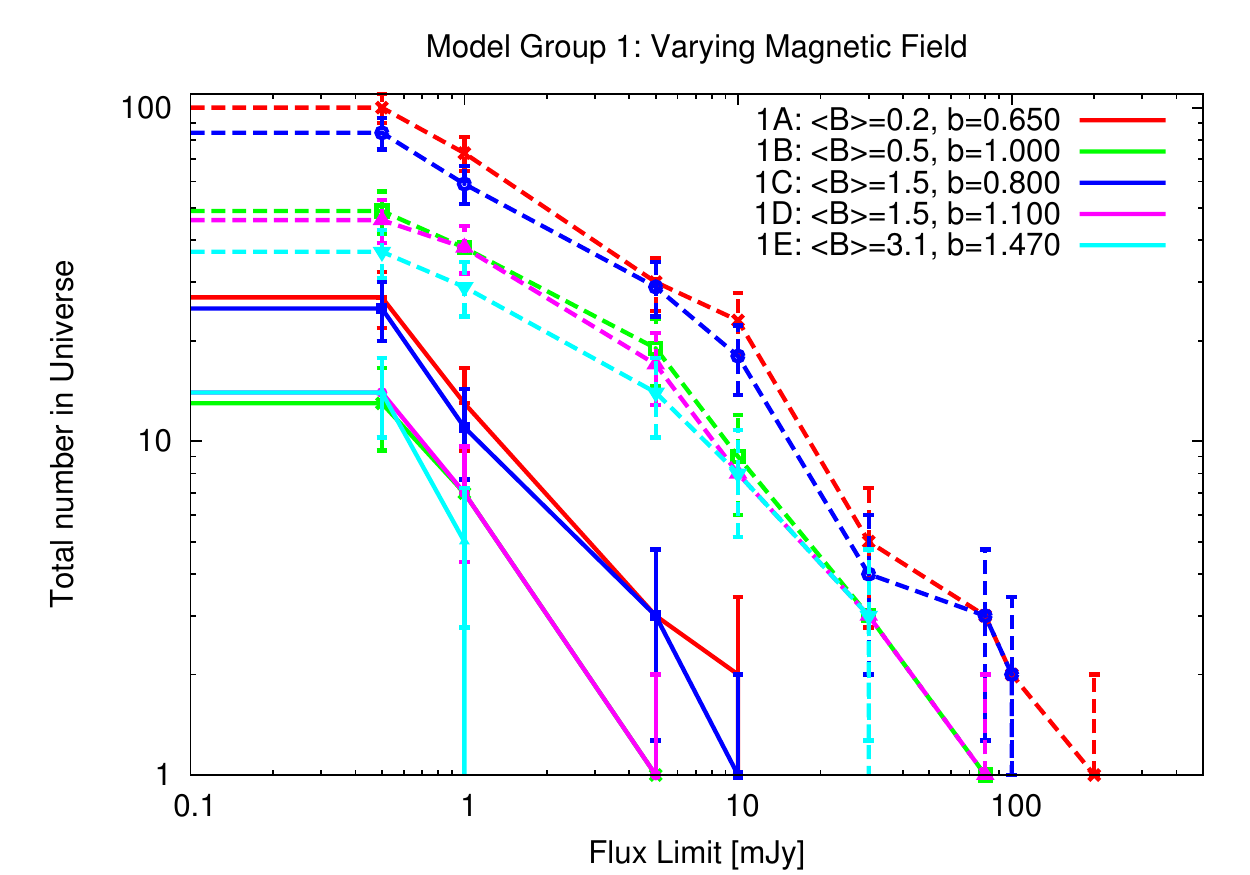}}
  {\includegraphics[width=0.49\textwidth]{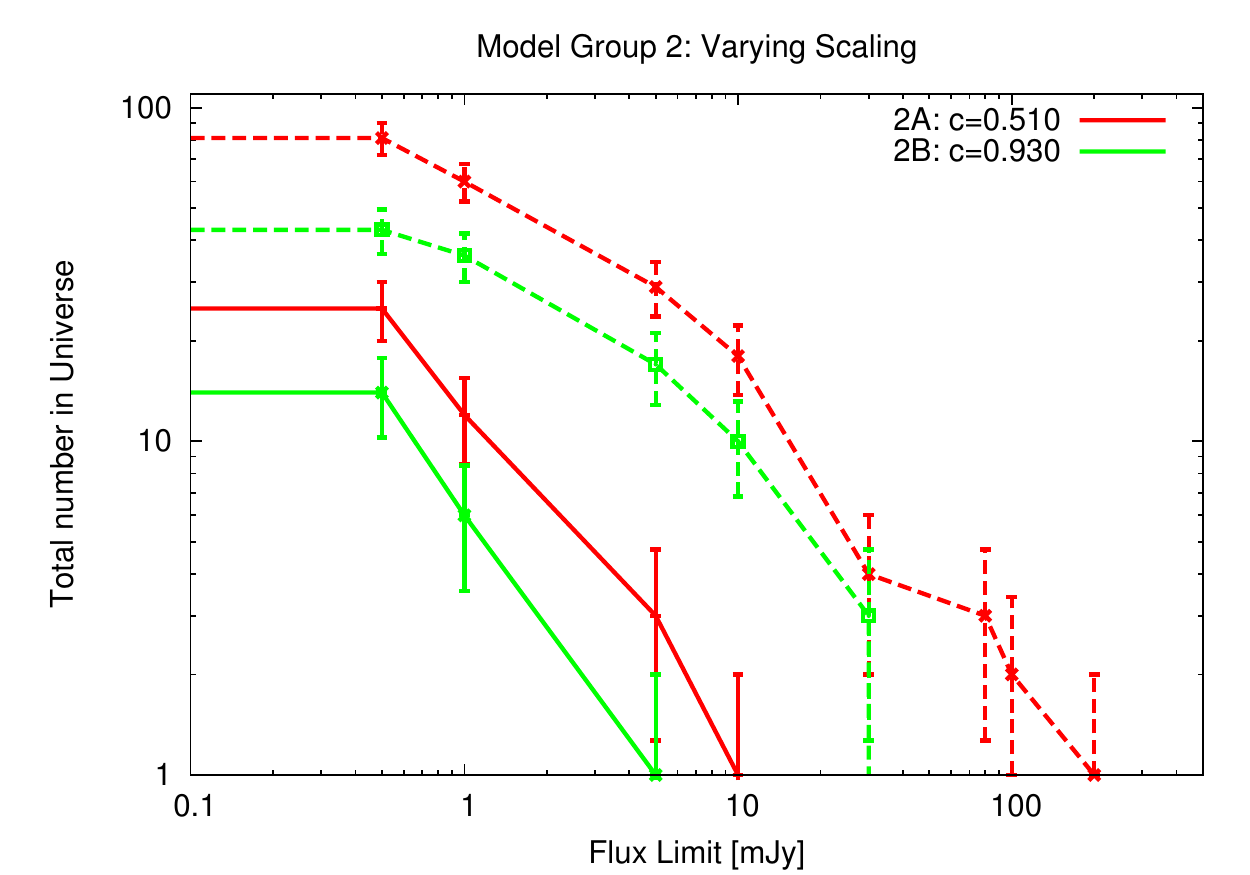}}
  {\includegraphics[width=0.49\textwidth]{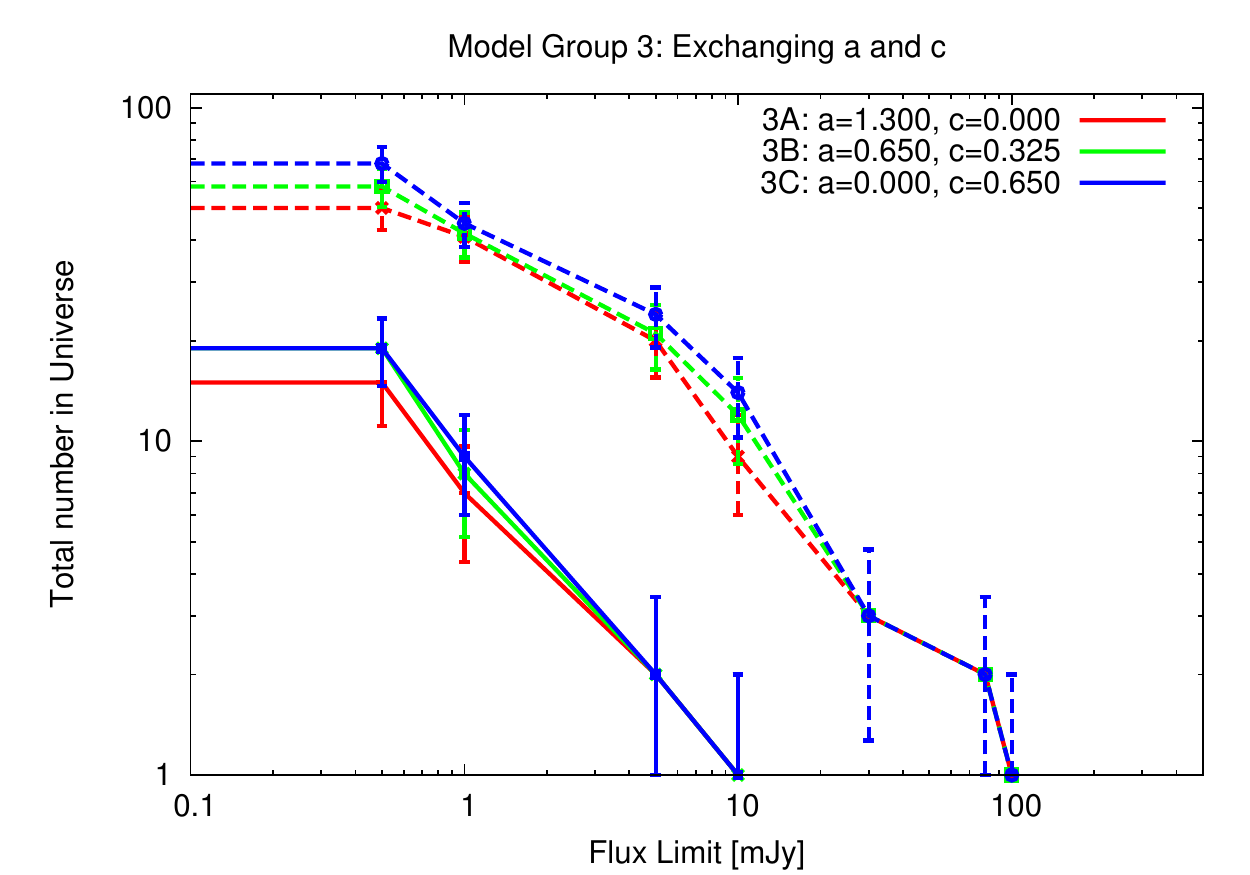}}
  {\includegraphics[width=0.49\textwidth]{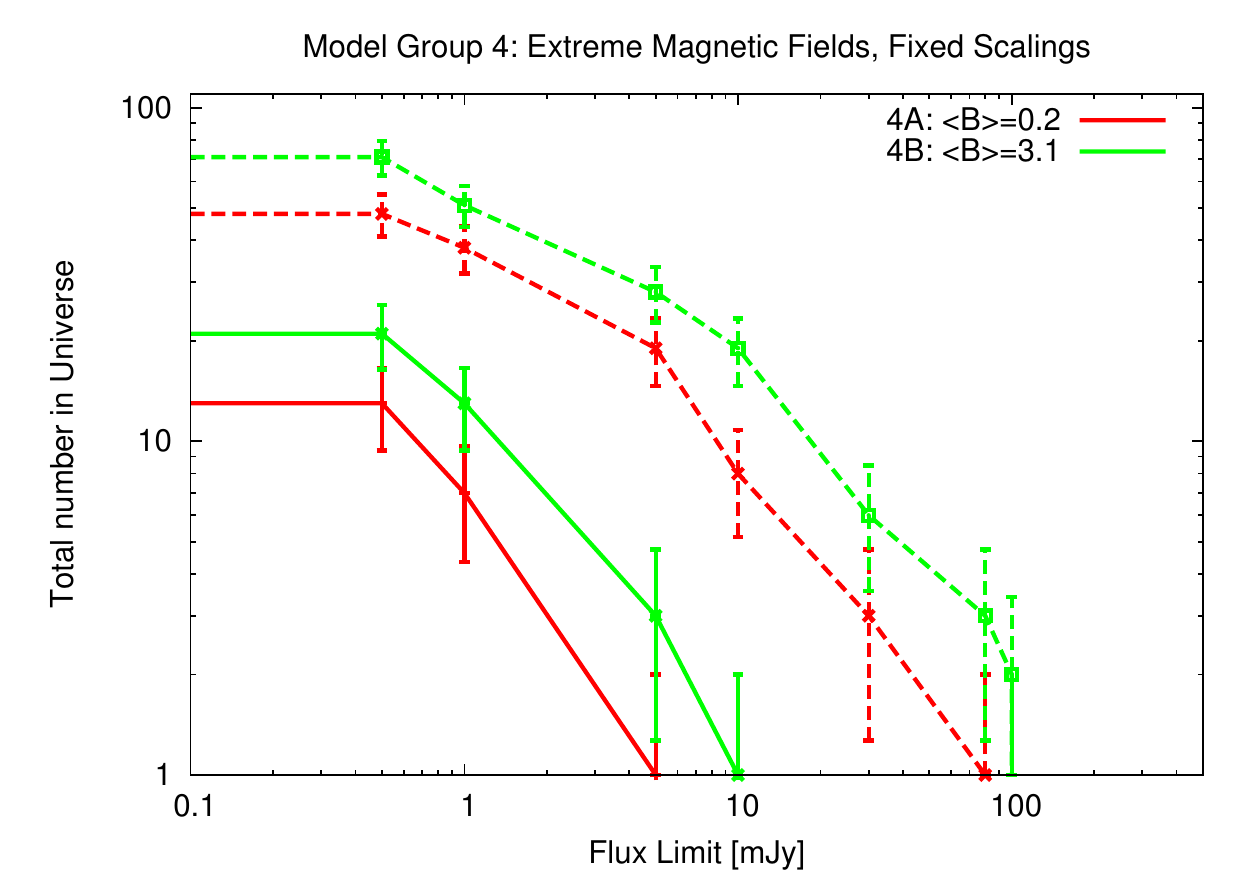}}
  {\includegraphics[width=0.49\textwidth]{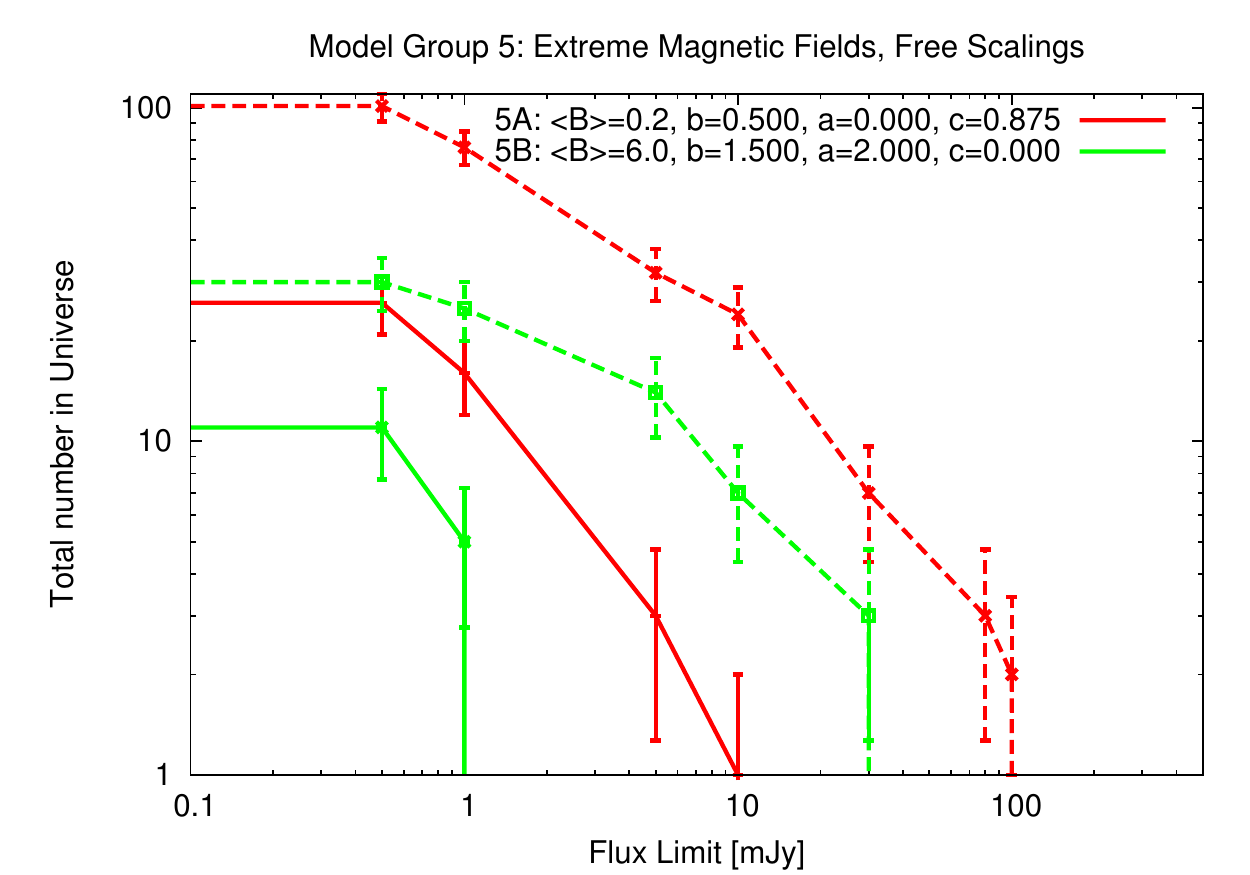}}
  {\includegraphics[width=0.49\textwidth]{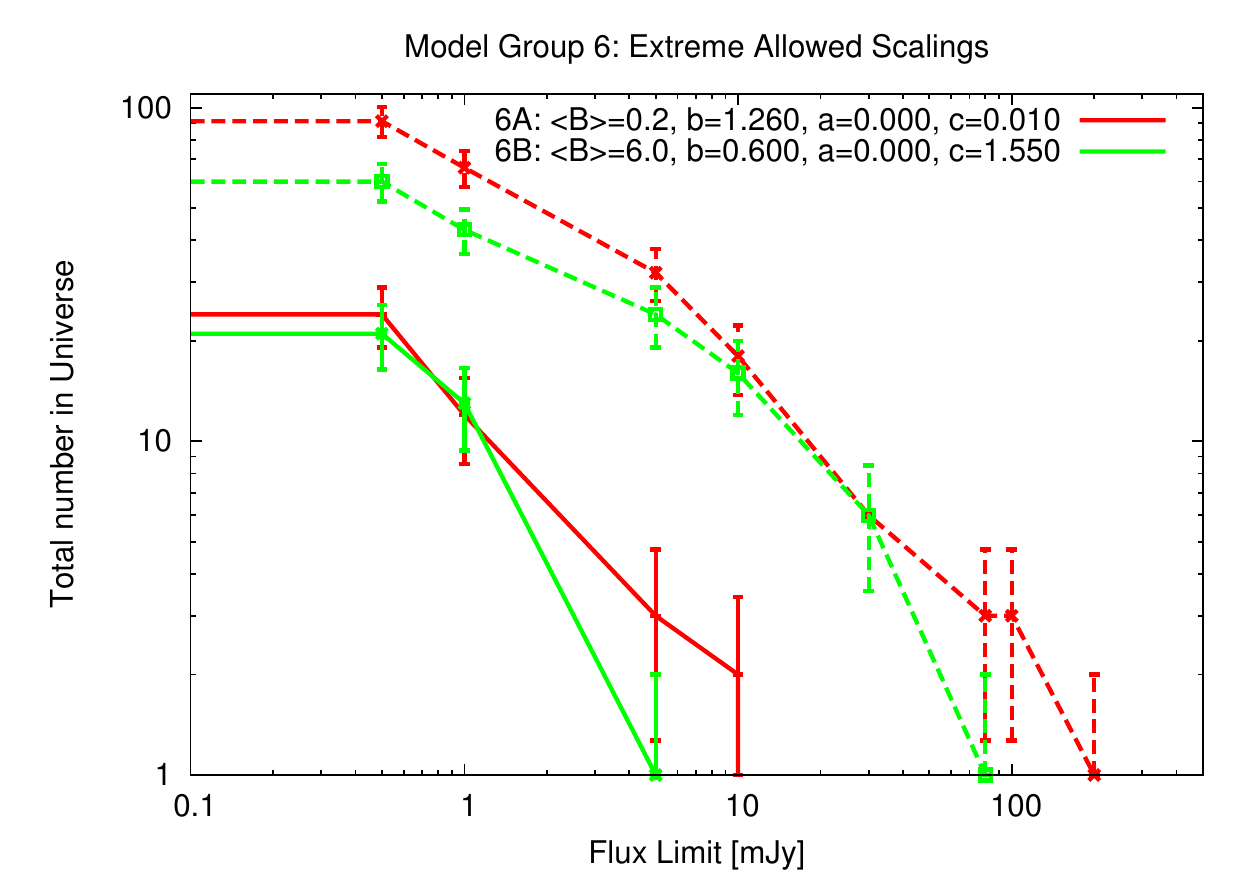}}  
  \caption[Radio halo total counts.]
           {Radio halo total counts at $1.4 \ghz$ (solid lines) 
            and $150 \mhz$ (dotted lines)
            versus flux limit in mJy.
            Error bars indicate $1\sigma$ Poisson uncertainties.
            We have identified each model with its designation 
            from~\tabref{\ref{tab:rh_rhModels}} and the portions of
            the model that change in the given model group. 
            $\aveb$ is given in units of
            ${\rm \mu G}$.
            Note that the counts given here do not include any corrections
            for small-box effects.
            } 
  \label{fig:rh_radtotalcounts} 
\end{figure*}

We begin with~\figref{\ref{fig:rh_radtotalcounts}}, where we show the total
counts of radio halos at $1.4 \ghz$ and $150 \mhz$ in the observable 
universe as a
function of flux limit in mJy. To generate error bars we propagate the
$1\sigma$ uncertainties in the derived $M_v$ - $M_{v,{\rm DM}}$ and
$\Gamma_v-M_v$ fits to generate a minimum and maximum radio power for each
cluster. Our number counts are bound by the resolvability limit of our
simulation. 

Also, we are unable to count all of the most massive clusters 
due to our limited simulation volume. 
However, we can estimate the magnitude of these effects in a simple 
way. 
By extrapolating our mass function, we estimate that we are missing $\sim$40
clusters with $M_v > 1.2 \times 10^{15} \hmsol$ in our simulation volume. 
We can use the fits to the $\pmvir$ relation to find the radio power of these
missing halos, which for all models leads 
to $P_{1.4} > 3 \times 10^{24} h_{70}^{-1} \whz$ for the missing clusters.
We assign radio halos to 30\% of these most massive clusters, which gives 
us an additional 12 halos.
If we assume that these clusters are evenly distributed within a $\sim$1~Gpc 
volume, then even the least luminous radio halo has flux $\gtrsim$ 100~mJy, 
so essentially all of these radio halos contribute to the number counts.
This increases our 1.4~GHz number counts to $\sim$12 objects above the 
10~mJy flux limit, roughly in line with known 
observations~\citep{GIOVANNINI1999, Cassano2006, Cassano2010}.
Since we expect these high-mass objects to host roughly the same proportion 
of 150~MHz radio halos, a
similar number contributes to our 150~MHz number counts.

While the model trends continue from the above analysis, we find
that at high flux limits ($>100$~mJy) and high frequencies, we have too few
radio halos to strongly distinguish several models, even those with large
discrepancies in either assumed average magnetic field or scalings with virial
mass or total turbulent pressure. This is due to the suppression of radio halos
at high redshift, meaning that the integrated counts depend most strongly on
high-luminosity objects, where the counts are nearly the same. At $150$~MHz and
an assumed LOFAR sensitivity limit of $30$~mJy, we find that although some
models, such as Model Sets 2A and 2B, produce an almost factor of two
difference in the total counts, the large uncertainties preclude any clean
distinction.

\begin{figure*}
  \centering
  {\includegraphics[width=0.49\textwidth]{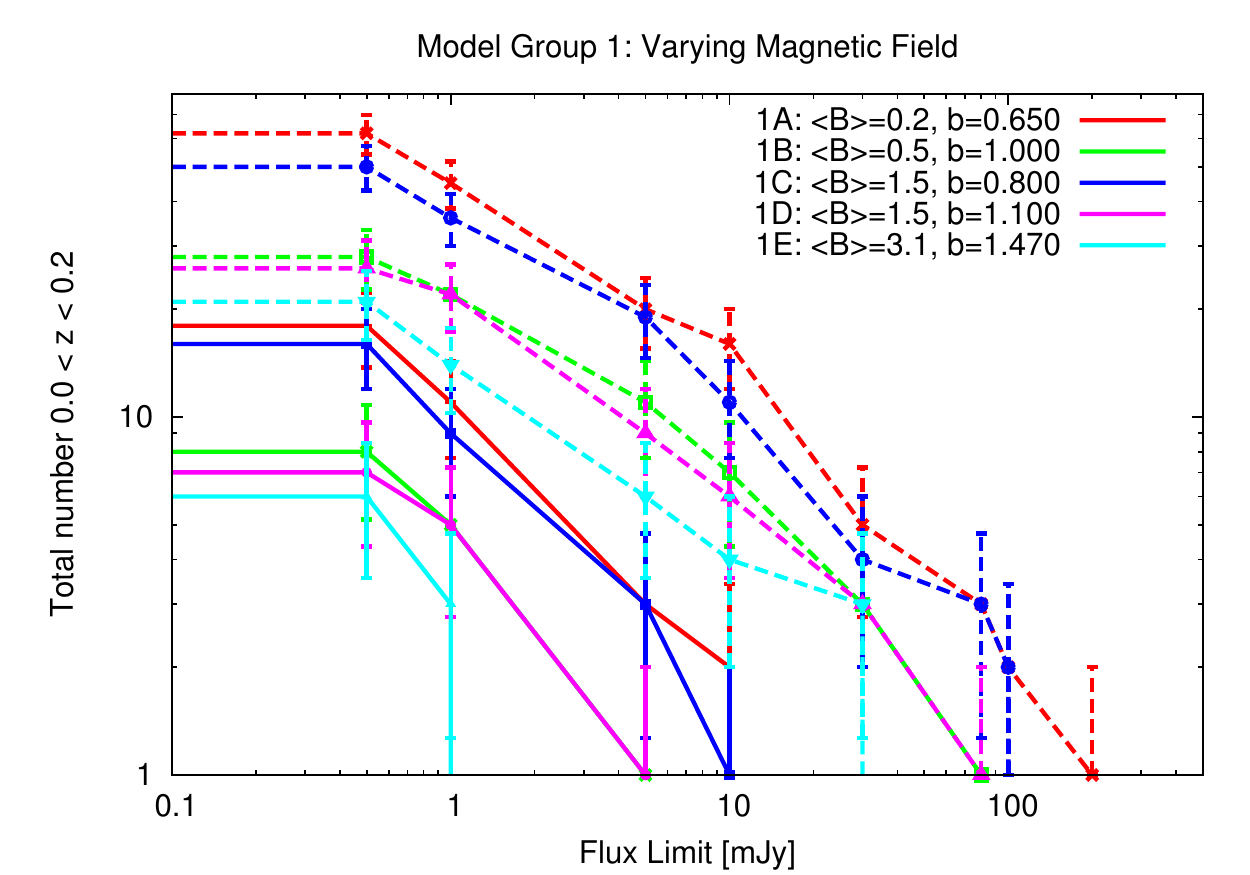}}
  {\includegraphics[width=0.49\textwidth]{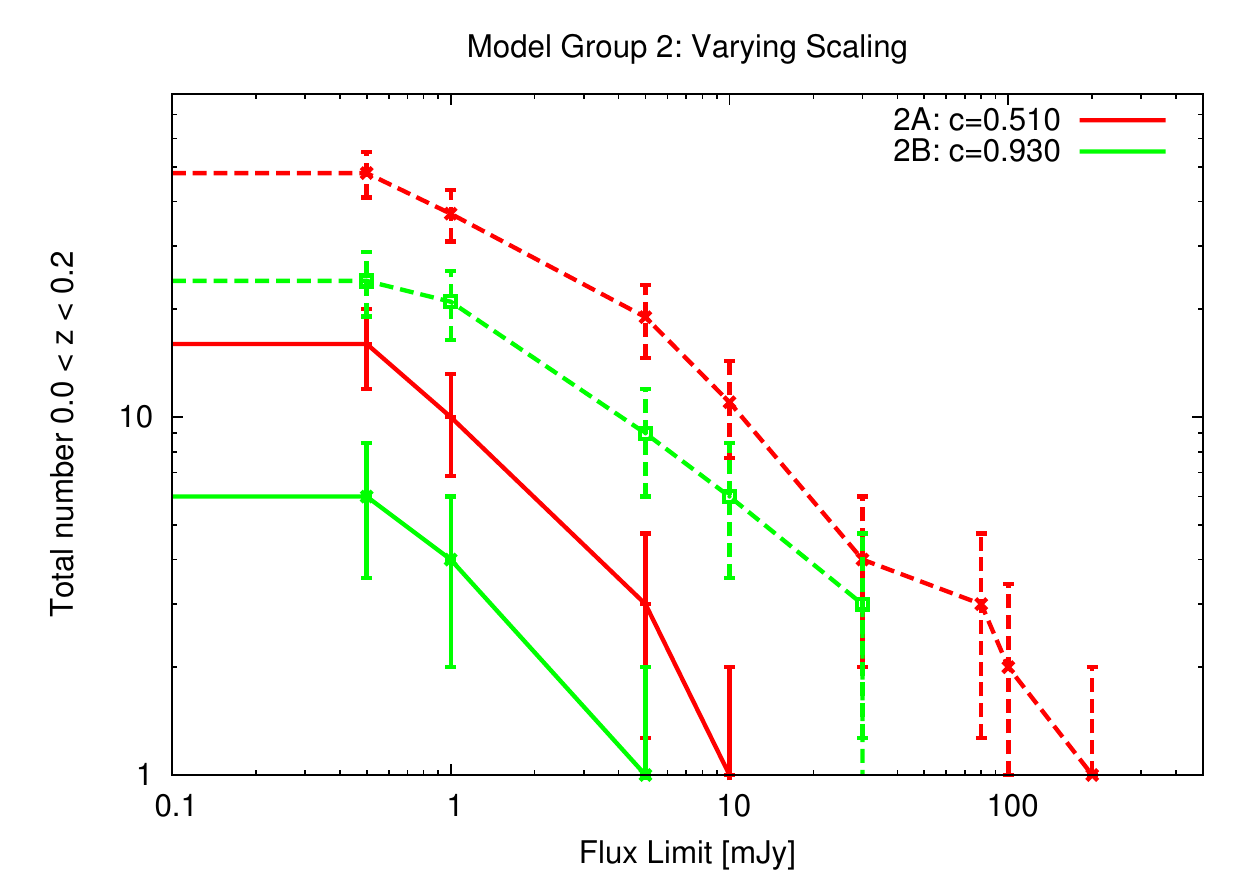}}
  {\includegraphics[width=0.49\textwidth]{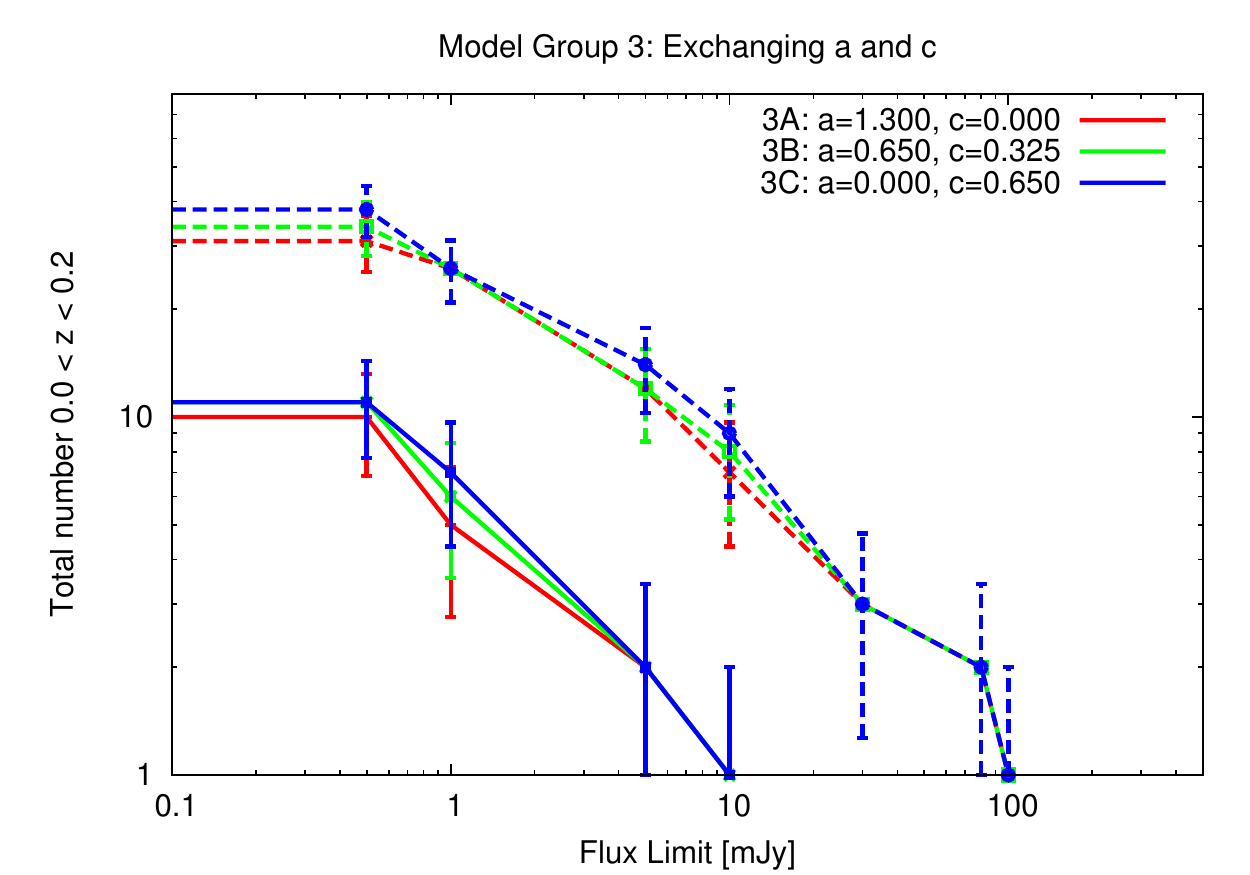}}
  {\includegraphics[width=0.49\textwidth]{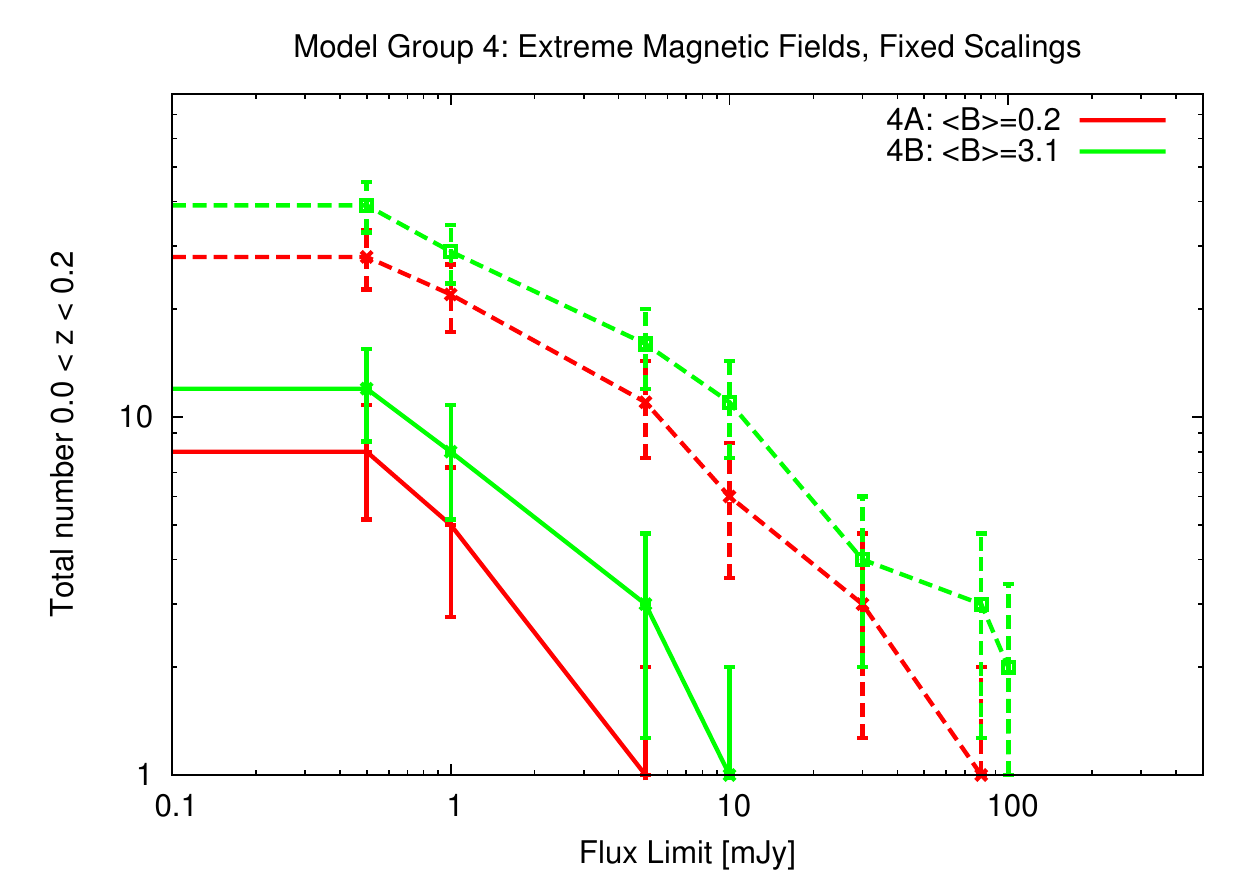}}
  {\includegraphics[width=0.49\textwidth]{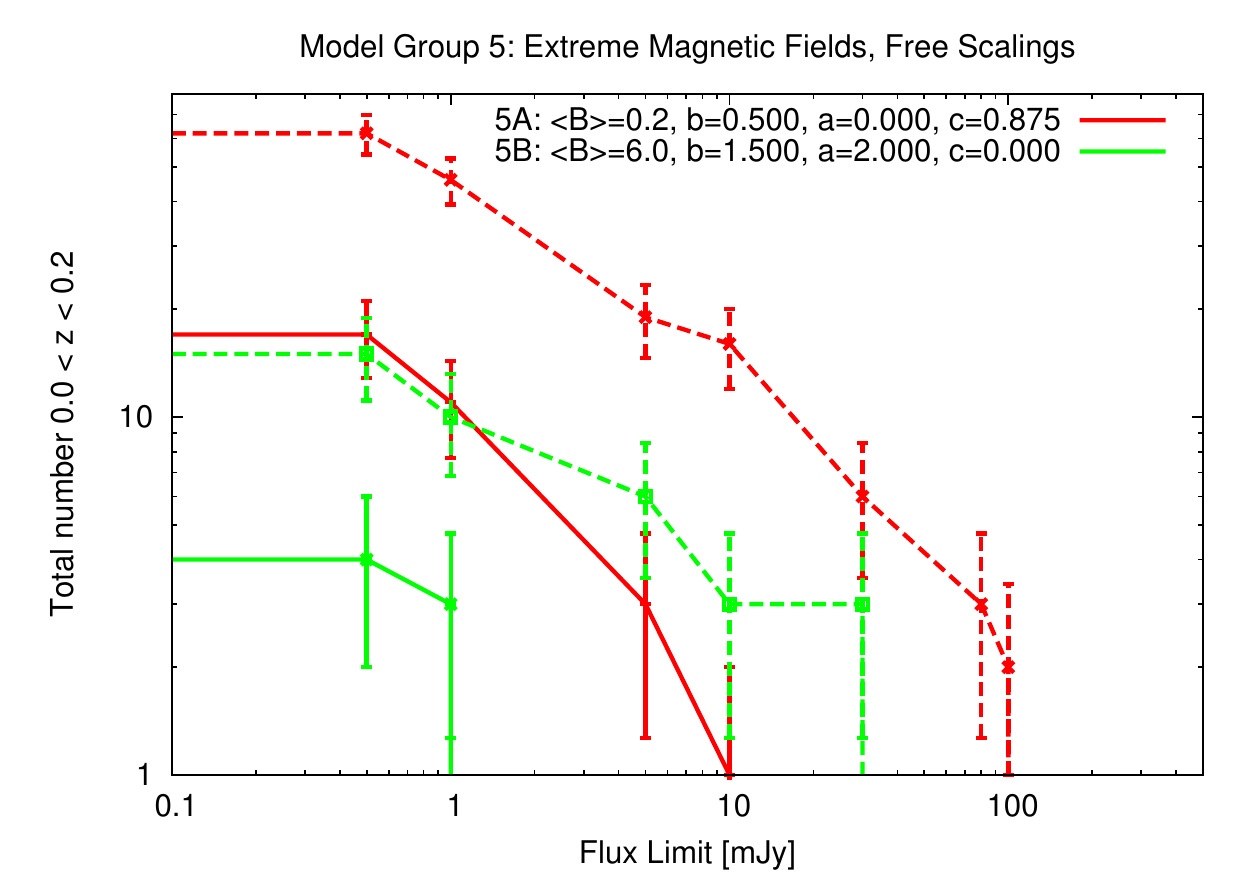}}
  {\includegraphics[width=0.49\textwidth]{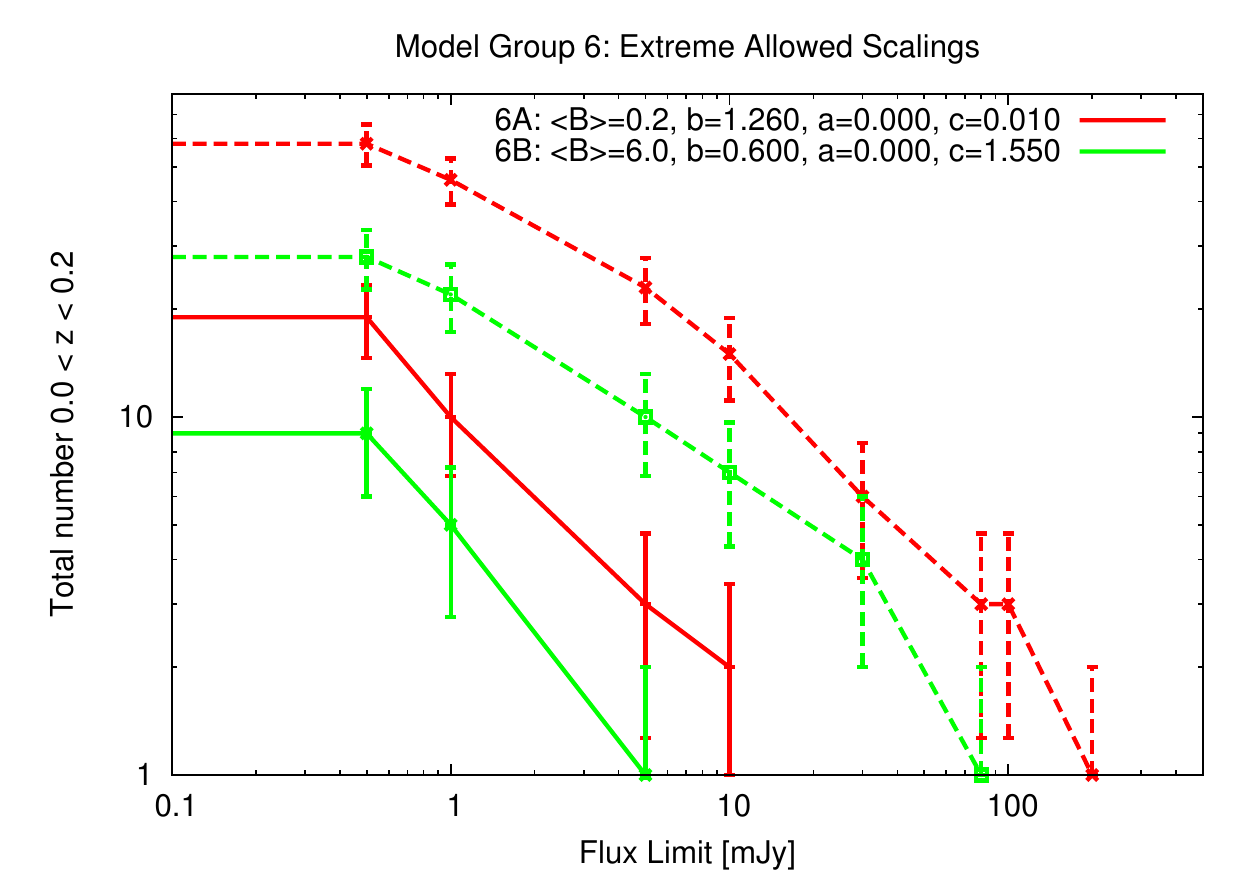}}
  \caption[Radio halo counts for $z < 0.2$.]
           {Radio halo counts for $z < 0.2$ at $1.4 \ghz$ (solid lines) 
            and $150 \mhz$ (dotted lines)
            versus flux limit in mJy.
            Error bars indicate $1\sigma$ Poisson uncertainties.
            We have identified each model with its designation 
            from~\tabref{\ref{tab:rh_rhModels}} and the portions of
            the model that change in the given model group.  
            $\aveb$ is given in units of
            ${\rm \mu G}$.
            Note that the counts given here do not include any corrections
            for small-box effects.
            } 
  \label{fig:rh_radz0.2counts}  
\end{figure*}

In~\figref{\ref{fig:rh_radz0.2counts}} we show the total 
counts of radio halos within redshift $z<0.2$ 
at $1.4 \ghz$ and $150 \mhz$. This redshift range fits largely 
within our computational volume without the need for 
periodic replication of the domain and is more easily accessible to
observers. Although we find little degradation in the total number counts in
the LOFAR-accessible regime ($>30$~mJy), the models remain indistinguishable.

\section{Simulated radio sky maps}
\label{sec:rh_skymap}

While we could in principle produce mock sky maps within any frequency range, we
choose LOFAR-like parameters since low-frequency instruments are able to survey
large portions of the sky and hence collect many halo images for use in
statistical comparison. We generate raw mock sky maps in the $20
- 240$~MHz LOFAR bandpass by following a similar strategy of interpolating and
redshift-correcting clusters as used above. 
Appropriate cosmological dimming and redshift are then applied to determine the
contribution of the slice to the sky observed at $z = 0$.  We generate a radio
image for each cluster by projecting its density and turbulent pressure onto
the sky map and computing the relevant radio intensity using a given set of
radio model parameters, ensuring that the integrated radio power across the
projected cluster is equal to the value obtained using $M_v$ and $\Gamma_v$ in
the above sections. We only project gas values within $R_v$.  For halos not
within the high-resolution sample, we identify the nearest high-resolution
cluster in mass and copy that high-resolution image to the location of the
low-resolution halo. Also, since we do not have imaging information for 
missing high-mass halos due to our limited simulation volume, 
these are not included in the mock skies. 
While this procedure is admittedly somewhat crude, it
does allow us to explore some of the observational consequences of these models
and demonstrates a method of generating radio maps in the future using more
sophisticated and realistic simulated data.  

\begin{figure*}
  \centering \includegraphics[width=\textwidth]{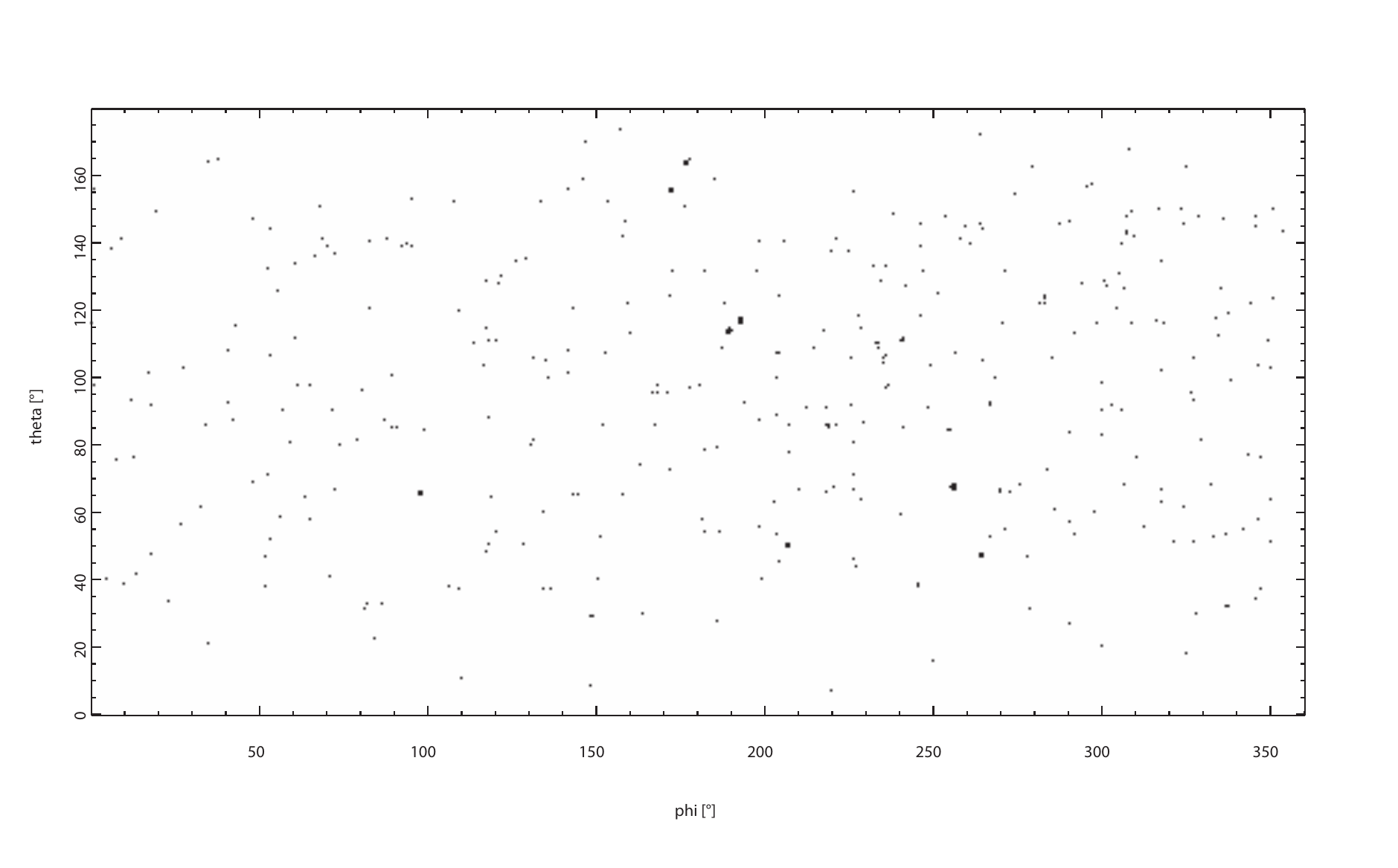}
  \caption[Example radio halo all-sky map.]
           {Example radio halo all-sky map. This map assumes $0$~mJy 
           sensitivity threshold and a resolution of $120$~arcsec.}
\label{fig:rh_catmapGlobal}
\end{figure*}

\figref{\ref{fig:rh_catmapGlobal}} shows the entire radio sky containing our
simulated clusters at $120$~arcsec resolution assuming no background (i.e.,
a threshold sensitivity of $0$~mJy). This resolution best approximates the
LOFAR beam at an average frequency of $\sim 120$~MHz and a longest baseline of
$L \sim 2$~km.  For this example we have chosen Model Set A1.  This
map particularly highlights the paucity of radio halos in the universe, even at
low sensitivity thresholds, but it is useful for providing a mock all-sky map
for linking simulations to observations. 

\begin{figure}
  \centering {\includegraphics[width=\columnwidth]{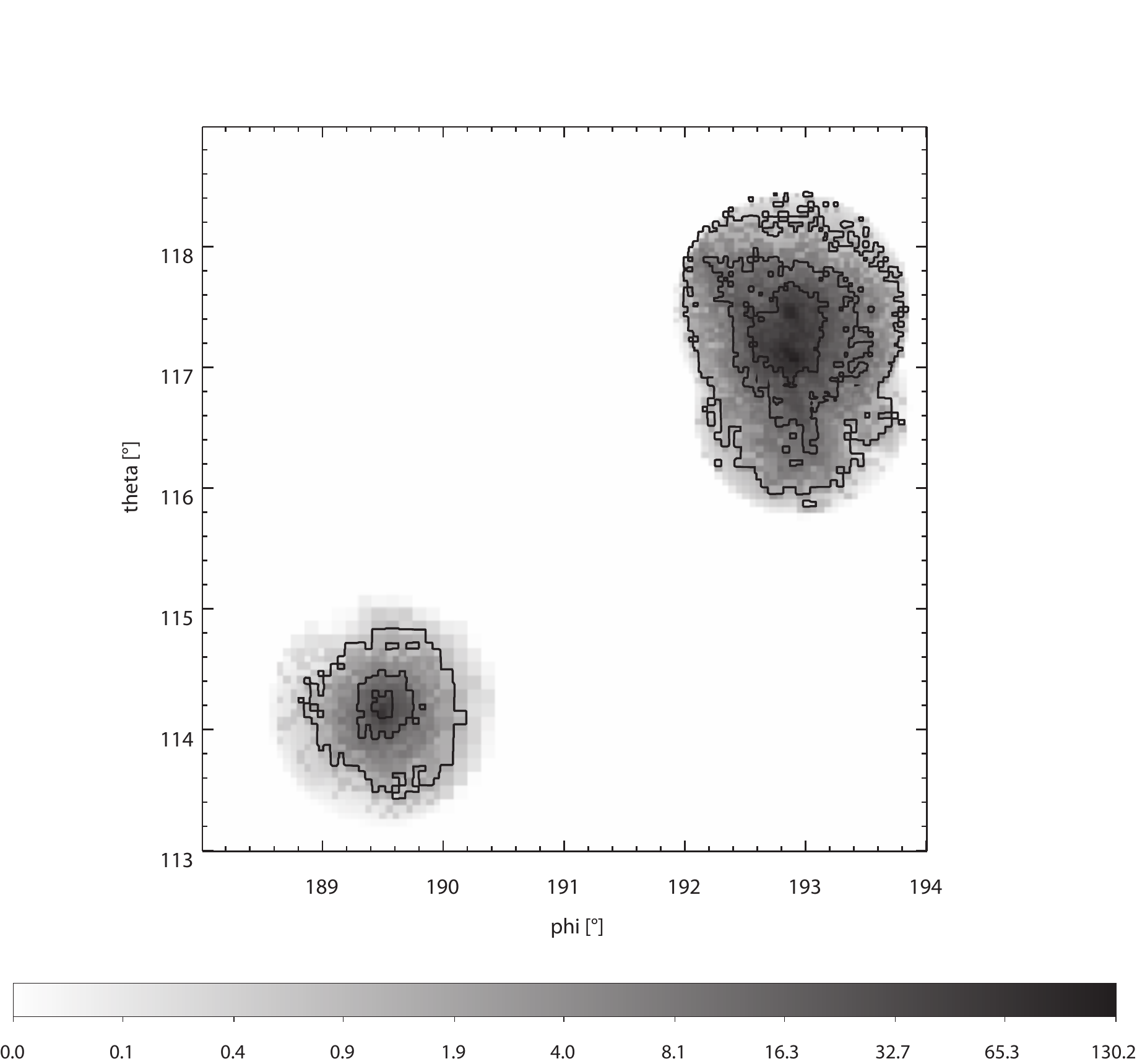}}
  \caption[Example radio halo partial-sky maps at various resolutions.]
           {Example radio halo partial-sky map at $10$~arcsec resolution.
            The color scale is the logarithm of radio power in mJy.
            Contours are drawn at $1$, $10$, and $30$~mJy levels.}
\label{fig:rh_catmapPatchRes}
\end{figure}

\figref{\ref{fig:rh_catmapPatchRes}} highlights a region of the sky $6$~degrees
on a side at a resolution of $10$~arcsec, representing  the high-resolution
capability between $20$ and $240$~MHz at the longest baseline configuration of
LOFAR.  We also draw contour levels at varying sensitivities: $1$, $10$, and
$30$~mJy.  These sensitivities represent different configurations of the LOFAR
array. At high resolution and peak sensitivity, we are able to clearly
distinguish several substructures and features within the two radio halos,
indicating that LOFAR may be able to cleanly distinguish various radio power
models based on their dependence on local gas density or local turbulent
pressure, which can have different characteristic structures in the cluster
atmosphere (Figure~\ref{fig:rh_projgamma}).  At lower sensitivities, we can
still distinguish features in the cluster cores, and early LOFAR images of
nearby and bright radio halos may also provide useful distinguishing results.
We will present a detailed radio morphological study, which requires knowledge
of the spatial dependence of the magnetic field, in a future paper.

\begin{figure}
  \centering
{\includegraphics[width=\columnwidth]{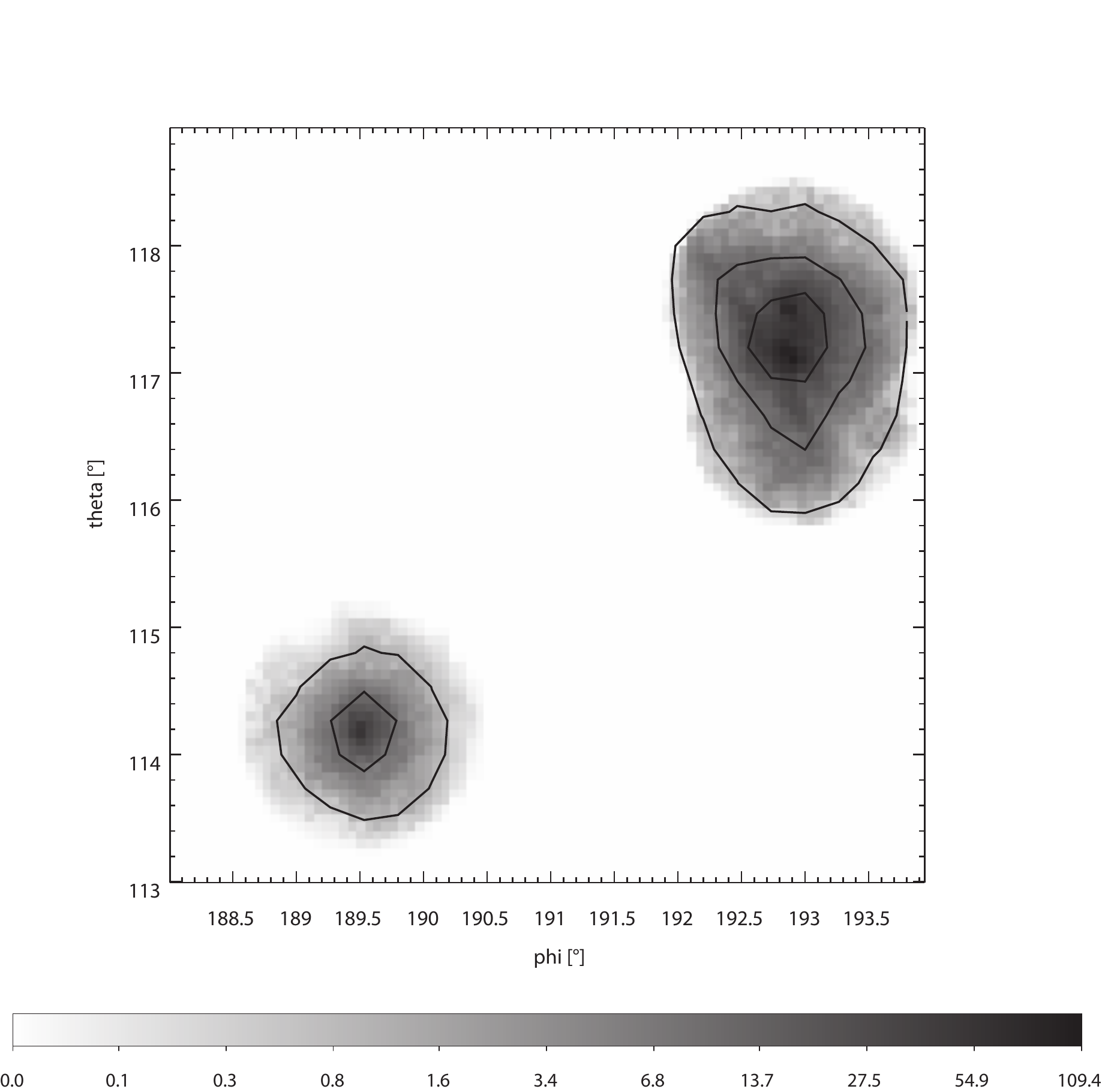}}
  \caption[Example radio halo partial-sky map at various sensitivities.]
           {Example radio halo partial-sky maps at $240$~arcsec resolution.
                        The color scale is the logarithm of radio power in mJy.
            Contours are drawn at $1$, $10$, and $30$~mJy levels}
\label{fig:rh_catmapPatchSens}
\end{figure}

\figref{\ref{fig:rh_catmapPatchSens}} shows the same region of the sky as above
with a much lower resolution of $240$~arcsec.  The contours are the same
as above.  While we lose significant information about distant and small
clusters, some larger clusters, such as the one shown, still show significant
structure even at lower resolutions.  We see that we can still identify
substructure within the large cluster, and the effects of higher sensitivity
thresholds are limited to distant clusters and the outer regions of nearby
objects. These results are encouraging, since they indicate that LOFAR may be
able to give detailed radio maps of many radio halos.

\section{Conclusions}
\label{sec:rh_conclusions}

We have introduced the first set of radio halo statistics derived entirely from
large-scale cosmological simulation.  Our radio power model is sufficiently
broad to encompass many viable and more realistic models of CR
generation and synchrotron emission in clusters of galaxies.  Our approach
demonstrates the viability of using large-scale simulation to bridge
simulations and observations, both by deriving radio halo statistics from the
simulated data to constrain possible radio power models and by producing mock
radio sky maps that can be directly compared to observations.

From our analysis we have determined that the slope and normalization of the
$\pmvir$ and $P_{1.4} - L_x$ relations are potentially key probes of the
various models of CR generation. They also allow us to place limits on
the average cluster magnetic field strength and the scaling of magnetic fields
with cluster mass. With the uncertainties from only $131$ observed radio halos
we can significantly constrain the scaling of radio power with cluster mass and
turbulent pressure. With the $131$ objects of our high-resolution sample we are
able to clearly separate some models with strong statistical significance. The
evolution with redshift of these relations also allows us to potentially
distinguish various models. Future low-frequency missions, which will surveys
large portions of the sky, may then potentially capture enough objects to
perform a similar analysis and obtain these constraints.

We predict an order of magnitude fewer
high-frequency radio halos at low mass than the analysis of CBS06
and~\citet{Cassano2010b}.  
Some of this discrepancy might be due to our lack of
steep-spectrum halos, which get counted via inclusion of the calculation of the
synchrotron break frequency, $\nu_b$.  Instead, we just assign a 
radio halo to $5\%$ of our clusters.  
We found that adjusting the spectral index to
$1.9$ (i.e., the average spectral index found by~\citealt{Cassano2010b}) only
increased the number counts by roughly $50 \%$, which is not nearly enough to
explain the differences, 
a higher probability of hosting a radio halo at 150~MHz as predicted 
in reacceleration 
models could explain the differences.
Also, since our limited simulation volume precludes us from counting all of the 
most massive clusters, we will systematically underestimate our total number 
counts at both 1.4~GHz and 150~MHz. 
However, our estimates in Section~\ref{sec:rh_counts} suggest that 
these under-counting effects only account for roughly $10 \%$ 
of the discrepancy in estimates of low-luminosity 150~MHz radio halos.
While the combined effects of the lack of steep-spectrum halos and our 
limited simulation volume may together be enough to explain the differences in number count predictions between our analysis and that of CBS06, 
this issue must be more carefully examined in subsequent work.

We have uncovered many degeneracies among the scalings of radio power with
cluster mass and turbulent pressure and the mass-dependence of cluster magnetic
fields.  These degeneracies can be broken by several methods. For example, a
better understanding of the relationship between cluster mass and magnetic
field will constrain our $\aveb$ and $b$ parameters, allowing us to make more
conclusive statements about the observational limits placed upon the $a$ and
$c$ parameters.  On the other hand, more high-resolution radio and X-ray images
of clusters may constrain the effectiveness of the various mechanisms of
generating CRs, which would further constrain our $a$ and $c$
parameters. In particular, measurements of intracluster 
gas velocities (such as may be possible with future X-ray 
spectroscopic missions) would enable mapping of the projected 
turbulent pressure, which could then be compared with the 
projected mass distribution.

We find that low-frequency surveys are potentially capable of distinguishing
and constraining the scaling of radio power with cluster mass and turbulent
pressure, since the radio halo number counts are much higher at lower
frequencies. Even though low-frequency surveys have relatively low
sensitivity, their ability to map large portions of the sky to moderately high
redshift means that they can gather many more objects than high-frequency (and
more sensitive) observations. However, our estimates indicate that LOFAR will
only see on the order of 10 radio halos within our studied mass range, 
and that future missions with more sensitivity
will be required in order to cleanly distinguish models. Future radio
observations are especially important since the gamma ray emission associated
with the production of CRs from hadronic secondary interactions from
clusters might be too small for Fermi to detect, which means that we may not be
able to use this instrument to distinguish models~\citep{Brunetti2009a}.  

Similarly, since hadronic secondary models of CR production are highly
degenerate with reacceleration models when only considering total counts and
integrated cluster quantities, high-resolution low-frequency radio images are
required in order to effectively distinguish these models.  Although
high-frequency observations can also (and do) produce similar maps, surveys
such as LOFAR have the unique capability of capturing many such images,
potentially providing a statistically significant means of distinguishing
models based on morphological differences. Our simulated radio cluster and sky
maps are freely available upon request to the authors or via the project
Web site \footnote{\url{http://sipapu.astro.illinois.edu/}} under the section
~\url{Projects/RadioHaloMaps}.  We have produced images at a variety of sky
coverage areas, sensitivity limits, and resolutions for all of the models
described above. These images are simple \texttt{FITS} files. The images are
straightforward to produce, allowing us to explore further refinements to the
models and more sophisticated instrument modeling.

Since this initial work is highly preliminary, we have room for many
improvements and modifications to make stronger connections with observations.
As an immediate improvement we may perform simulations with larger volumes than
our $1 \hgpc$ box (or, equivalently, perform multiple realizations of the same
volume) in order to capture more massive objects.  With larger volumes we can
also capture more low- and moderate-mass objects to obtain better statistics
for the $\pmvir$ and other relations. These simulations will allow us employ
more sophisticated models of CR generation and evolution as well as
enable us to calculate statistics of radio halo morphologies. With more
simulations, we may begin to investigate the dependence of radio halo counts on
cosmological parameters.  We may also begin to self-consistently include
magnetic fields, although our results in this approach would be tied to a
specific model of magnetic field injection and growth. Similarly, we can begin
to investigate generating and propagating CRs in the simulation,
although it is difficult to scale current methods to large volumes and high
resolutions. Since the expected low-luminosity radio halo number counts are so
sensitive to the level of turbulence in cluster atmospheres, we must
incorporate more careful techniques for estimating this. 
A future crucial test of various models is the shape of the 
spectrum~\citep{Cassano2006,Brunetti2008}, and future simulations must 
be able to reliably predict such spectra.
However, our results
demonstrate an important first step in bridging simulations and observations to
more fully understand the large-scale radio universe. 

\section*{Acknowledgments} 

The authors acknowledge support under a Presidential Early Career Award from
the U.S. Department of Energy, Lawrence Livermore National Laboratory (contract
B532720).  Additional support was provided by a DOE Computational Science
Graduate Fellowship (DE-FG02-97ER25308) and the National Center for
Supercomputing Applications.  The software used in this work was in part
developed by the DOE-supported ASC / Alliance Center for Astrophysical
Thermonuclear Flashes at the University of Chicago.  This research used
resources of the National Center for Computational Sciences at Oak Ridge
National Laboratory under the AST019 Director's Discretionary allocation. The
computing resources used are supported by the Office of Science of the US
Department of Energy under Contract No.\ DE-AC05-00OR22725. 

\bibliography{ms}		
\bibliographystyle{apj}	
\nocite{*}

\end{document}